\documentclass[authoryear,preprint,review,12pt]{elsarticle}

\usepackage{amssymb}

\def\spose#1{\hbox to 0pt{#1\hss}}

\def\deg{\ifmmode^\circ\else$\null^\circ$\fi}
\def\lta{\mathrel{\spose{\lower 3pt\hbox{$\mathchar "218$}}\raise 2.0pt\hbox{$\mathchar"13C$}}}
\def\gta{\mathrel{\spose{\lower 3pt\hbox{$\mathchar "218$}}\raise 2.0pt\hbox{$\mathchar"13E$}}}
\def\scinot#1.{\hbox{$\, \times \, 10^{#1}$}}
\def\CHfour{{\fam0 CH_4}}
\def\CHthree{{\fam0 CH_3}}

\def\CH{{\fam0 CH}}
\def\COtwo{{\fam0 CO_2}}
\def\CO{{\fam0 CO}}
\def\CtwoHtwo{{\fam0 C_2H_2}}

\def\C{{\fam0 C}}
\def\threeCHtwo{{\fam0 ^{3}CH_2}}

\def\HCO{{\fam0 HCO}}
\def\HtwoCO{{\fam0 H_2CO}}
\def\CtwoHtwoOH{{\fam0 C_2H_2OH}}
\def\HtwoO{{\fam0 H_2O}}
\def\Htwo{{\fam0 H_2}}
\def\Hatom{{\fam0 H}}
\def\M{{\fam0 M}}
\def\Net{{\fam0 Net:}}
\def\OH{{\fam0 OH}}

\def\Oatom{{\fam0 O}}

\journal{Icarus}

\begin{document}

\begin{frontmatter}



\title{Dust Ablation on the Giant Planets: Consequences for Stratospheric Photochemistry}


\author[label1]{Julianne I.~Moses}
\author[label2]{Andrew R.~Poppe}

\address[label1]{Space Science Institute, 4750 Walnut Street, Suite 205, Boulder, CO 80301, USA}

\address[label2]{Space Sciences Laboratory, 7 Gauss Way, University of California, Berkeley, CA 94720, USA}

\begin{abstract}
Ablation of interplanetary dust supplies oxygen to the upper atmospheres of 
Jupiter, Saturn, Uranus, and Neptune. Using recent dynamical model predictions for the dust influx rates 
to the giant planets (Poppe, A.R.~et al.~[2016], Icarus 264, 369), we calculate the ablation profiles 
and investigate the subsequent coupled oxygen-hydrocarbon neutral photochemistry in the stratospheres of 
these planets. We find that dust grains from the Edgeworth-Kuiper Belt, Jupiter-family comets, and Oort-cloud 
comets supply an effective oxygen influx rate of 1.0$^{+2.2}_{-0.7} \scinot7.$ O atoms cm$^{-2}$ s$^{-1}$ to 
Jupiter, 7.4$^{+16}_{-5.1} \scinot4.$ cm$^{-2}$ s$^{-1}$ to Saturn, 8.9$^{+19}_{-6.1} \scinot4.$ cm$^{-2}$ 
s$^{-1}$ to Uranus, and 7.5$^{+16}_{-5.1} \scinot5.$ cm$^{-2}$ s$^{-1}$ to Neptune. The fate of the ablated 
oxygen depends in part on the molecular/atomic form of the initially delivered products, and on the 
altitude at which it was deposited. The dominant stratospheric products are CO, H$_2$O, and CO$_2$, 
which are relatively stable photochemically. Model-data comparisons suggest that interplanetary dust 
grains deliver an important component of the external oxygen to Jupiter and Uranus but fall far short of 
the amount needed to explain the CO abundance currently seen in the middle stratospheres of Saturn and 
Neptune. Our results are consistent with the theory that all of the giant planets have experienced large 
cometary impacts within the last few hundred years. Our results also suggest that the low background H$_2$O 
abundance in Jupiter's stratosphere is indicative of effective conversion of meteoric oxygen to CO during 
or immediately after the ablation process ---  photochemistry alone cannot efficiently convert the H$_2$O 
into CO on the giant planets.
\end{abstract}


\end{frontmatter}


\section{Introduction}\label{sec:intro}

Small interplanetary dust grains are continually showering down into the atmospheres of solar-system planets.  
This dust originates from the disruption and outgassing of comets, from impacts and collisions between 
objects of various sizes (particularly from mutual collisions within the asteroid belt and Edgeworth-Kuiper 
belt), from particles ejected from active plumes on satellites such as Io, Enceladus, and Triton, and from 
interstellar dust particles streaming into the solar system.  Beyond Jupiter's orbit, the main progenitors 
are the Edgeworth-Kuiper belt, long-period Oort-cloud comets, and short-period Jupiter-family and Halley-type comets 
\citep[e.g.,][]{stern96,yamamoto98,landgraf02,poppe15,poppe16}, 
and the particles likely contain ices, along with silicate and organic material.  As the dust grains spiral 
in through the outer solar system, they are affected by gravity from the Sun and planets, solar wind and 
Poynting-Robertson drag, stellar radiation pressure, and collisions \citep[e.g.,][]{burns79,gustafson94,horanyi96,liou97}.

\citet{poppe16} recently developed a comprehensive model for the dynamical evolution and density distribution 
of dust grains in the outer solar system.  The model considers the dominant interplanetary dust sources for the 
outer solar system described above, and includes the relevant physics for the dynamical and collisional evolution 
of the grains.  \textit{In situ} spacecraft measurements from the \textit{New Horizons} Student Dust Counter, the 
\textit{Galileo} Dust Detection System, and the \textit{Pioneer 10} meteoroid detector are used to constrain 
the model.  One important byproduct of the \citet{poppe16} model is a prediction of the total mass influx 
rate of dust grains to the giant planets.

Gravitational focusing by the planets will cause the incoming dust particles to enter the upper atmospheres 
of Jupiter, Saturn, Uranus, and Neptune at high velocities \citep{poppe16}, leading to full or partial ablation 
of the grains \citep{moses92abl,moses97,moses01,pryor94,moses00b,kim01}.  This ablation introduces gas-phase metals and 
water to the thermospheres and stratospheres of these planets; such species would otherwise not be present in the
upper atmosphere because of condensation and sequestering of the intrinsic water and metals in the deeper 
troposphere.  
The unablated or recondensed refractory component provides a source of high-altitude haze and condensation nuclei 
that can facilitate condensation of stratospheric hydrocarbons \citep{moses92nucl} and  
can alter atmospheric radiative and scattering properties \citep[e.g.,][]{rizk90,pryor94,moses95c}.
The ablated metals and water can affect the chemistry and structure of the ionosphere 
\citep[e.g.,][]{connerney84,connerney86,majeed91,cravens94,lyons95,mosesbass00,kim01,grebow02,moore04,molinacuberos08}, 
while water and the other oxygen species can affect the neutral photochemistry and aerosol structure in the 
stratosphere \citep[e.g.,][]{moses92abl,moses00b,moses05,ollivier00}.  

Sublimation from H$_2$O, CO, and CO$_2$ ices in the grains as they are heated during atmospheric entry 
releases these molecules directly into the atmosphere, while thermochemical reactions within the meteor trail, 
energetic collisions with atmospheric molecules, or subsequent photochemical 
interactions within the stratosphere can further process the oxygen-bearing component.
For example, the ablated water can be photolyzed by ultraviolet 
radiation from the Sun to produce hydroxyl radicals (OH), which can react with methane photochemical products 
to produce CO \citep[e.g.,][]{prather78,strobel79,moses00b,moses05,ollivier00}, potentially diminishing the abundance 
of unsaturated hydrocarbon molecules such as C$_2$H$_2$ and C$_2$H$_4$ in the process \citep[e.g.,][]{moses00b}.  
The H$_2$O introduced from the icy component of the grains will condense at relatively high altitudes 
on all the giant planets, affecting the stratospheric aerosol 
structure and properties, while the CO$_2$ will condense on colder Uranus and Neptune.  

Although CO has been observed in giant-planet stratospheres and is a major end product of the 
chemistry of the ablated vapor (see sections 3.2-3.6), it is the most volatile of the major oxygen-bearing species 
on the giant planets and 
is not expected to condense.   Stratospheric CO has additional potential sources, both external and internal to 
the giant planets, such as large cometary impacts and/or thermochemical quenching and convective transport from the 
deep troposphere
\citep{prinn77,fegley94,lodders02,bezard02,lellouch02,lellouch05,lellouch06,lellouch10,visscher05,visscher10co,hesman07,cavalie08co,cavalie09,cavalie10,cavalie13,cavalie14,cavalie17,luszczcook13,wang15,wang16};
water and carbon dioxide can also be delivered from cometary impacts \citep[e.g.,][]{lellouch96,lellouch02}.
Accurately predicting the fate of the oxygen from interplanetary dust sources therefore has important implications for the 
bulk elemental oxygen abundance on the giant planets, the strength of convective mixing from the deep atmosphere, and 
the impact rates of large comets in the outer solar system.  These implications, combined with the recent improved 
predictions for the incoming dust fluxes to Jupiter, Saturn, Uranus, and Neptune \citep{poppe16} and new constraints 
on the abundance of stratospheric oxygen species from \textit{Spitzer} and \textit{Herschel} observations
\citep{meadows08,lellouch10,fletcher12spire,cavalie13,cavalie14,orton14temp,orton14chem}, motivate us to  
theoretically track the fate of the volatiles released from the ablation of interplanetary dust on the giant 
planets.

To determine how the dust-delivered oxygen affects stratospheric photochemistry on the outer planets, we first 
run an ablation code \citep[see][]{moses92abl,moses97} with the interplanetary dust fluxes, mass distributions, and 
velocity distributions from the \citet{poppe16} dynamical model as input.  After 
making assumptions about the bulk composition of the grains based on cometary dust and nucleus compositions 
\citep{greenberg99,lisse06,lisse07}, we then use the ablation model to calculate the mass loss and vapor 
release as a function of altitude from the incoming grains (see section~\ref{sec:ablresults}).  
The resulting gas production rate 
profiles from the ablation process are then included as a source of oxygen species to stratospheric photochemical 
models \cite[e.g.,][]{moses00b,moses05,moses15} that consider coupled hydrocarbon-oxygen chemistry (see sections 
3.2-3.5).  In sections 3.2-3.5, we compare the photochemical model results with observations and discuss the implications 
with respect to the origin of the observed oxygen species on each planet, and in section \ref{sec:thermo} we discuss 
the likely importance of thermochemistry and high-energy collisions during the meteor phase in securing the high 
inferred CO/H$_2$O ratio in the stratospheres of these planets.

\section{Theoretical model description}\label{sec:model}

Two main theoretical models are used for these calculations.  The first is the meteoroid ablation code 
described in \citet{moses92abl}, with updates from \citet{moses97}.  The second is the Caltech/JPL 
one-dimensional (1D) KINETICS photochemical model developed by Yuk Yung and Mark Allen 
\citep[e.g.,][]{allen81,yung84}, most recently updated for the giant planets by \citet{moses15}. 

The physics of meteoroid ablation has been understood for decades, at least in an idealized sense 
\citep[e.g.,][]{opik58}.  The interplanetary dust grains being considered here are typically much 
smaller than the mean free path of the atmosphere 
in the region in which they ablate, which puts the physics in the free-molecular-flow regime.  Under such 
conditions, the incoming grains collide directly with individual air molecules, leading to deceleration
and heating of the grains.  The heating is offset by radiative and evaporative cooling and by the change in 
internal energy of the grains.  As in \citet{moses92abl} and \citet{moses97}, we assume the incoming 
grains are solid compact spheres (and remain spherical throughout their flight), have a uniform composition, 
are heated uniformly throughout their volume, and are not affected by sputtering, fragmentation, or thermal 
diffusion within the solid.  For more sophisticated treatments, see \citet{vondrak08}.

The physics in this case is reduced to a set of four coupled differential equations that track the evolution of an 
incoming dust grain's mass, velocity, temperature, and position within the atmosphere \citep[see equations (1)-(4)
in][]{moses92abl}.  We assume that the entry angle is 45\deg\ and remains constant throughout the 
particle's flight.  The grains are assumed to be composed of either pure water ice, ``silicates'', or ``organics'', 
with the incoming dust mass flux being divided such that 26\% of the grains are silicate, 32\% are refractory organic, 
and 42\% are ices, based roughly on their corresponding mass fractions within cometary nuclei and dust 
\citep{greenberg99}.  The water ice and silicate material properties are taken from \citet{moses92abl} (and 
references therein), except for the vapor pressures, which follow the recommendations of \citet{moses97}.  
For the organic grains, which 
were not considered by \citet{moses92abl,moses97}, we take the material properties somewhat arbitrarily from 
benzo(a)pyrene (C$_{20}$H$_{12}$), as a refractory organic that has a vaporization temperature in the 
appropriate 400-600 K range.  The vapor pressure of the representative organic material is log$_{10}\ p(atm) = 
9.110 - 7100/T(K)$ \citep{murray74}, with an assumed bulk density of 1.24 g cm$^{-3}$, an assumed latent 
heat of sublimation of 118 kJ/mol, and an assumed specific heat of 254.8 J mol$^{-1}$ K$^{-1}$ \citep{roux08}.
For the purposes of converting mass loss into the amount of organic vapor ``molecules'' injected into the 
atmosphere, we assume a mean molecular mass of 77 amu for the organic vapor (e.g., a single C$_6$H$_5$ 
organic ring), although the exact choice is unimportant, as the organic vapor is ignored in the subsequent 
photochemical calculations because the carbon released is a small fraction of the carbon already in the 
atmosphere.

Grains smaller than the wavelength of the peak emission in the Planck blackbody function do not radiate their 
heat efficiently (i.e., they have emissivities less than 1).  If the grain materials are relatively transparent 
at infrared wavelengths, with a low imaginary refractive index, emissivities can drop much lower than unity, 
increasing the overall ablation rate \citep[e.g.,][]{rizk91}.  We therefore calculate the emissivity from 
the absorption efficiency determined from Mie theory at the wavelength of maximum emission (Wien's law) 
for the particle's temperature and size at each time step in the calculations.  The optical properties for 
the representative silicate material are taken from the ``olivine with iron'' case of \citet{rizk91}, the 
water-ice values are from \citet{warren84}, and the ``organic'' values are from \citet{ligreen97}.

\begin{figure*}[!htb]
\begin{center}
\includegraphics[clip=t,width=5.4in]{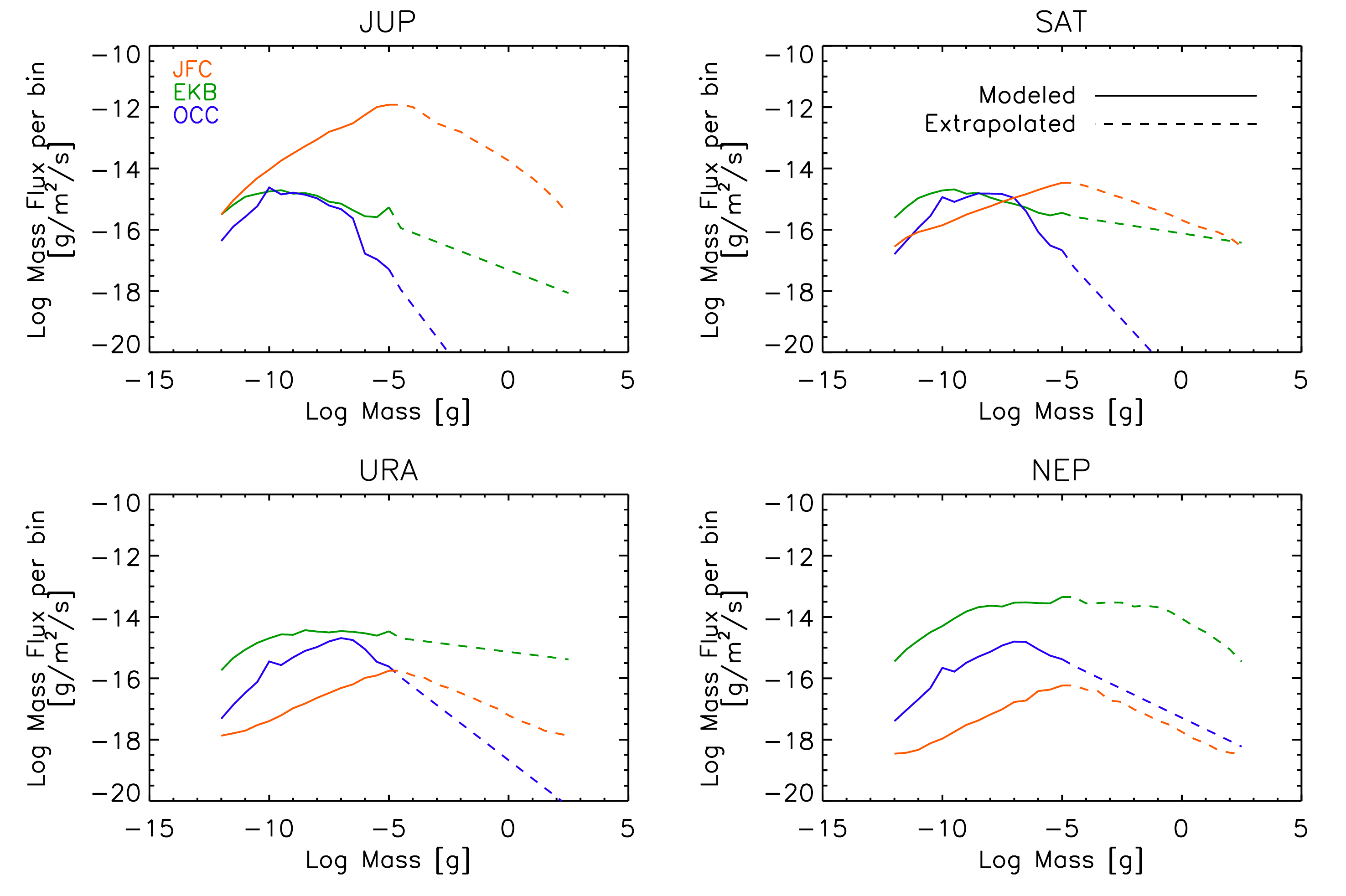}
\end{center}
\vspace{-0.5cm}
\caption{Total particle mass flux (g m$^{-2}$ s$^{-1}$) in each mass bin encountering Jupiter (Top left), 
Saturn (Top right), Uranus (Bottom left), and Neptune (Bottom right) at the planet's exobase, for dust 
populations from Jupiter-family comets (orange), the Edgeworth-Kuiper belt (green), and Oort-Cloud comets (blue).  
Particles were grouped into bins based on mass, with 2 bins per decade of mass.
The solid curves are from \citet{poppe16}, and the dashed curves were extrapolated as described in the text.
(For interpretation of the references to color in this figure legend, the reader is referred to the web 
version of this article.)\label{figmassflux}}
\end{figure*}

The initial mass and velocity distributions of the incoming grains are taken from \citet{poppe16}.  The 
velocities vary with the particle size, source population, and the planet in question.  Gravitational focusing 
of the interplanetary dust velocity distribution was appropriately taken into account by dynamically tracing 
the dust-grain trajectories from the Hill radius to the planetary exobase and recomputing the dust velocity 
distribution immediately before entry into the planetary atmosphere.  The particle mass flux distributions 
for the different dominant populations --- Edgeworth-Kuiper belt grains, Jupiter-family comet grains, 
and Oort-cloud comet grains --- encountering the different planets are shown in Figs.~\ref{figmassflux}, 
\ref{figcontmassflux}, \& \ref{contmassfluxnep}, as well as are provided in tables in the Supplementary 
Material.  The contribution from Halley-type cometary grains was determined by \citet{poppe16} to be a 
much less significant source of dust in the outer solar system and is not considered here.  
Due to the computational constraints associated with modeling the dynamics of large grains 
(which have progressively larger lifetimes), \citet{poppe16} considers grain masses only up to 10$^{-5}$~g 
(or $\sim$100 $\mu$m radius for an assumed particle density of 2.5 g cm$^{-3}$).  Larger particles will 
also be present in the outer solar system, and the mass flux from these particles could constitute an 
important fraction of the incoming total mass flux to the planets.  We therefore crudely estimate the 
flux from these larger particles in the following 
manner.  First, if the mass-flux-versus-mass curve for a given population/planet possesses a well-defined peak 
in the 10$^{-12}$ to 10$^{-5}$~g range, then we simply extrapolate linearly in log-space to higher masses (see, 
for example, the Oort cloud distributions in Fig.~\ref{figmassflux} for any of the planets).  Secondly, if the 
mass-flux-versus-mass curve does not have a well-defined peak in the lower-mass range, we make the assumption 
that the peak of the mass function occurs right at 10$^{-5}$~g, and we ``reflect'' the mass curve about 
10$^{-5}$~g to make a symmetric mass curve versus mass (see, for example, the Jupiter-family comet grain 
curves at any planet, or the Edgeworth-Kuiper belt grains at Neptune in Fig.~\ref{figmassflux}).  This 
assumption, which must remain crude as a result of the lack of constraints from actual data, is nevertheless
motivated by the fact that the mass flux of interplanetary grains at 1 AU is observed to peak at 
$\sim$10$^{-5}$~g \citep[e.g.,][]{grun85}.  For the velocity distribution of 
the extrapolated large grains, we simply adopt the velocity distribution from the 10$^{-5}$~g grains from 
the \citet{poppe16} model.  Note that the velocity distributions vary only moderately with particle size, 
so this assumption should not be too problematic.  One can see from Fig.~\ref{figmassflux} that different 
source populations dominate the total incoming mass flux at different planets, with Edgeworth-Kuiper belt 
grains strongly dominating at Neptune, Jupiter-family comet grains strongly dominating at Jupiter, and 
multiple sources contributing at Saturn and Uranus. 
Figs.~\ref{figcontmassflux} \& \ref{contmassfluxnep} further show how the mass flux varies with incoming 
particle velocity.

\begin{figure*}[!htb]
\includegraphics[clip=t,scale=0.8]{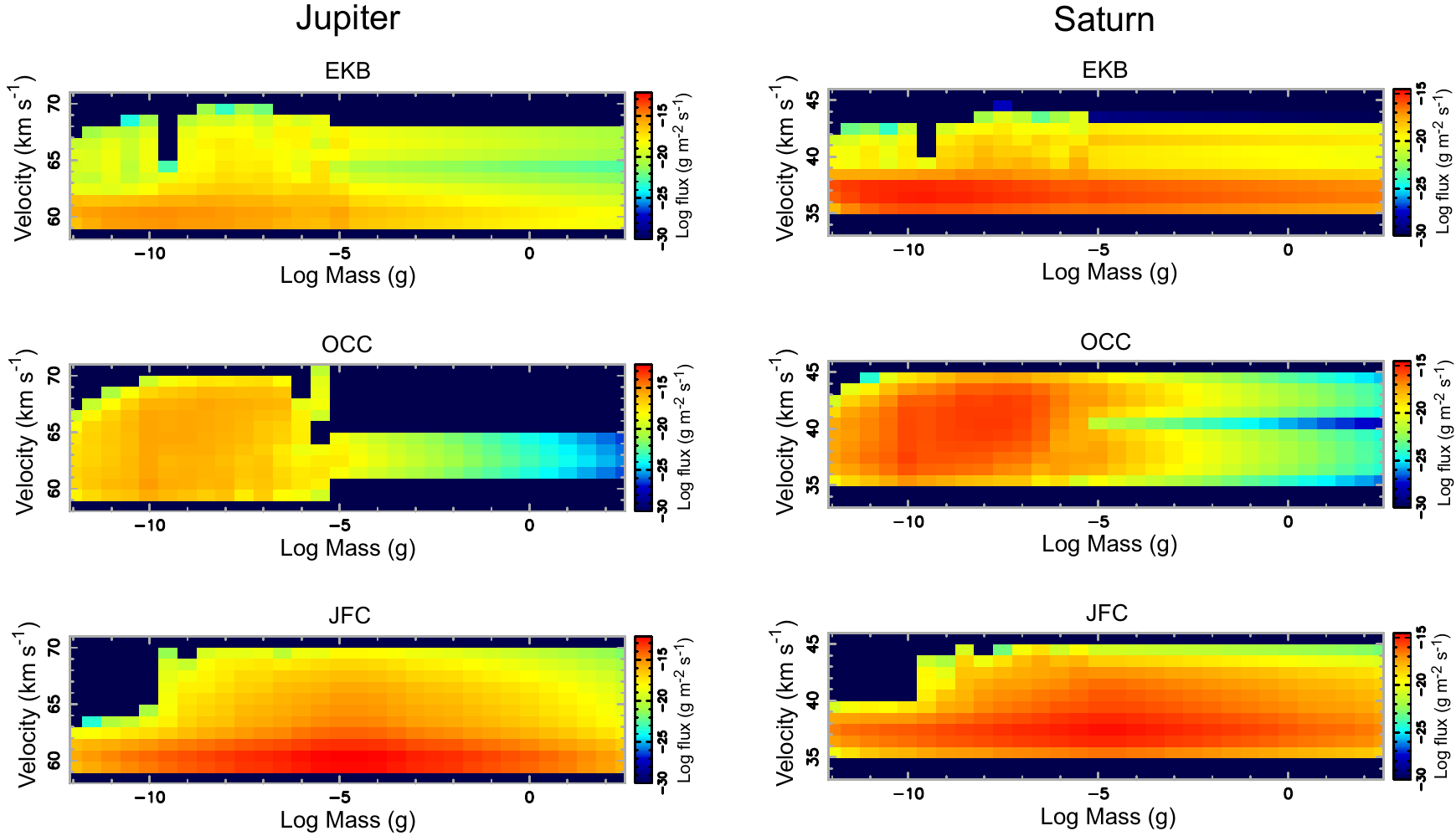}
\vspace{-0.5cm}
\caption{Differential particle mass flux (g m$^{-2}$ s$^{-1}$) as a function of mass and velocity encountering the 
top of the atmosphere (exobase) for Jupiter (Left) and Saturn (Right), for dust populations from the Edgeworth-Kuiper 
belt (EKB, Top), Oort-Cloud comets (OCC, Middle), and Jupiter-family comets (JFC, Bottom).  Particle fluxes of zero 
were assigned to 10$^{-30}$ g m$^{-2}$ s$^{-1}$ (dark blue) for plotting purposes on this logarithmic 
scale.  Gravitational focusing is considered in these calculations, such that the lower limit to the particle entry 
velocity is the escape velocity of the planet at the exobase.
(For interpretation of the references to color in this figure legend, the reader is referred to the web 
version of this article.)
\label{figcontmassflux}}
\end{figure*}

\clearpage

\begin{figure*}[!htb]
\vspace{-1.5cm}
\includegraphics[clip=t,scale=0.8]{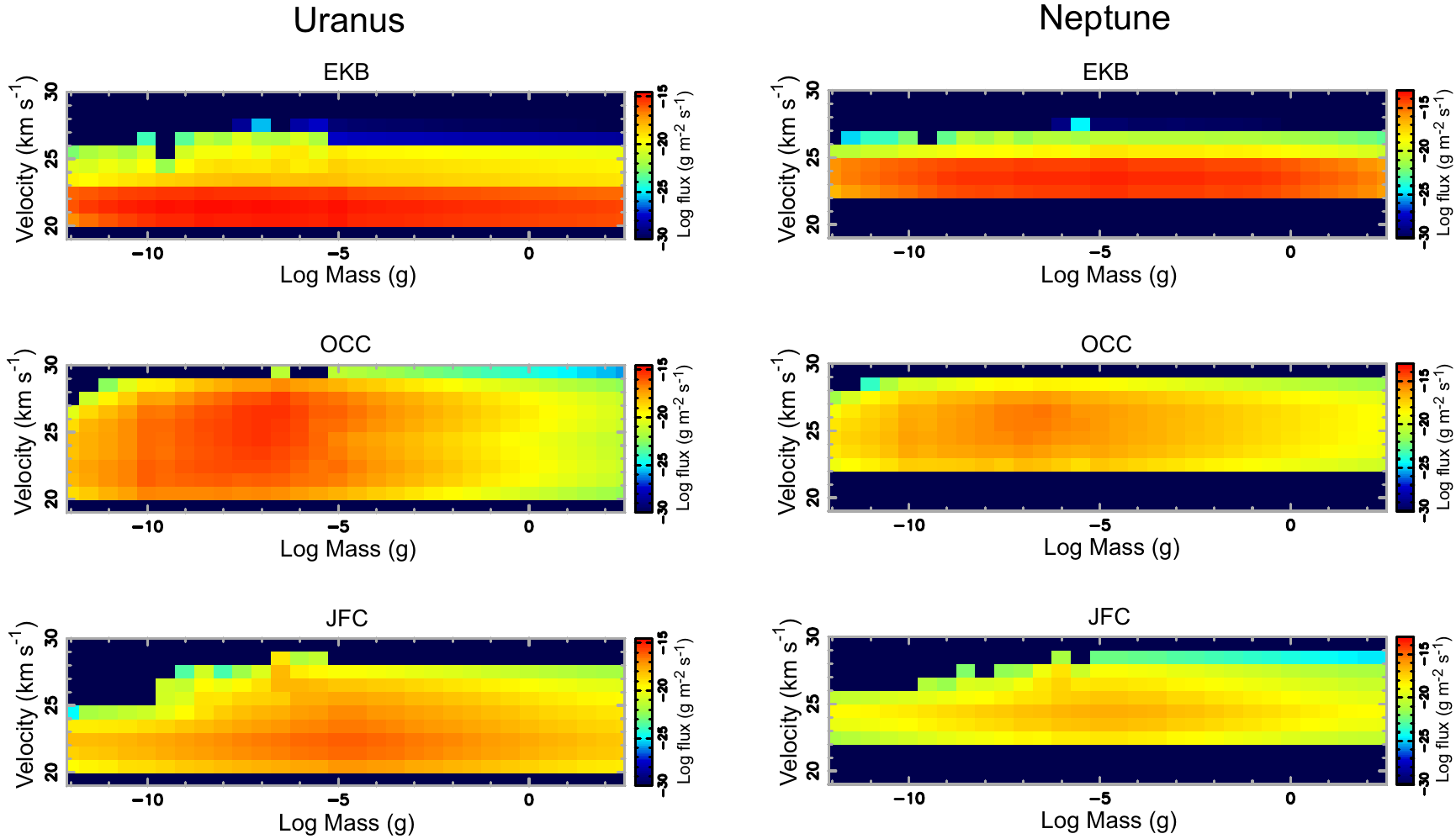}
\vspace{-0.5cm}
\caption{Same as Fig.~\ref{figcontmassflux}, except for Uranus (Left) and Neptune (Right).
\label{contmassfluxnep}}
\end{figure*}

The coupled ablation equations are solved using a fourth-order Runge Kutta technique \citep{press92}.  The 
background atmospheric structure is taken from \citet{moses05} for Jupiter and Neptune, \citet{moses15} for Saturn, 
and \citet{orton14temp} for Uranus; the same background structure is also used in the photochemical models.  

The vapor released from the ablation process described above is then added as a source term to a steady-state, 
diurnally averaged, one-dimensional (1D) photochemical model for the outer planets.  The photochemical model, 
which is based on the Caltech/JPL KINETICS code \citep{allen81,yung84}, solves the continuity equations for over 
60 hydrocarbon and oxygen species as they interact via $\sim$500 chemical reactions and are transported 
vertically via molecular and eddy diffusion.  The use of a 1D model is justified by the fact that, to first 
order, the incoming dust has no preferred latitude distribution, and transport over longitudes within the atmosphere 
is very rapid compared with vertical settling time scales.  The chemical reaction list and photolysis cross 
sections are taken largely from \citet{moses15}; the rate coefficients for the non-photolysis reactions are 
included in the Supplementary Material.  The cross sections for the photolysis reactions are discussed more fully 
in \citet{moses00a,moses05} and references therein, along with some additional recent updates \citep[e.g.,][]{sander11,hebrard13}.  
The photochemical models consider neutral chemistry only (no ion chemistry) and 
are designed to represent global-average conditions.  Tropospheric nitrogen and phosphorus photochemistry 
is also omitted, as our main goal is to better understand the stratospheric chemistry.  

The eddy diffusion 
coefficient profiles and other details of the models can be found in \citet{moses05} for Jupiter and Neptune, 
and \citet{moses15} for Saturn, as well as in the full model output in the Supplementary Material.  For Uranus, we 
started with the eddy diffusion coefficient profile from the nominal model of \citet{orton14chem}, but the chemical 
reaction rate coefficients for that Orton et al. model were chosen in such a way as to optimize the high C$_2$H$_2$/C$_2$H$_6$ 
ratio observed for Uranus.  Those rate-coefficient choices did not provide as good a fit to the C$_2$H$_2$ and 
C$_2$H$_6$ abundances on the other giant planets. The reaction mechanism adopted in this paper provides a 
better compromise for all the giant planets, although the fit to the Uranus \textit{Spitzer} data is not as 
good as with the \citet{orton14chem} nominal model.  Our resulting adopted eddy diffusion coefficient for Uranus is 
5000 cm$^2$ s$^{-1}$, independent of altitude, for a tropopause CH$_4$ mixing ratio of 1$\scinot-5.$.  The choice of 
the eddy diffusion coefficient profile has a minor effect on the shape of the vertical profiles for the oxygen species, the 
vertical diffusion time scales, and (potentially, but not necessarily) the inferred influx rates needed to fit 
the oxygen-species observations, but the choice does not affect any of our conclusions regarding the chemistry 
of the oxygen species.

Carbon monoxide is the only oxygen-bearing constituent that is assumed to have a non-negligible source from below 
our 5--7 bar lower model boundaries.  In our nominal models, we assume a fixed CO lower-boundary mole fraction (volume 
mixing ratio) of 1$\scinot-9.$ for Jupiter \citep{bezard02}, 1$\scinot-9.$ for Saturn \citep{noll90,cavalie09}, 
5$\scinot-10.$ for Uranus \citep[below the upper limit of][]{teanby13}, and 8$\scinot-8.$ for Neptune 
\citep{luszczcook13}.  If the external source of oxygen were ignored, the upward flux of CO from the interior 
would supply some oxygen to the stratospheres of the giant planets, but in amounts insufficient to explain the 
observed abundances of H$_2$O, CO$_2$, and CO in the stratosphere 
\citep[cf.][]{beer75,beer78,larson78,bjoraker86,noll86co,noll88co,noll97,rosenqvist92,marten93,marten05,guilloteau93,encrenaz96,encrenaz04,courtin96,feuchtgruber97,feuchtgruber99,degraauw97,moses00b,moses05,bergin00,bezard02,lellouch02,lellouch05,lellouch06,lellouch10,moreno03,burgdorf06,hesman07,meadows08,cavalie08,cavalie09,cavalie10,cavalie12,cavalie13,cavalie14,fletcher10akari,fletcher12spire,abbas13,luszczcook13,orton14chem,irwin14}.  
The tropospheric CO mixing ratio has only been firmly established for Jupiter \citep{bezard02}, and the vertical 
profile has not been uniquely determined for any of the planets.  However, recent observations have made it clear 
that the distribution of CO is not vertically uniform on any of the giant planets --- the CO mixing ratio increases 
from the troposphere to the stratosphere, indicating an external source of CO to these planets 
\citep[e.g.,][and more recent observations listed above]{bezard02,encrenaz04,lellouch05,cavalie09}.

A source of external oxygen is introduced to the stratosphere using the ablation profiles from the ``ice'' 
component of the incoming interplanetary dust grains.  Although the ``silicate'' portion of the grains also contains 
a non-trivial amount of oxygen, we assume that the oxygen from silicate ablation eventually ends up back in 
condensed silicates once it is released in the atmosphere (e.g., the oxygen is ablated via vapor species such as SiO, 
which are more likely to recondense than be photolyzed to release the O).  If this sequestering back into silicates 
does not occur or is only partially occurring, there could be an additional deeper release of oxygen that is not being 
considered in the models.
Once the ablated oxygen-bearing vapor is released from the grain, it can be photolyzed by ultraviolet radiation from 
the Sun and by solar Lyman 
alpha photons scattered from atomic hydrogen in the interplanetary medium --- the latter source (assumed isotropic 
in the model) becomes more important to the overall methane photochemistry the farther the planet is from the Sun 
\citep[e.g.,][]{strobel90,bishop92}.  The oxygen species can also react with hydrocarbons produced from methane 
photochemistry.  Water vapor is recycled fairly efficiently in giant-planet stratospheres, but coupled water-methane 
photochemistry can also lead to the production of CO and CO$_2$ \citep{moses00b,moses05}.
Initially, we assume the ablated vapor is 100\% water.  
However, because that assumption provides a poor fit to the stratospheric H$_2$O, CO$_2$, and CO observations for 
all of the giant planets, we also scale the overall influx rate and/or adjust the speciation of the ablated vapor in 
later models to produce a better fit.  These model-data comparisons provide insight into the possible chemical 
processing of the oxygen that might be occurring in the meteor phase before the further processing that occurs 
from photochemistry.

The external oxygen species diffuse down from their high-altitude ablation source region, where they eventually 
encounter lower-stratospheric regions that are cold enough to cause the H$_2$O (all planets) and CO$_2$ (Uranus and 
Neptune) to condense.  Condensation is included in the photochemical model in the manner described in 
\citep{moses00a,moses00b}.  Note that we neglect methane condensation in the photochemical models because it 
tends to cause annoying numerical instabilities.  Because CH$_4$ \textit{does} actually condense on Uranus and 
Neptune, we simply adopt the observed stratospheric mixing ratios at the lower boundary of these models in order 
to have appropriate stratospheric CH$_4$ abundances.  This assumption will lead to inaccurate chemical abundances 
within the methane condensation region and below (particularly for CH$_4$ in the troposphere), so we only focus 
on the stratosphere when showing results for Uranus and Neptune.  However, because CH$_4$ is not photochemically 
active in the tropopause region or below, this assumption has little effect on the hydrocarbon and oxygen 
photochemistry itself.

\section{Results and Discussion}\label{sec:results}

\subsection{Ablation profiles}\label{sec:ablresults}

\begin{figure*}[!htb]
\includegraphics[clip=t,width=5.8in]{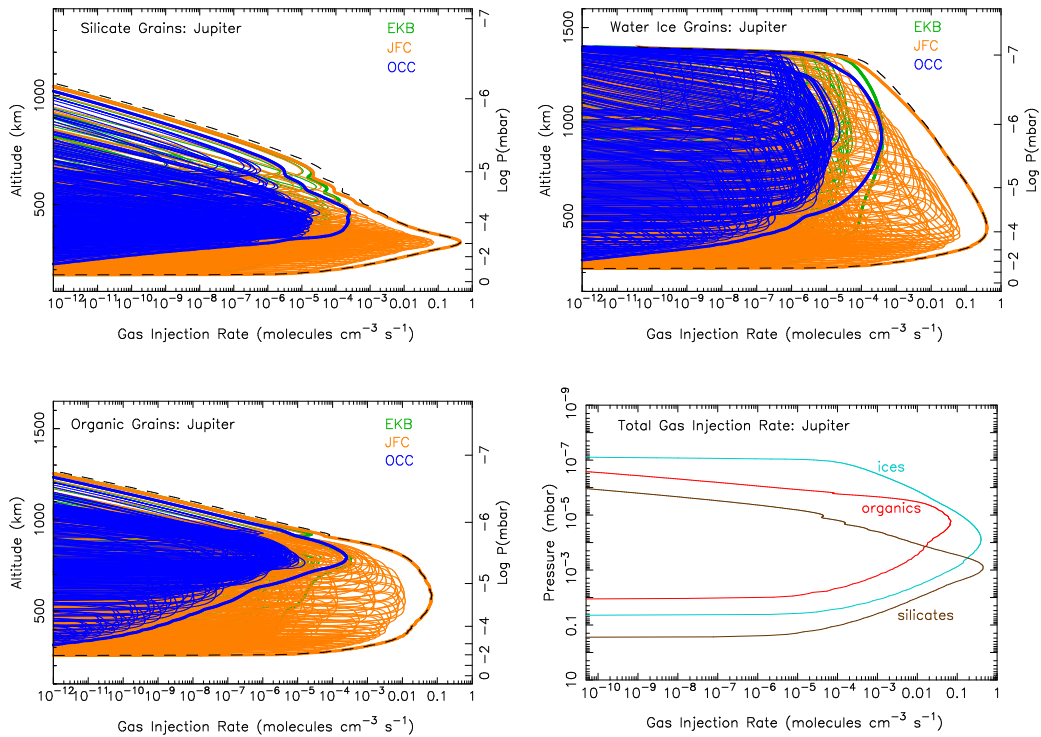}
\vspace{-0.5cm}
\caption{Ablation rate profiles at Jupiter for silicate grains (Top left), water-ice grains (Top right), organic 
grains (Bottom left), and the total ablation rate from each component (Bottom right).  The colored lines in the 
first three panels show the ablation profiles from each individual entry velocity and mass bin for the Edgeworth-Kuiper 
belt grains (green), Jupiter-family comet grains (orange), and Oort-cloud comet grains (blue).  The thicker colored 
lines show the sum of all grains within each of these three populations, and the black dashed lines show 
the sum of all grains from all populations.  These totals from all populations are shown as a function of pressure 
instead of altitude in the bottom right panel. 
(For interpretation of the references to color in this figure legend, the reader is referred to the web 
version of this article.)
\label{prodmjup}}
\end{figure*}

\begin{figure*}[!htb]
\includegraphics[clip=t,width=5.8in]{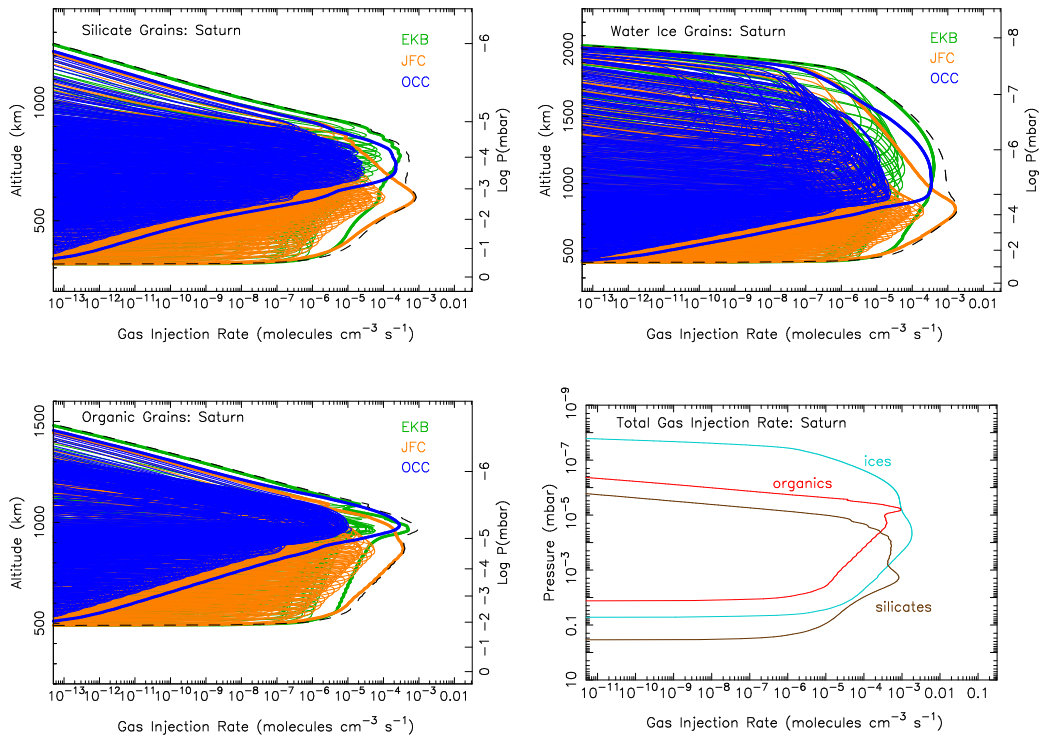}
\vspace{-0.5cm}
\caption{Same as Fig.~\ref{prodmjup}, except for Saturn.
\label{prodmsat}}
\end{figure*}

\begin{figure*}[!htb]
\includegraphics[clip=t,width=5.8in]{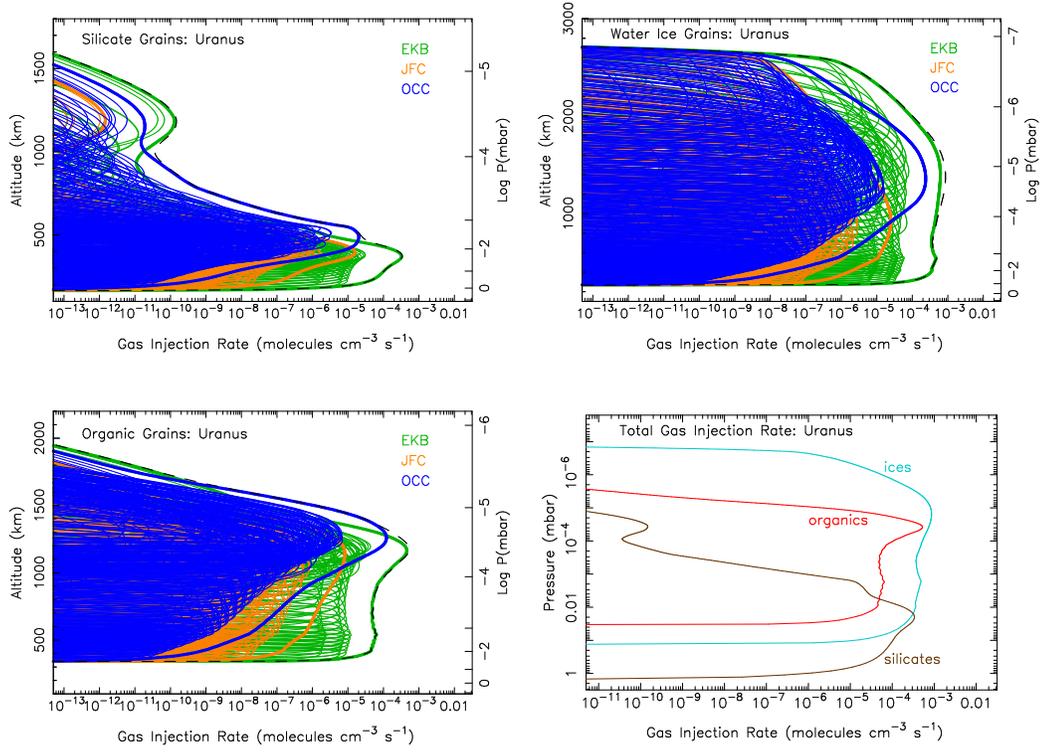}
\vspace{-0.5cm}
\caption{Same as Fig.~\ref{prodmjup}, except for Uranus.
\label{prodmuran}}
\end{figure*}

\begin{figure*}[!htb]
\includegraphics[clip=t,width=5.8in]{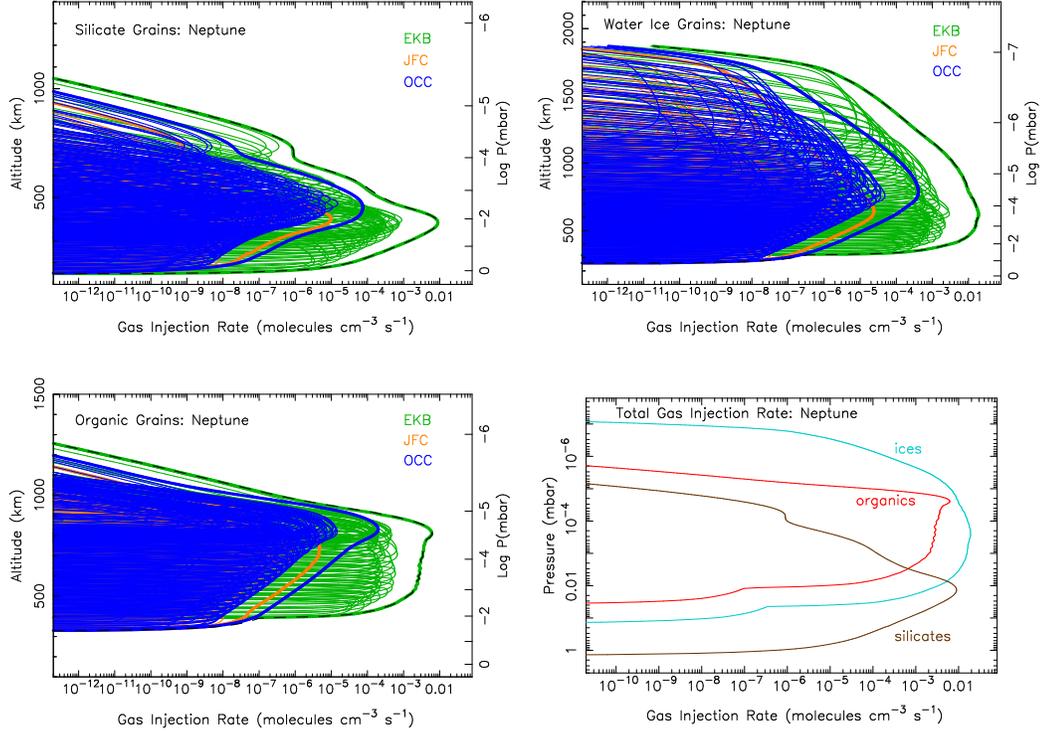}
\vspace{-0.5cm}
\caption{Same as Fig.~\ref{prodmjup}, except for Neptune.
\label{prodmnep}}
\end{figure*}

The calculated ablation profiles for each giant planet are shown in Figs.~\ref{prodmjup} (Jupiter), 
\ref{prodmsat} (Saturn), \ref{prodmuran} (Uranus), \& \ref{prodmnep} (Neptune).  Jupiter-family 
comet grains clearly dominate the dust ablation source of external material at Jupiter, while 
Edgeworth-Kuiper belt grains clearly dominate at Uranus and Neptune.  This result is largely a 
reflection of the incoming mass flux for the different populations from the \citet{poppe16} 
model, as shown in Fig.~\ref{figmassflux}.  The incoming mass fluxes for the different populations 
are more balanced at Saturn, so all three populations contribute notably to the ablation profile in 
Fig.~\ref{prodmsat}, with the Jupiter-family comet grain population dominating at the lower altitudes 
and Edgeworth-Kuiper belt grains at higher altitudes.  The larger grains in the Jupiter-family 
comet population are not decelerated as efficiently as the smaller grains in the Edgeworth-Kuiper belt 
population and so tend to penetrate deeper before being ablated.

In general, larger grains penetrate deeper than smaller grains for any given incoming velocity and 
material properties.  In addition, because the larger grains experience a greater number of collisions 
with air molecules, they are heated to higher temperatures and ablate more fully than smaller 
particles.  Entry velocity also has an effect.  Grains entering the atmosphere at higher velocities 
heat up faster than those at lower velocities because of the larger kinetic energy being converted 
into heating the grain.  Faster particles therefore tend to ablate at higher altitudes and reach higher 
maximum temperatures, allowing them to ablate more efficiently.  

Because of gravitational focusing by the planet, 
entry velocities at Jupiter are larger than those for the other planets (see Figs.~\ref{figcontmassflux} 
\& \ref{contmassfluxnep}).  For the case of the relatively refractory silicate grains, the greater 
incoming velocities on Jupiter and Saturn lead to greater maximum temperatures during atmospheric 
entry, allowing all of the grains to completely ablate before they are decelerated.  The slower entry 
velocities at Uranus and Neptune lead to smaller maximum temperatures, such that some of the silicate 
grains do not completely ablate before they are decelerated and cool through radiative and evaporative 
cooling \citep[see][]{moses92abl}.  The residual grains continue to fall through the atmosphere at the 
terminal settling velocity and contribute to the overall aerosol burden of the atmosphere.  For Neptune, 
only 53\% of the incoming silicate grain mass flux is ablated, releasing silicon and metal vapor into
the atmosphere.  Entry velocities at Uranus are even lower, and only 22\% of the silicate mass flux 
ends up being released as vapor.  In fact, even some of the slowest and smallest organic particles will 
not fully ablate, leading to a 95\% ablation efficiency of the organic grains on Uranus.  The organic 
grains fully ablate on the other three planets, and the ice grains ablate fully on all four planets.
Tables of the total ablation rate profile for each composition at each planet as a function of pressure 
are provided in the Supplementary Material.

\clearpage

\clearpage

\clearpage

The difference in material properties also affects the ablation profile.  The ablation peak for silicate 
grains is deeper in the atmosphere than that of the water-ice or organic grains because of the much higher 
vaporization temperature of the silicates, whereas the water-ice grains begin to ablate at very high altitudes 
because of their low vaporization temperatures.  Organic grains fall in between.  The latent heat of 
vaporization and the emissivity of the material also affect the overall shape of the ablation profile.  If the 
latent heat is low for any given material vapor pressure, the particles do not cool as efficiently during entry, so 
they heat up faster, ablate at higher altitudes, and have an overall narrower profile with a greater maximum 
ablation rate at the peak.  Water (both ice and liquid) has a higher latent heat of vaporization than that of 
the organics assumed here, which explains why the water-ice ablation profile is so much broader in altitude 
than that of the organic grains.  The water begins to sublimate from the grains at low temperatures 
high up in the atmosphere, and the relatively large latent heat causes evaporative cooling to be effective 
at keeping the grains at low temperatures throughout their 
flight through the atmosphere.  For organic grains, on the other hand, the latent heat cooling term is less 
efficient, so although the organics begin to ablate at lower altitudes than the water ice, the ablation is 
complete before that of the water-ice grains.  Similarly, if we had assumed the grain emissivity were constant at 1.0, the 
grains would have radiated the heat away more efficiently, so they would have survived to deeper altitudes, 
but would have had narrower ablation peaks with greater maximum peak ablation rates because of the increased 
heating in the higher-density deeper atmosphere.

Given that real particles consist of a mixture of materials, are not spherical, and are subject to other 
physical processes like sputtering and fragmentation during entry, our ablation calculations 
are simple approximations of the real situation.  Observations of metals in the Earth's atmosphere and in 
residual micrometeoroids suggest that ``differential ablation'' does occur in real atmospheres, such that 
the more volatile components can ablate at higher altitudes and more completely than less volatile components 
\citep[e.g.,][]{mcneil98,vonzahn99,vondrak08,janches09}; however, fragmentation and simple ablation also 
occur.  The overall shape of the ablation profiles for the different vapors being released is complicated and 
cannot be entirely captured with simplified models and their assumptions \citep{malhotra11}.
Fortunately, the chemical consequences of the oxygen vapor being released have little sensitivity to the 
actual shape of the ablation profile, and a much greater dependency on the overall integrated flux of the 
oxygen-bearing vapor being released \citep{moses00b,mosesbass00} and the molecular/atomic form of the 
released vapor (section \ref{sec:thermo}).  That lack of sensitivity to the details of the ablation 
profile results from the fact that most of the interesting (i.e., non-recycling) oxygen chemistry occurs in the 
$\sim$10--10$^{-2}$ mbar region, which is below the peak ablation region for the icy component, as well as to the
fact that diffusion time scales in the upper atmosphere are shorter than the chemical lifetimes for the oxygen 
species, and the overall total available amount of chemically active oxygen is an important factor affecting the 
resulting chemistry.

Given the various uncertainties in the modeling of both the dust dynamics and the ablation process, as well as 
uncertainties in the dust composition, we estimate that our predicted oxygen influx rates from ablation are 
uncertain by about an order of magnitude.

\begin{figure*}[!htb]
\vspace{-1.5cm}
\includegraphics[clip=t,width=5.5in]{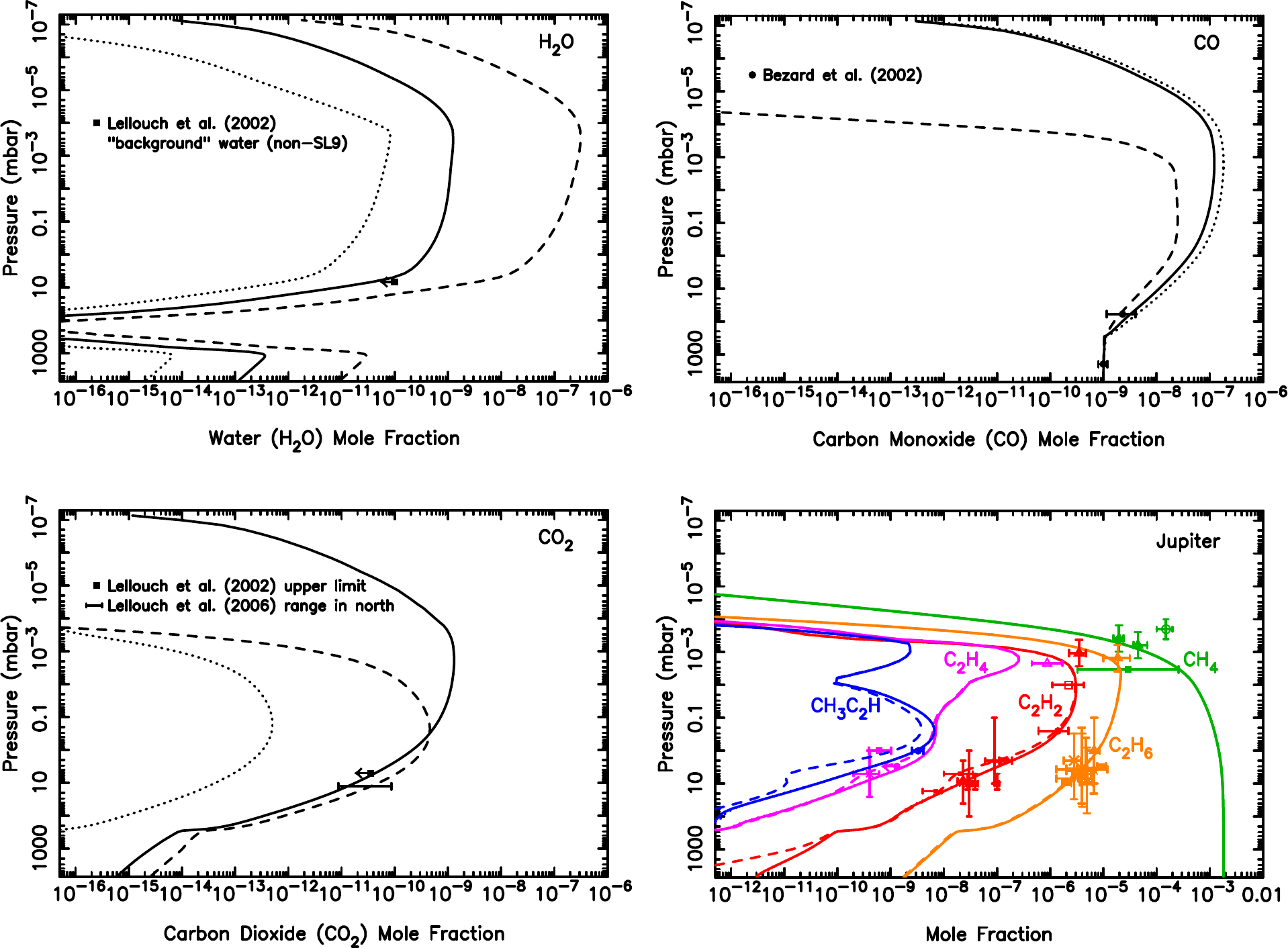}
\vspace{-0.5cm}
\caption{Mixing ratio profiles for H$_2$O (Top left), CO (Top right), CO$_2$ (Bottom left), and several 
hydrocarbons (Bottom right), as labeled, in Jupiter's atmosphere as a result of the ablation of oxygen-rich 
icy grains.  The dashed lines represent a model in which all the ablated icy component is released as water 
(integrated flux of 1.0$\scinot7.$ H$_2$O molecules cm$^{-2}$ s$^{-1}$), the dotted lines represent a model in 
which all the ablated icy component is released as carbon monoxide (integrated flux of 1.0$\scinot7.$ CO molecules 
cm$^{-2}$ s$^{-1}$), and the solid lines represent a model in which the relative influx rates (4.0$\scinot4.$ H$_2$O 
molecules cm$^{-2}$ s$^{-1}$, 7.0$\scinot6.$ CO molecules cm$^{-2}$ s$^{-1}$, 1.0$\scinot5.$ CO$_2$ molecules 
cm$^{-2}$ s$^{-1}$) are scaled to fit the H$_2$O, CO, and CO$_2$ observations of \citet{lellouch02,lellouch06} 
and \citet{bezard02} for the regions the least influenced by the Shoemaker-Levy 9 impacts.  The data points 
with error bars represent various observational constraints (see text).
(For interpretation of the references to color in this figure legend, the reader is referred to the web 
version of this article.)
\label{figjupoxy}}
\end{figure*}

\subsection{Jupiter results and comparisons with observations}\label{sec:jup}

The effect of the ablated oxygen vapor on the composition of the Jovian atmosphere is shown in 
Fig.~\ref{figjupoxy}.  The ablation of the icy grains, as calculated in section~\ref{sec:ablresults}, 
peaks near 8$\scinot-5.$ mbar, which is just above the methane homopause.  The full ice ablation profile 
extends decades in pressure in both directions away from this peak, with a total integrated column influx 
of oxygen of 1.0$^{+2.2}_{-0.7}\scinot7.$ O atoms cm$^{-2}$ s$^{-1}$.  If we assume that the gas released from the icy 
grains is all in the form of water, then the photochemical model produces the results shown by the dashed lines 
in Fig.~\ref{figjupoxy}.  The total stratospheric water column abundance in this model is 1.9$\scinot16.$ cm$^{-2}$. 
This H$_2$O column abundance is $\sim$5--20 times greater than 
the observed global-average water abundance on Jupiter \citep{feuchtgruber99,bergin00,cavalie08,cavalie12}, the 
bulk of which actually derives from the 1994 impact of Comet Shoemaker-Levy 9 (SL9) with Jupiter 
\citep[e.g.,][]{lellouch02,lellouch06,cavalie13}.  The ``background'' water influx rate from interplanetary dust 
is observationally constrained by \citet{lellouch02} to be less than 8$\scinot4.$ H$_2$O molecules cm$^{-2}$ s$^{-1}$, 
and probably as low as 4$\scinot4.$ H$_2$O molecules cm$^{-2}$ s$^{-1}$.

Thus, as first discussed by \citet{poppe16}, we have an interesting situation in which the dust dynamical 
model delivers seemingly too much water to Jupiter.  Photochemistry cannot resolve this problem.  While some 
of the water is photochemically converted to CO and CO$_2$, the conversion is simply too ineffective to 
remove enough water to explain the observations --- kinetic recycling of H$_2$O is efficient 
in Jupiter's hydrogen-dominated atmosphere (see also \citealt{moses00b,moses05} and section \ref{sec:oxychem}).  
It seems unlikely that the incoming dust, which is dominated by Jupiter-family comet grains,  
is two orders of magnitude less oxygen-rich than we have assumed, so oxygen-depleted dust is also unsatisfactory 
as a possible explanation (see \citealt{bockeleemorvan11} and \citealt{dellorusso16} for reviews of cometary composition).  
Although Jupiter-family comet grains do spend more time at smaller heliocentric distances than their Edgeworth-Kuiper 
Belt counterparts, such that the ice components within the grains have a greater chance of being sublimated before 
atmospheric entry, other sources of oxygen and H$_2$O in the grains should remain intact, such as hydrated silicates.  

The overall large CO/H$_2$O ratio in Jupiter's atmosphere has led 
\citet{bezard02} to conclude that small comets are responsible for supplying Jupiter's external oxygen, and this 
explanation has been reinforced by recent observations and modeling efforts 
\citep{lellouch02,lellouch06,cavalie08,cavalie12,cavalie13}.  The prevailing theory is that oxygen from 
cometary impacts is thermochemically converted to CO during the energetic impact and plume splashback 
phases \citep[e.g.,][]{zahnle96}, whereas it is unclear whether this conversion to CO can happen during dust 
ablation.  However, it seems unlikely that the \citet{poppe16} dust flux predictions are over 
two orders of magnitude too high, especially given the \textit{in situ} observational constraints provided by 
the \textit{Galileo} Dust Detection System, the \textit{Pioneer 10} meteoroid detector, and the 
\textit{New Horizons} Student Dust Detector.  Instead, we suggest that H$_2$O can be kinetically or 
thermochemically converted to CO during the meteoric entry phase, a topic that will be discussed further in section
\ref{sec:thermo}.

Therefore, we also show in Fig.~\ref{figjupoxy} what happens if CO rather than H$_2$O were the only oxygen-bearing 
product released from the icy grain ablation (see dotted model curves).  Carbon monoxide, like water, is 
relatively stable photochemically in Jupiter's atmosphere because it is shielded from short-wavelength UV radiation 
by other atmospheric gases, and its strong carbon-oxygen bond makes it kinetically unreactive at Jovian atmospheric 
temperatures once the CO is thermalized.  A small amount of the CO is photochemically converted to H$_2$O and CO$_2$ 
(see Fig.~\ref{figjupoxy} and section \ref{sec:oxychem}), but most remains as CO.  This model scenario fits the data 
better, with a predicted CO column abundance consistent with the observations of \citet{bezard02} and a predicted 
H$_2$O column abundance well within the upper limit for the background (non-SL9) water abundance \citep{lellouch02}.  
This model, however, predicts too little carbon dioxide to explain the CO$_2$ abundance in the northern hemisphere 
of Jupiter, which is believed to have been relatively unaffected by the SL9 impacts at the time of the observations 
\citep{lellouch06}.

We therefore freely adjust the relative influx rates of CO, H$_2$O, and CO$_2$ to better match all the oxygen species 
observations, keeping the shape of the ablation profile 
the same, but adjusting the magnitude of the different gases released.  The solid lines in Fig~\ref{figjupoxy}
show model results that assume a column-integrated influx rate of 7$\scinot6.$ CO molecules cm$^{-2}$ s$^{-1}$, 
1$\scinot5.$ CO$_2$ molecules cm$^{-2}$ s$^{-1}$, and 4$\scinot4.$ H$_2$O molecules cm$^{-2}$ s$^{-1}$ (i.e., a relative 
influx rate of 98\% CO, 1.4\% CO$_2$, and 0.6\% H$_2$O).  This low water value is the favored ``background'' water 
influx rate from interplanetary dust as derived by \citet{lellouch02} (see also \citealt{lellouch06}, 
\citealt{cavalie08,cavalie12,cavalie13}), and the CO influx rate is within the (1.5--10)$\scinot6.$ molecules cm$^{-2}$ s$^{-1}$ 
range determined by \citet{bezard02} for the external source on Jupiter.  The influx rates for all three species 
in this model produce CO and CO$_2$ column 
abundances consistent with the \citet{bezard02} high-resolution ground-based infrared observations of CO, and 
the \citet{lellouch06} northern-hemisphere \textit{Cassini} Composite Infrared Spectrometer (CIRS) observations 
of CO$_2$.  The total oxygen influx rate from this model, 7.2$\scinot6.$ O atoms cm$^{-2}$ s$^{-1}$, is 
$\sim$30\% smaller than that predicted from our dust ablation model, which is well within our estimated order of 
magnitude uncertainty due to the \citet{poppe16} dynamical modeling, our extrapolation to larger grain sizes, and 
our assumptions about the grain composition and ablation process.

From these model-data comparisons, we conclude that dust grains supply a major component of the external 
oxygen on Jupiter, in addition to what is being supplied by large comets such as  SL9 or smaller, more frequent 
cometary impacts \citep[e.g.,][]{bezard02}.  In fact, 
interplanetary dust grains could be the dominant source of external oxygen on Jupiter when averaged over long 
time scales, provided that the ablated oxygen is released predominantly in the form of CO, or if the oxygen is 
converted to CO during the meteor phase (i.e., to explain the low H$_2$O abundance in Jupiter's 
stratosphere).  We explore this topic further in section \ref{sec:thermo}.

The bottom right panel in Fig.~\ref{figjupoxy} shows how the external oxygen species affect the mixing ratios 
of several hydrocarbons.  The observational data for the hydrocarbons are from 
\citet{gladstone83}, \citet{wagener85}, \citet{noll86}, \citet{kostiuk87}, \citet{morrissey95}, 
\citet{yelle96,yelle01}, \citet{sada98}, \citet{fouchet00hc}, \citet{bezard01dps}, \citet{moses05}, 
\citet{romani08}, \citet{greathouse10}, \citet{nixon10}, and \citet{kim14}.  For the model with a water influx
rate of 1$\scinot7.$ H$_2$O molecules cm$^{-2}$ s$^{-1}$ (dashed line), the coupled water-hydrocarbon 
photochemistry causes a notable reduction in the abundance of unsaturated hydrocarbons such as C$_2$H$_2$, 
C$_2$H$_4$, CH$_3$C$_2$H, and (not shown) C$_4$H$_2$.  Much of the carbon removed from these species ends up 
in CO, and to a much lesser extent CO$_2$ \citep[see][and section \ref{sec:oxychem}]{moses00b,moses05}.  However, 
for the more realistic water 
influx rate of 4$\scinot4.$ H$_2$O molecules cm$^{-2}$ s$^{-1}$ (solid model curves), the oxygen photochemistry has 
little effect on the hydrocarbon abundances because the resulting H$_2$O mixing ratio is much less than that of 
C$_2$H$_2$, the main hydrocarbon with which the OH reacts, and because the dominant oxygen species in that model --- 
CO --- is less photochemically active.

\begin{figure*}[!htb]
\begin{center}
\includegraphics[clip=t,width=5.4in]{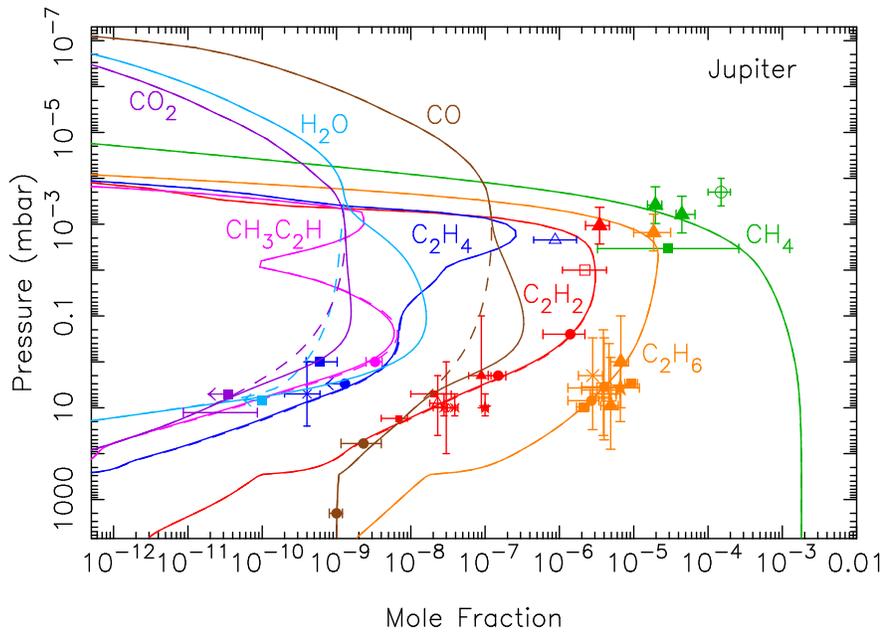}
\end{center}
\vspace{-0.5cm}
\caption{Mixing ratio profiles for several species (as labelled) in Jupiter's atmosphere for a model (solid line) that 
includes a source of oxygen from the Comet Shoemaker-Levy 9 impacts, along with the background steady dust 
influx rate.  This model is more representative of the Jovian southern hemisphere today than the pure dust 
ablation model (dashed line in the figure here, solid line in Fig.~\ref{figjupoxy}). \label{figjupcomet}}
\end{figure*}

On top of the steady background dust influx described above and in Fig.~\ref{figjupoxy}, the Shoemaker-Levy 9 
impacts delivered a large amount of oxygen to Jupiter's middle stratosphere in 1994 \citep[see the review of][]{lellouch96}, 
much of which is still concentrated in the southern hemisphere \citep[e.g.,][]{moreno03,lellouch02,lellouch06,cavalie13}.  
To illustrate what the H$_2$O, CO, and CO$_2$ profiles might look like today from this recent cometary source, we 
ran an additional time-variable model using observed comet-derived profiles as initial conditions, along with the 
steady background dust influx from our best-fit model (solid line) shown in Fig.~\ref{figjupoxy}.  This model that 
includes the cometary source is more representative of conditions in the southern hemisphere at the present time.  
Although we cannot capture the full 3D atmospheric behavior with this simple 1D model, much of the horizontal spreading 
of the comet debris occurred in the first few years after the impacts \citep{lellouch02,moreno03}, so we use 
observations a couple years or more after the impacts to set our initial conditions.  Based on the analysis of 
\citet{lellouch02}, we assume that the initial mixing ratios of H$_2$O and CO$_2$ are a constant 4$\scinot-8.$ 
and 7$\scinot-9.$, respectively, above 0.5 mbar.  Based on \citet{moreno03}, we assume that the initial CO 
mixing ratio is 9$\scinot-7.$ above 0.3 mbar.  The resulting abundances after 23 years (representing the time 
since the 1994 impacts) are shown in Fig.~\ref{figjupcomet} and can be directly compared with Fig.~\ref{figjupoxy}.  
The middle-stratospheric bulges in all three oxygen-bearing molecules are readily apparent in the 10--10$^{-2}$ mbar 
region more than 20 years after the impacts.  However, more recent observations by \citet{cavalie12} suggest that 
the rate of diffusion or spreading of H$_2$O may be greater than is indicated by these models, leading to some 
overpredictions of the oxygen species abundances in the $\sim$0.1-1 mbar region in our model.  In any case, the SL9 impacts 
strongly perturbed the background oxygen-bearing species abundances in the middle stratosphere; however, as can be 
seen from a comparison of Figs.~\ref{figjupoxy} \& \ref{figjupcomet}, these 
comet-delivered species today have little effect on the profiles below $\sim$5 mbar (because the cometary oxygen 
species have not yet been transported to the lower stratosphere) or above 1 microbar (where dust delivery of oxygen 
dominates the source, and molecular diffusion time scales are short).

Another interesting consequence of the predicted large dust influx rate to Jupiter is the relatively large 
resulting CO mixing ratio in the upper atmosphere.  The CO will influence ionospheric chemistry by reacting 
with H$_3$$^+$ (which dominates below the main electron-density peak) to produce HCO$^+$, and the 
HCO$^+$ will rapidly recombine with an electron to produce CO + H \citep[see][]{mosesbass00}, potentially 
causing a reduction in the local electron density in the process.  Ablated metal vapor could also affect 
the lower ionospheric chemistry and structure \citep[e.g.,][]{moses92abl,mosesbass00,lyons95}, as could unablated 
dust or recondensed ablation products.  Future Jovian ionospheric models should therefore consider the 
potential effects of CO, metals, and other debris resulting from meteoric input.

\subsection{Saturn results and comparisons with observations}\label{sec:sat}

Our models predict that volatile oxygen is released from icy-grain ablation in Saturn's atmosphere with a 
column-integrated influx rate of 7.4$^{+16}_{-5.1} \scinot4.$ atoms cm$^{-2}$ s$^{-1}$.  As discussed in \citet{poppe16}, 
this flux is more than an order of magnitude too small to explain the stratospheric H$_2$O, CO, and CO$_2$ 
observed on Saturn \citep{feuchtgruber97,feuchtgruber99,moses00b,bergin00,cavalie09,cavalie10,fletcher12spire,abbas13}.  
\textit{Cassini}'s discovery of plumes on Enceladus \citep{dougherty06,porco06} that are spewing water molecules into 
the Saturnian system has helped reveal the likely source of the large stratospheric water abundance on Saturn 
\citep[e.g.,][]{cassidy10,hartogh11,fleshman12}, but Saturn's rings \citep{connerney84,tseng10,moore15} 
and large cometary impacts \citep{cavalie10} could also be major contributors.

\begin{figure*}[!htb]
\includegraphics[clip=t,width=5.5in]{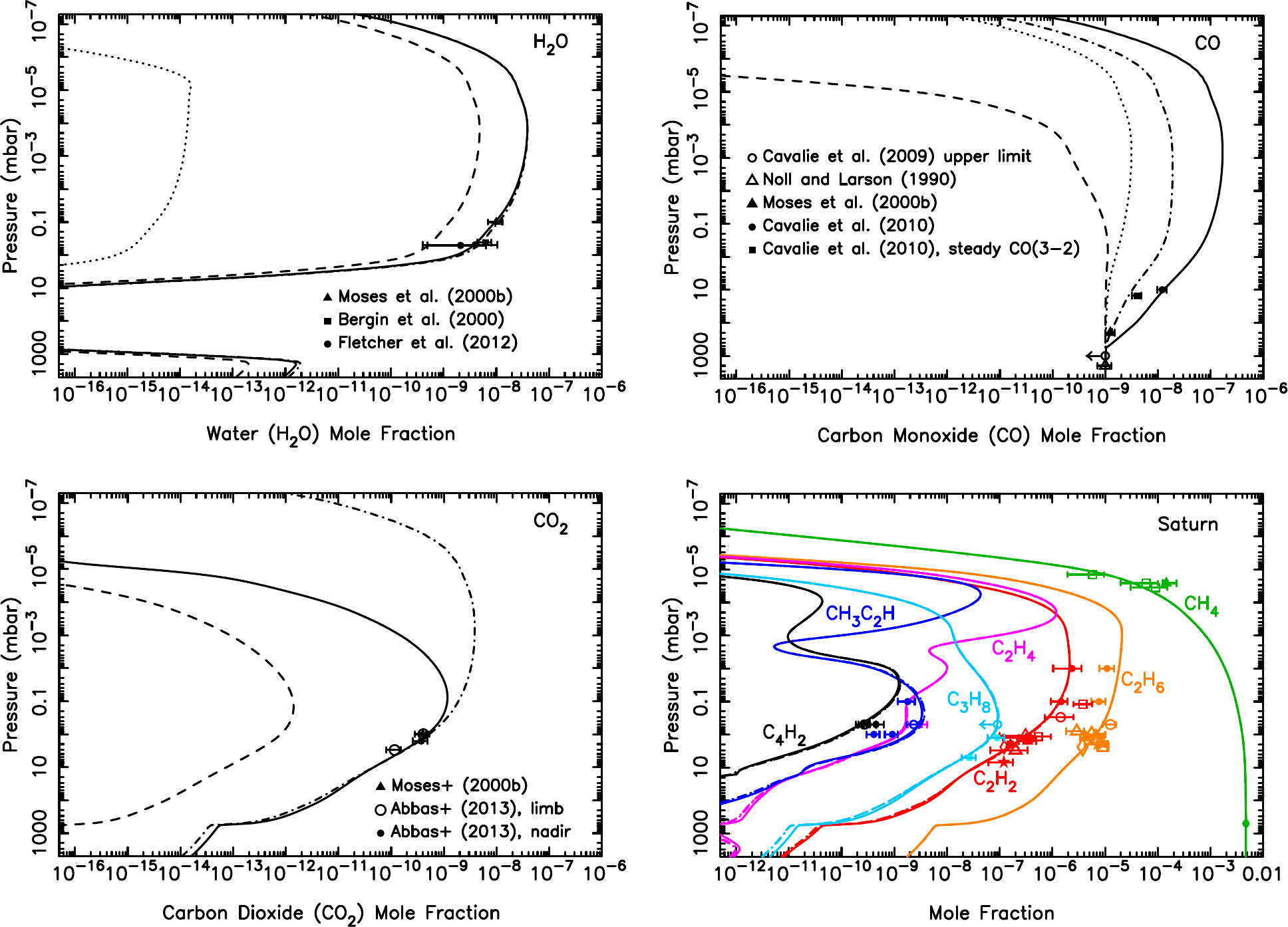}
\vspace{-0.5cm}
\caption{Same as Fig.~\ref{figjupoxy}, but for Saturn.  The dashed lines represent a model in which all the 
ablated icy component is released as water (integrated flux of 7.4$\scinot4.$ H$_2$O molecules cm$^{-2}$ s$^{-1}$); 
the dotted lines represent a model in which all the ablated icy component is released as carbon monoxide 
(integrated flux of 7.4$\scinot4.$ CO molecules cm$^{-2}$ s$^{-1}$); the dot-dashed lines represent a model 
in which the relative influx rates are scaled to 6.2$\scinot5.$ H$_2$O molecules cm$^{-2}$ s$^{-1}$, 4.2$\scinot5.$ 
CO molecules cm$^{-2}$ s$^{-1}$, and 1.2$\scinot5.$ CO$_2$ molecules cm$^{-2}$ s$^{-1}$; and the solid lines 
represent a model in which the relative influx rates are scaled to 6.2$\scinot5.$ H$_2$O molecules cm$^{-2}$ s$^{-1}$ 
and 4.1$\scinot6.$ CO molecules cm$^{-2}$ s$^{-1}$, with no direct CO$_2$ injection.  
The data points with error bars represent various observational constraints (see text).
(For interpretation of the references to color in this figure legend, the reader is referred to the web 
version of this article.)
\label{figsatoxy}}
\end{figure*}

Figure~\ref{figsatoxy} shows the vertical mixing ratio profiles of the major oxygen and hydrocarbon species 
on Saturn for different assumptions about the oxygen influx rate.  If we assume that the gas released from the 
ablated icy grains is all in the form of H$_2$O (or CO) with an integrated influx rate of 7.4$\scinot4.$ molecules 
cm$^{-2}$ s$^{-1}$, as indicated by our modeling, then the photochemical model produces the results shown by 
the dashed (or dotted) lines in Fig.~\ref{figsatoxy}.  Both models fall grossly short in explaining the observed 
stratospheric abundance of H$_2$O and CO$_2$ from the \textit{Infrared Space Observatory} (ISO), the \textit{Submillimeter 
Wave Astronomy Satellite} (SWAS), the \textit{Herschel} SPIRE, and the \textit{Cassini} CIRS limb and nadir observations 
\citep[see][]{feuchtgruber97,moses00b,bergin00,fletcher12spire,abbas13}.  A model in which the ablation profile is 
scaled such that the column-integrated influx rates of H$_2$O, CO, and CO$_2$ are 6.2$\scinot5.$, 4.2$\scinot5.$, and 
1.2$\scinot5.$ cm$^{-2}$ s$^{-1}$, respectively, fits the H$_2$O and CO$_2$ observations well (dot-dashed line in 
Fig.~\ref{figsatoxy}), and produces a CO column abundance 
above 400 mbar that is consistent with the ground-based infrared CO observations of \citet{noll90} (see 
\citealt{moses00b} for further details).  The relative influx rates in this model are 53\% H$_2$O, 36\% CO, 
and 10\% CO$_2$.  This model, however, has insufficient middle-stratospheric carbon monoxide to explain the large 
CO mixing ratios derived from ground-based limb observations of the emission core of the CO(6-5) 
rotational line at submillimeter wavelengths \citep{cavalie10}, and falls slightly short of the CO mixing 
ratio inferred from analysis of the CO(3-2) rotational line, using model profiles that assume a steady background 
influx of CO \citep{cavalie10}.

We therefore also test a model in which the integrated CO influx rate is 4.1$\scinot6.$ molecules cm$^{-2}$ 
s$^{-1}$ (favored by \citealt{cavalie10} for the assumption of a steady background influx from their analysis 
of the CO(6-5) line), the H$_2$O influx rate is 6.2$\scinot5.$ molecules cm$^{-2}$ s$^{-1}$, and the CO$_2$ influx 
rate is zero (i.e., CO$_2$ is not released as a separate component but is formed by coupled H$_2$O-CO photochemistry 
only).  The relative influx rates in this model are 87\% CO, 13\% H$_2$O, and 0\% CO$_2$.  Figure~\ref{figsatoxy} 
demonstrates that this model (solid lines) fits the H$_2$O and CO$_2$ observations well, and the results are also consistent 
with the CO mixing ratios needed to explain the CO(6-5) line, but the CO column abundance in the lower stratosphere 
and upper troposphere is too high to explain the \citet{noll90} infrared observations \citep{moses00b} or the 
\citet{cavalie10} CO(3-2) submillimeter observations.  In fact, \citet{cavalie10} find that they can only fit 
both the CO(3-2) and CO(6-5) lines if the CO is concentrated at relatively high altitudes, and so they favor a 
scenario in which a large cometary impact 220 $\pm$ 30 years ago deposited (2.1 $\pm$ 0.4)$\scinot15.$ g of CO 
above 0.1 mbar on Saturn.  Although it remains to be seen whether such a scenario can be consistent with the infrared 
observations of \citet{noll90} --- and note that such cometary models would be more consistent with the infrared 
observations if the internal tropospheric CO source were smaller than we have assumed at our lower boundary 
--- we concur that a relatively recent cometary impact is the most reasonable explanation for the submillimeter CO observations.

The bottom right panel in Fig.~\ref{figsatoxy} shows that for the fluxes considered here, the external 
oxygen species have little effect on the hydrocarbon mixing ratios.  In this plot, the observational data for 
the hydrocarbons derive from \citet{festou82}, \citet{smith83}, \citet{courtin84}, \citet{noll86}, \citet{chen91},
\citet{sada96c2h6,sada05}, \citet{moses00a,moses15}, \citet{bezard01dps}, \citet{greathouse05,greathouse06},
\citet{fletcher09}, \citet{guerlet09,guerlet10}, and \citet{sinclair13}.

Based on the \citet{cavalie10} scenario, the CO in Saturn's stratosphere most likely derives from a large 
cometary impact that occurred a couple hundred years ago.  This putative impact may also have been responsible for some 
fraction of the currently observed water on Saturn.  The diffusion time scale from an assumed 0.1 mbar plume-splashback 
deposition region to the $\sim$3 mbar H$_2$O condensation region in our model is about 150 years, so cometary water 
would be removed from the stratosphere faster than its corresponding cometary CO counterpart (note that this time scale 
is an order of magnitude longer than a previous quote from \citet{moses00b} due to an apparent typographical or 
calculation error in the previous paper).  However, if the shocked cometary material maintains the 
same CO/H$_2$O influx ratio of $\gta$ 100 as the SL9 impacts on Jupiter \citep[see][and section \ref{sec:jup} 
above]{zahnle96,lellouch96,bezard02}, then the bulk of the water currently in Saturn's stratosphere must derive 
from an additional external source, such as Enceladus plume vapor \citep{guerlet10,cassidy10,hartogh11,fleshman12}.  In that 
situation, the vertical profiles for the oxygen species shown in Fig.~\ref{figsatoxy} could be quite different.  
The comet-derived CO would be more concentrated in the middle stratosphere and less abundant elsewhere, while the 
H$_2$O could be more prevalent in the thermosphere than is shown in Fig.~\ref{figsatoxy}.  

\begin{figure*}[!htb]
\begin{center}
\includegraphics[clip=t,width=5.4in]{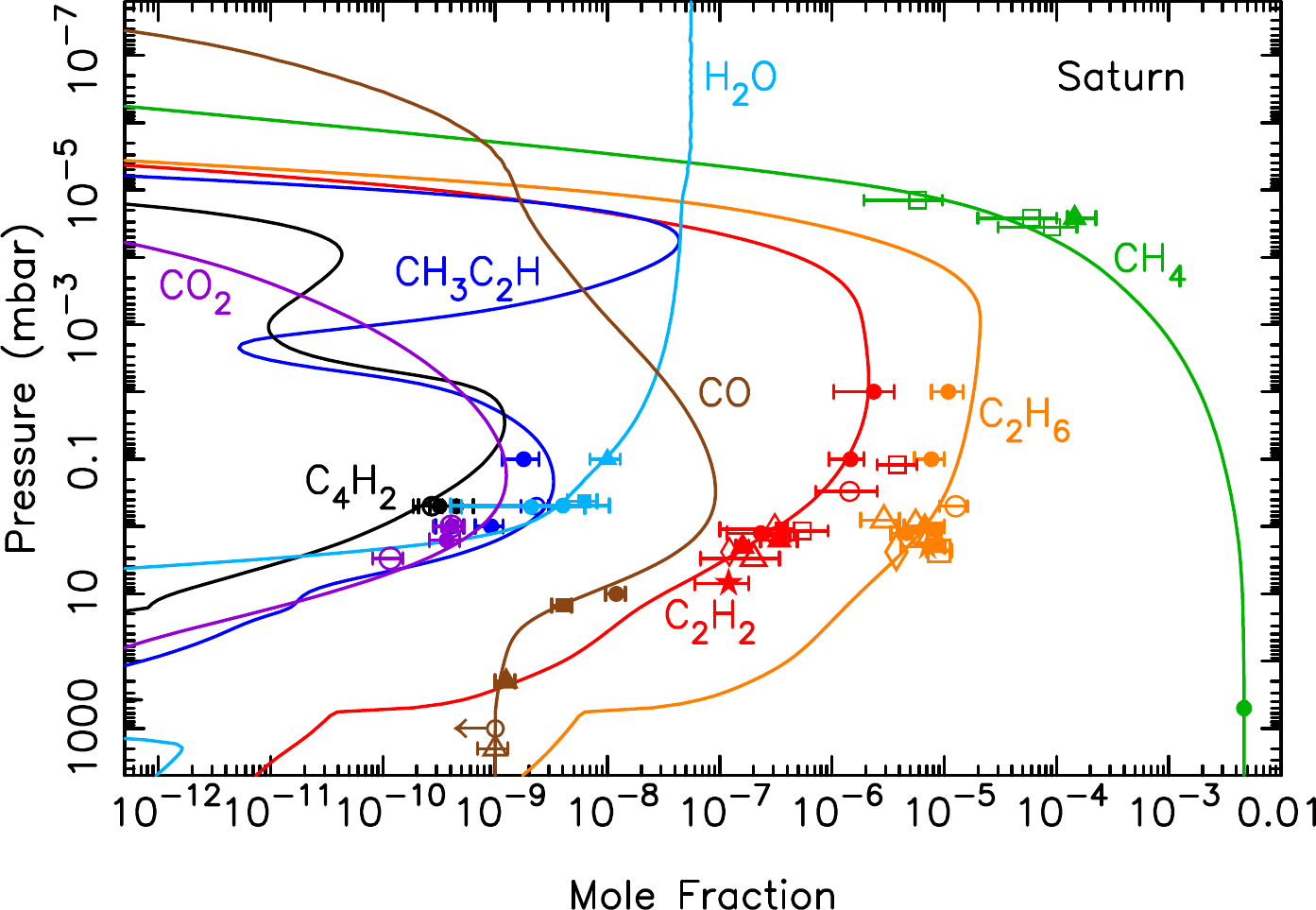}
\end{center}
\vspace{-0.5cm}
\caption{Mixing ratio profiles for several species (as labelled) in Saturn's atmosphere for a model that 
includes a source of H$_2$O flowing in at the top of the atmosphere from a local source such as Enceladus, and a 
cometary source from an impact that occurred 400 years ago, along with a background injection rate 
of 6.7$\scinot4.$ CO molecules cm$^{-2}$ s$^{-1}$ and 7.2$\scinot3.$ H$_2$O molecules cm$^{-2}$ s$^{-1}$ from 
the ablation of icy grains.\label{figsatcomet}}
\end{figure*}

We therefore consider an additional model in which H$_2$O vapor flows in from the top of the atmosphere, as with 
the possible Enceladus source, and the CO derives from an historical cometary impact (see Fig.~\ref{figsatcomet}), 
along with our predicted oxygen influx rate due to the ablation of icy grains.  
The H$_2$O flux at the top of the atmosphere due to the ``Enceladus'' source is set to 6.2$\scinot5.$ cm$^{-2}$ s$^{-1}$ 
to remain roughly consistent with the ISO, SWAS, and \textit{Herschel}/SPIRE observations \citep{moses00b,bergin00,fletcher12spire} 
and with model predictions \citep[e.g.,][]{cassidy10,hartogh11}.  
The dust ablation source is assumed to supply an integrated 6.7$\scinot4.$ CO molecules cm$^{-2}$ s$^{-1}$ and 
7.2$\scinot3.$ H$_2$O molecules cm$^{-2}$ s$^{-1}$, with no separate CO$_2$ source.  The cometary 
source of CO is assumed to have an initial mixing ratio of 3$\scinot-6.$ confined to pressures less than 0.1 mbar, 
based on \citet{cavalie10}.  Because our eddy diffusion coefficient in the relevant middle stratospheric region 
is smaller than that adopted by \citet{cavalie10}, we find that the model has to evolve for a longer time than 
the 220 $\pm$ 30 years derived by \citet{cavalie10} to bring sufficient CO down to the altitude regions in which 
the (sub)-millimeter observations are most sensitive --- the model results shown in Fig.~\ref{figsatcomet} are for 
400 years after the impact.  This model provides a reasonable fit to all the available observations of the oxygen 
species, but it is not unique.  Other combinations of the amount of cometary CO deposited, the height at which it 
was deposited, the time elapsed since the impact, and the eddy diffusion coefficient profile could provide similar 
results.

In any case, we conclude that the ablation of interplanetary dust plays a very minor role in delivering oxygen to Saturn, 
based on the observed abundances of CO and H$_2$O in comparison with our relatively low predicted dust influx rates.
Better observational determinations of the CO and H$_2$O vertical profile will be critical for determining the relative 
roles of various external and internal sources in supplying oxygen to Saturn's stratosphere.  High-spectral-resolution 
observations in the near- and far-infrared, in particular \citep[see the Jupiter observations of][]{bezard02,feuchtgruber99}, 
would be useful to have in hand before the planning stage of any future entry probe mission such as is described in 
\citet{atkinson12} or \citet{mousis14,mousis16}.

\subsection{Uranus results and comparisons with observations}\label{sec:uran}

Sluggish atmospheric mixing on Uranus prevents methane from being carried to high altitudes, resulting in a unique 
situation in which much of the ablation occurs above the CH$_4$ homopause.  In fact, water delivered by 
interplanetary dust particles will largely condense before it can photochemically interact with hydrocarbons, 
leading to less photochemical production of CO and CO$_2$ from coupled H$_2$O-CH$_4$ photochemistry.  Therefore, 
the relative abundance of H$_2$O, CO, and CO$_2$ on Uranus could provide a ``cleaner'' representation of the source 
itself and could help us to better understand the initial chemical form of the ablated vapor.  


\begin{figure*}[!htb]
\vspace{-1.5cm}
\includegraphics[clip=t,width=5.5in]{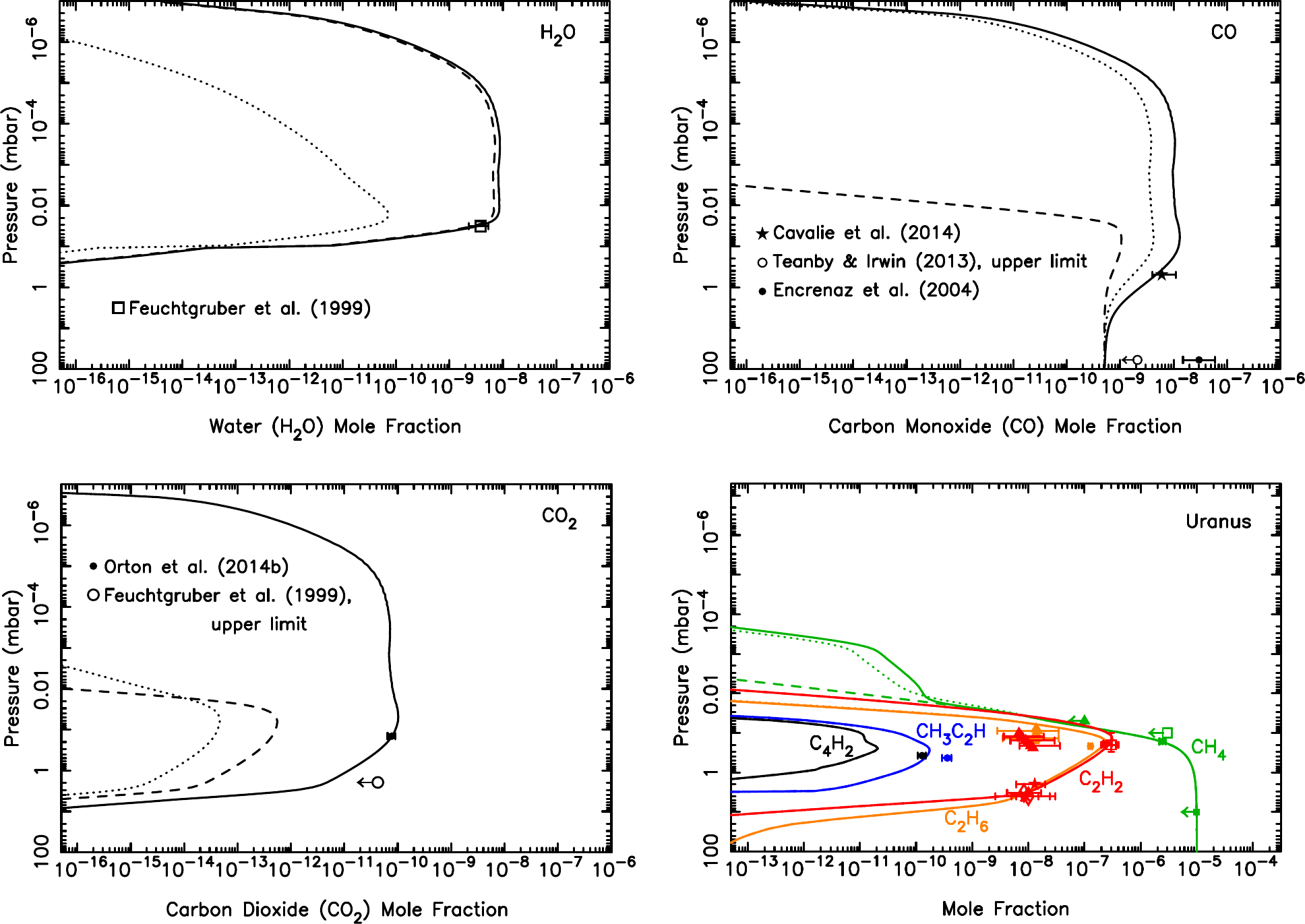}
\vspace{-0.5cm}
\caption{Mixing ratio profiles for H$_2$O (Top left), CO (Top right), CO$_2$ (Bottom left), and several 
hydrocarbons (Bottom right), as labeled, in Uranus' atmosphere as a result of the ablation of oxygen-rich 
icy grains.  The dashed lines represent a model in which all the ablated icy component is released as water 
(integrated flux of $\sim$9$\scinot4.$ H$_2$O molecules cm$^{-2}$ s$^{-1}$), the dotted lines represent a model in 
which all the ablated icy component is released as carbon monoxide (integrated flux of 9.0$\scinot4.$ CO molecules 
cm$^{-2}$ s$^{-1}$), and the solid lines represent a model in which the relative influx rates (1.2$\scinot5.$ H$_2$O 
molecules cm$^{-2}$ s$^{-1}$, 2.7$\scinot5.$ CO molecules cm$^{-2}$ s$^{-1}$, 3.0$\scinot3.$ CO$_2$ molecules 
cm$^{-2}$ s$^{-1}$) are scaled to fit the H$_2$O, CO, and CO$_2$ observations of \citet{feuchtgruber99}, 
\citet{cavalie14}, \citet{teanby13}, and \citet{orton14chem}.  The data points with error bars represent various 
observational constraints (see text).
(For interpretation of the references to color in this figure legend, the reader is referred to the web 
version of this article.)
\label{figuranoxy}}
\end{figure*}


Our dust-ablation model predicts an integrated influx of 8.9$^{+19}_{-6.1} \scinot4.$ oxygen atoms cm$^{-2}$ s$^{-1}$ to Uranus 
from the ablation of icy grains.  Fig.~\ref{figuranoxy} demonstrates that while the external delivery of this amount of 
H$_2$O could explain the infrared water observations of \citet{feuchtgruber97,feuchtgruber99}, that scenario (i.e., dashed curves 
in Fig.~\ref{figuranoxy}) cannot explain the relatively large amount of CO and CO$_2$ observed in the Uranian 
stratosphere \citep{cavalie14,orton14chem}.  On the other hand, if all that oxygen were introduced as CO (dotted curves in 
Fig.~\ref{figuranoxy}), our models predict much less H$_2$O and CO$_2$ than is observed on Uranus.  We therefore 
freely scale the relative influx rates of these three species, keeping the shape of the ablation profile the same, but 
adjusting the magnitude in order to provide a better fit to the observations.  Our best-fit model has an integrated 
influx rate of 1.2$\scinot5.$ H$_2$O molecules cm$^{-2}$ s$^{-1}$ \citep[which is fully consistent with the range 
determined by][from ISO observations and modeling]{feuchtgruber99}, 2.7$\scinot5.$ CO molecules cm$^{-2}$ s$^{-1}$ 
(the same CO influx rate derived by \citealt{cavalie14} when considering the case of a steady background influx for 
the same \citealt{orton14chem} thermal structure), and 3$\scinot3.$ CO$_2$ molecules cm$^{-2}$ s$^{-1}$ (the same 
influx rate derived by \citealt{orton14chem} from their \textit{Spitzer} spectral analysis).  The corresponding 
relative influx rates for this best-fit model are 31\% H$_2$O, 69\% CO, and 0.8\% CO$_2$.  

The total oxygen influx rate in this best-fit model is 4$\scinot5.$ oxygen atoms cm$^{-2}$ s$^{-1}$, which is roughly 
a factor of 4 greater than our original predictions from the ablation of icy grains.  This model-data mismatch 
could simply represent uncertainties in our modeling procedure, such as the extrapolation of the mass flux to 
larger grains, or it could indicate an additional external source of oxygen to Uranus, such as satellite/ring 
debris or cometary impacts.  One possible source is the interaction of the extended Uranian exosphere with the 
inner ring system \citep{esposito89}.  Based on the observed or inferred cometary impact source of CO on 
Jupiter, Saturn, and Neptune \citep{lellouch96,lellouch05,cavalie10}, a cometary source of CO is not an unexpected
possibility for Uranus, as well.  In fact, based on the outer solar system impact-rate calculations of \citet{levison97} 
and \citet{zahnle03}, comets might supply an external amount of oxygen that is of the same magnitude as the dust 
influx \citep{poppe16}.  Note, however, that our inferred CO/H$_2$O influx ratio of $\sim$2 on Uranus 
is much less than that inferred for Jupiter.  It is unclear at this point whether this difference is due to 
(1) the slower entry velocity at Uranus leading to different relative chemical processing of CO vs H$_2$O during 
cometary impacts and/or meteor entry, (2) whether there is a fundamental difference in chemical composition of 
Edgeworth-Kuiper belt dust, which dominates at Uranus, in comparison with Jupiter-family comet dust, which dominates 
at Jupiter, or (3) whether there is an additional local source of H$_2$O on Uranus, as with Enceladus on Saturn.

It should also be kept in mind that the eddy diffusion coefficient ($K_{zz}$) profile is not well constrained on Uranus due to 
uncertainties in the thermal structure, stratospheric methane profile, and related degeneracies in the modeling of infrared 
emission features \citep{orton14temp,orton14chem}.  We have not explored the sensitivity of the results to different 
eddy diffusion coefficient profiles in this paper.  \citet{orton14chem} performed numerous such sensitivity tests, including 
sloped $K_{zz}$ profiles and different combinations of tropopause CH$_4$ mixing ratio and $K_{zz}$ values.  Although not 
discussed in their paper, the Orton et al. sensitivity tests suggest that changes in the eddy diffusion coefficient profile 
have only a minor effect on the profiles of H$_2$O and CO$_2$, which both condense in the middle-to-upper stratosphere of 
Uranus, while the mixing ratio of CO could be affected by a factor of a few in the $\sim$0.03--3 mbar region.

Figure~\ref{figuranoxy} demonstrates that CO and H$_2$O are the dominant molecules (other than H$_2$) at pressures 
less than $\sim$10 $\mu$bar on Uranus.  As such, molecular ions such as HCO$^{+}$ and H$_{3}$O$^{+}$ will 
dominate over hydrocarbon ions in the lower portion of the extended Uranian ionosphere.  The effect of H$_2$O 
leading to reduced peak electron densities in giant-planet ionospheres has been well studied \citep[e.g.,][]{connerney84,nagy09}, 
and CO could play a similar role in the lower ionosphere. 
It remains to be seen whether our relatively moderate predicted oxygen influx rate 
of $\lta$ 4$\scinot5.$ cm$^{-2}$ s$^{-1}$ can provide sufficient CO and H$_2$O (and/or solid particles) in the 
thermosphere to help explain the very low electron densities observed on Uranus \citep{lindal87}, but earlier 
modeling of the process suggests not \citep{waite87,shinagawa89}.  We note that our predicted dust-derived oxygen 
influx rate of 9$\scinot4.$ cm$^{-2}$ s$^{-1}$ and the observationally inspired 4$\scinot5.$ cm$^{-2}$ s$^{-1}$ 
influx rate discussed above are both comfortably below the upper limit of 1$\scinot6.$ H$_2$O molecules cm$^{-2}$ 
s$^{-1}$ required to explain the lack of H$_2$O absorption in the \textit{Voyager 2} Ultraviolet Spectrometer 
occultation observations \citep{herbert87,shinagawa89}.

\subsection{Neptune results and comparisons with observations}\label{sec:nep}

The integrated influx rate from the ablation of icy grains supplies 7.5$^{+16}_{-5.1} \scinot5.$ O atoms cm$^{-2}$ 
s$^{-1}$ to Neptune, according to our ablation models.  This flux is consistent with the ISO observational analysis of 
\citet{feuchtgruber99}, who conclude that an H$_2$O flux of (1.2-150)$\scinot5.$ molecules cm$^{-2}$ s$^{-1}$ and 
a CO$_2$ flux of (6-7)$\scinot4.$ molecules cm$^{-2}$ s$^{-1}$ are needed to reproduce the infrared observations 
of these species on Neptune.  However, the observed CO abundance on Neptune is enormous in comparison to that on
the other giant planets 
\citep[see][]{marten93,marten05,rosenqvist92,guilloteau93,naylor94,encrenaz96,courtin96,lellouch05,lellouch10,hesman07,fletcher10akari,luszczcook13,irwin14}.  
A deep-tropospheric source from Neptune's heavy-element-rich interior could potentially explain the large observed 
CO abundance \citep[e.g.][]{lodders94,luszczcook13,cavalie17}, but the most recent series of observations listed above
confirm that the CO mixing ratio is at least a factor or 2 larger in the stratosphere than the troposphere, 
unambiguously pointing to the existence of an external source of CO for Neptune that dominates over the internal 
source.  The required external flux, which is $\sim$1$\scinot8.$ CO molecules cm$^{-2}$ s$^{-1}$ \citep[e.g.,][]{lellouch05}, 
is well outside our estimated uncertainty for the interplanetary dust delivery, but is consistent with what might be 
expected from cometary impacts \citep{luszczcook13}.  In fact, the large abundance of CO, its vertical profile, 
the large observed CO/H$_2$O ratio, and the additional presence of HCN in Neptune's stratosphere 
\citep{marten93,marten05,rosenqvist92,lellouch94,rezac14} originally prompted \citet{lellouch05} to suggest a 
cometary impact within the last couple hundred years as the source of the observed CO.  High-resolution 
submillimeter and millimeter observations of CO \citep{hesman07,luszczcook13} on Neptune continue to support this 
possibility.  The fact that the \citet{poppe16} dust dynamical models fall many orders of magnitude short in 
explaining the observed amount of CO on Neptune also makes this comet impact hypothesis very likely.


\begin{figure*}[!htb]
\vspace{-1.5cm}
\includegraphics[clip=t,width=5.0in]{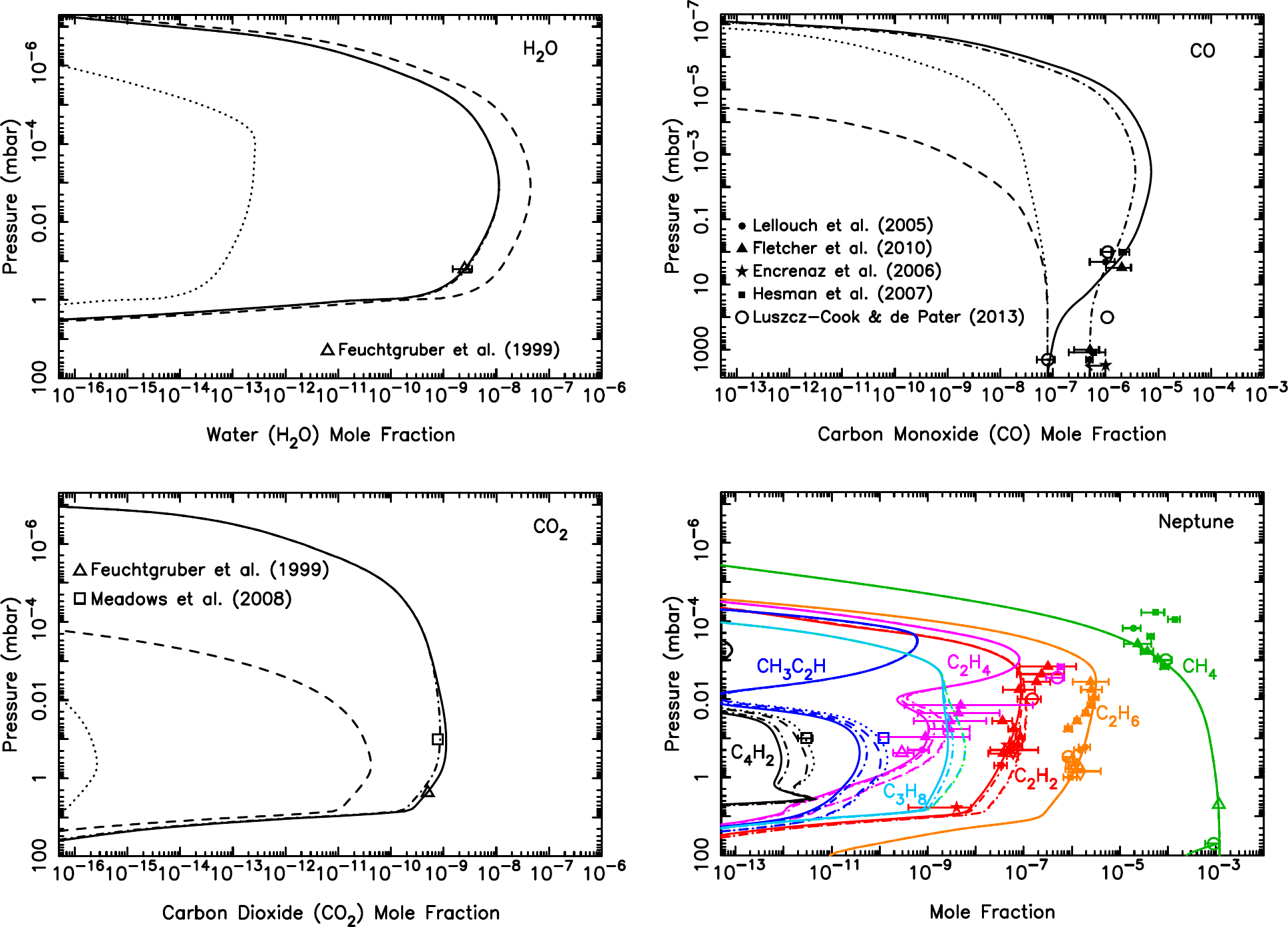}
\vspace{-0.5cm}
\caption{Mixing ratio profiles for H$_2$O (Top left), CO (Top right), CO$_2$ (Bottom left), and several 
hydrocarbons (Bottom right), as labeled, in Neptune's atmosphere as a result of the ablation of oxygen-rich 
icy grains.  The dashed lines represent a model in which all the ablated icy component is released as water 
(integrated flux of 7.5$\scinot5.$ H$_2$O molecules cm$^{-2}$ s$^{-1}$), the dotted lines represent a model in 
which all the ablated icy component is released as carbon monoxide (integrated flux of 7.5$\scinot5.$ CO molecules 
cm$^{-2}$ s$^{-1}$), and the solid lines represent a model in which the relative influx rates are scaled to 2$\scinot5.$ 
H$_2$O molecules cm$^{-2}$ s$^{-1}$, 2$\scinot8.$ CO molecules cm$^{-2}$ s$^{-1}$, and 2$\scinot4.$ CO$_2$ molecules 
cm$^{-2}$ s$^{-1}$ to fit the low tropospheric CO mixing ratio determined by \citet{luszczcook13} and the large 
stratospheric mixing ratio determined by \citet{hesman07} and \citet{fletcher10akari}.  For the dot-dashed model, 
we increased the CO mixing ratio at the lower boundary of the model and scaled the dust influx rates to 
2$\scinot5.$ H$_2$O molecules cm$^{-2}$ s$^{-1}$, 1$\scinot8.$ CO molecules cm$^{-2}$ s$^{-1}$, and 2$\scinot4.$ 
CO$_2$ molecules  cm$^{-2}$ s$^{-1}$ to compare better with the observations of \citet{lellouch05,lellouch10}.  
(For interpretation of the references to color in this figure legend, the reader is referred to the web 
version of this article.)
\label{fignepoxy}}
\end{figure*}
 

Figure \ref{fignepoxy} shows the results of the photochemical modeling for Neptune.  If we assume that all the 
vapor from the ablation of icy dust grains is released in the form of H$_2$O only (dashed curves, with an 
integrated H$_2$O influx rate of $\sim$7.5$\scinot5.$ molecules cm$^{-2}$ s$^{-1}$), then this model overestimates 
the stratospheric abundance of H$_2$O \citep{feuchtgruber99}, but underestimates 
the stratospheric abundances of CO$_2$ \citep{feuchtgruber99,meadows08} and CO
\citep{lellouch05,lellouch10,hesman07,fletcher10akari,luszczcook13}.  If we assume that all the vapor from icy 
grains is released in the form of CO only (dotted curves, with an integrated CO influx rate of 7.5$\scinot5.$ 
molecules cm$^{-2}$ s$^{-1}$), then this model underestimates the stratospheric abundance of all the observed 
oxygen species.  If we keep the vertical ablation profile the same but scale the relative magnitude of the 
influx rates such that we have 2$\scinot5.$ H$_2$O molecules cm$^{-2}$ s$^{-1}$, 2$\scinot8.$ CO molecules 
cm$^{-2}$ s$^{-1}$, and 2$\scinot4.$ CO$_2$ molecules cm$^{-2}$ s$^{-1}$ (solid curves in Fig.~\ref{fignepoxy}), 
then this model reproduces the observed H$_2$O and CO$_2$ abundances, the relatively small tropospheric CO mixing 
ratio determined by \citet{luszczcook13} (see also \citealt{irwin14}), and the relatively large stratospheric mixing 
ratio determined by \citet{hesman07} and \citet{fletcher10akari}.  However, some of the CO observations favor larger 
tropospheric CO mixing ratios and/or smaller stratospheric CO mixing ratios, so we also ran a model (dot-dashed curves in 
Fig.~\ref{fignepoxy}) with a lower-boundary CO mixing ratio of 5$\scinot-7.$ 
\citep[cf.][]{lellouch05,lellouch10,hesman07,fletcher10} and a CO influx rate of 1$\scinot8.$ CO molecules cm$^{-2}$ 
s$^{-1}$, with the fluxes of H$_2$O and CO$_2$ remaining the same as the solid-curve model.  This model also fits 
the observed stratospheric H$_2$O and CO$_2$ abundances \citep{feuchtgruber99,meadows08}.

Both the vertical profile and absolute stratospheric abundance of CO have not been firmly established to date, 
in part because of uncertainties in the atmospheric thermal structure, so we do not favor either of these ``best-fit'' 
models over the other.  Both, however, have the same inferred influx rate for CO$_2$ and H$_2$O, and both have 
very large inferred influx rates for CO.  In fact, the stratospheric CO mixing ratio in both these models exceeds
that of C$_2$H$_6$, the dominant methane photochemical product.  Such a large CO abundance has consequences for the 
hydrocarbon photochemistry, with larger CO mixing ratios leading to smaller abundances of C$_2$H$_2$ and several other 
higher-order hydrocarbons.  However, as is discussed in section \ref{sec:oxychem}, the sensitivity of the hydrocarbons 
to the CO abundance could be largely an artifact of the low-resolution ultraviolet cross sections used in the model.  When 
the CO abundance becomes large enough, it shields C$_2$H$_6$ from photolysis in the model, whereas that is 
unlikely to happen as effectively in the real atmosphere.  The hydrocarbon observational data points in 
Fig.~\ref{fignepoxy} derive from numerous ultraviolet, infrared, and sub-millimeter observations 
\citep{caldwell88,bezard91,bishop92,orton92,kostiuk92,yelle93,schulz99,burgdorf06,meadows08,fletcher10akari,greathouse11,lellouch15}.

Note that if the observed CO on Neptune derives from a large cometary impact that occurred roughly 200 years ago, 
as was suggested originally by \citet{lellouch05}, then the steady-state CO mixing-ratio profile shown in 
Fig.~\ref{fignepoxy} contains too much CO at the highest thermospheric altitudes.  The plume splashback phase of a large 
cometary impact can deposit shock-produced CO predominantly in the middle stratosphere
\citep{zahnle96,lellouch96,lellouch97}, and this CO will slowly settle through the atmosphere over 
time, causing the peak CO abundance to migrate downward and lessen in magnitude with time 
\citep[see][]{bezard02,lellouch02,lellouch06,moreno03,cavalie09,cavalie12}.  


\begin{figure*}[!htb]
\begin{center}
\includegraphics[clip=t,width=3.5in]{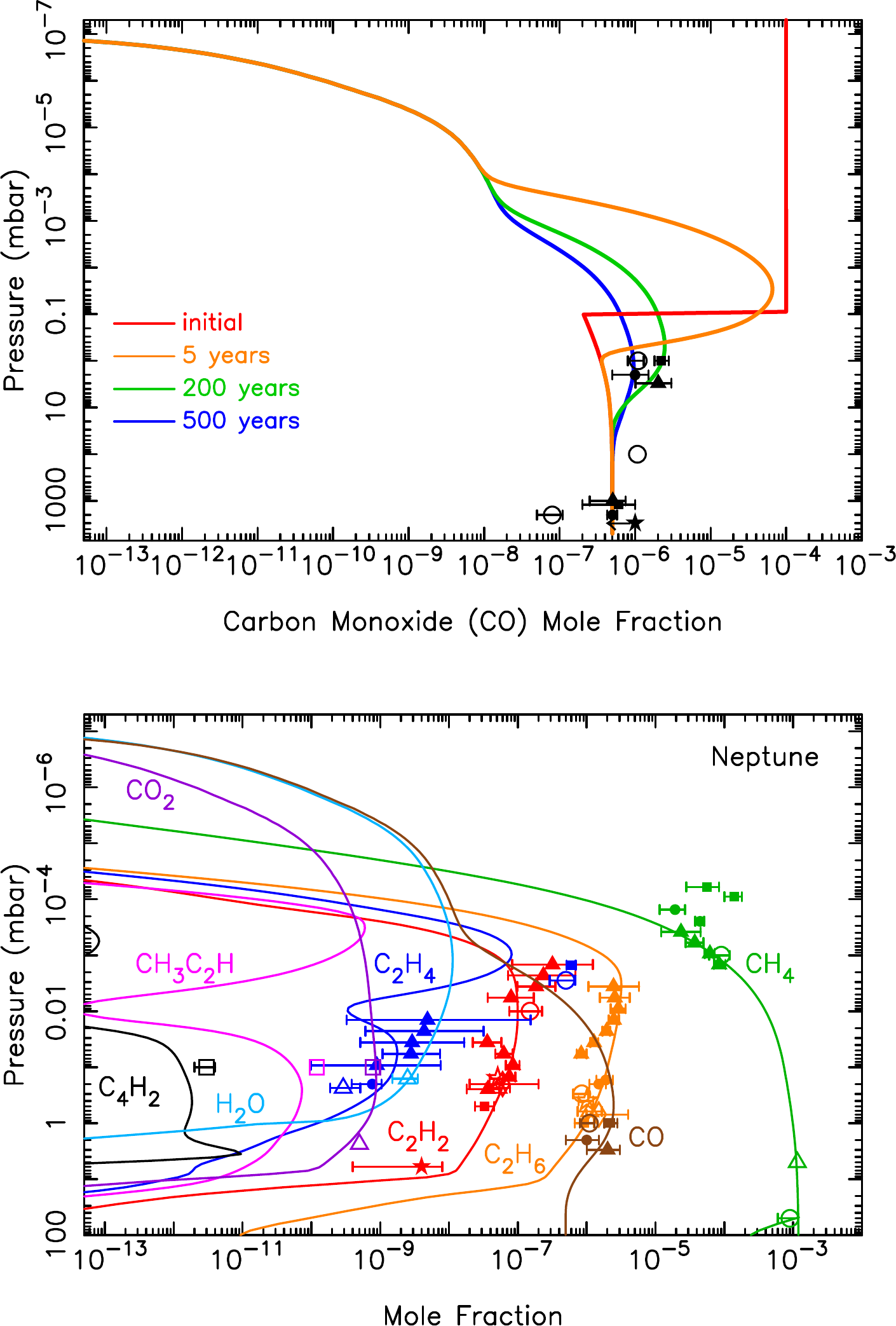}
\end{center}
\vspace{-0.5cm}
\caption{(Top) Time evolution of CO delivered from a large cometary impact, 
in combination with a smaller steady influx due to the ablation of icy grains (see text for details). (Bottom) 
Mixing ratio profiles for important hydrocarbon and oxygen species 200 years after the cometary impact 
described in the top panel.
\label{fignepcomet}}
\end{figure*}


We therefore investigate a separate case to illustrate what the CO profile would look like now from this 
scenario of a large cometary impact that occurred 200 years ago (Fig.~\ref{fignepcomet}).  
For this model, we assume that the ablation of interplanetary dust supplies a steady background influx rate of 
2$\scinot5.$ H$_2$O molecules cm$^{-2}$ s$^{-1}$, 5$\scinot5.$ CO molecules cm$^{-2}$ s$^{-1}$, and 2.3$\scinot4.$ 
CO$_2$ molecules cm$^{-2}$ s$^{-1}$ (consistent with our predicted overall oxygen influx rate); however, in addition to 
that background influx is a sudden comet-supplied CO amount with a initial mixing ratio of 1$\scinot-4.$ above 0.1 mbar 
that then evolves within the confines of the photochemical model. Figure \ref{fignepcomet} illustrates how the CO profile 
evolves with time, being lost quickly in the upper stratosphere and thermosphere due to molecular diffusion, and 
diffusing more slowly into the lower stratosphere due to eddy diffusion.  This model is presented purely for 
illustrative purposes --- the exact shape of the CO profile will depend on how much CO was originally deposited, 
the altitude at which it was deposited, the date at which it was deposited, and the eddy diffusion coefficient 
profile or other details about stratospheric circulation, none of which are well known.  The main point here is 
that although the comet-derived CO mixing ratio is very large in the middle and lower stratosphere, it is likely 
smaller in the upper stratosphere and thermosphere than is shown in Fig.~\ref{fignepoxy} for our simple dust-scaled 
case.  Because the presence of large amounts of CO would affect ionospheric chemistry, future investigations into 
aeronomical consequences of oxygen influx on the giant planets should keep the possible cometary source of this 
CO in mind.

Note also that the observed H$_2$O column abundance on Neptune is a couple orders of magnitude smaller than that of 
CO above a few millibar.  Although the very large amount of CO in Neptune's stratosphere, along with a vertical 
profile that increases with height, strongly suggests that the CO was supplied by a large cometary impact at some 
point in the recent past, the amount of H$_2$O delivered by that putative cometary impact must have either been much 
smaller than that of the CO, or any comet-delivered H$_2$O must have already diffused down from its deposition region 
to pressure levels where it would condense and be removed from the vapor phase.  In relation to this last point, the 
diffusion time scale from a potential deposition region near 0.1 mbar to the condensation region near 1 mbar in our model 
is roughly 60 years.  Cometary water would have already been removed from the stratosphere for any large impact 
that occurred much more than 60 years ago.

%

\subsection{Dominant oxygen reactions}\label{sec:oxychem}

The dominant chemical reactions influencing the oxygen species in giant-planet atmospheres are discussed in the
Saturn study of \citet{moses00b}. The key reactions are initiated by coupled water-methane photochemistry, i.e., 
the kinetics resulting from the photolysis of water and methane --- carbon monoxide is too kinetically stable 
to play a dominant role, and CO$_2$ is less abundant.  Jupiter, Saturn, and Neptune have very similar oxygen 
chemistry, whereas Uranus differs because of its low-altitude methane homopause.

On all the planets, water is lost primarily by photolysis throughout the middle and upper stratosphere and by 
condensation in the lower stratosphere.  In the background H$_2$-rich atmosphere, however, 
the OH released from water photolysis is efficiently recycled through the following dominant scheme:
\begin{eqnarray}
\HtwoO \, + \, h\nu \, & \rightarrow & \, \Hatom \, + \, \OH \nonumber \\
\OH \, + \, \Htwo \, & \rightarrow & \, \HtwoO \, + \, \Hatom \nonumber \\
\noalign{\vglue -10pt}
\multispan3\hrulefill \nonumber \cr
\Net \ \ \Htwo \, & \rightarrow & \, 2\, \Hatom  , \\
\end{eqnarray}
where $h\nu$ represents an ultraviolet photon.
Despite the relatively small rate coefficient for the reaction OH + $\Htwo$ $\rightarrow$ $\HtwoO$ + H \citep{baulch05} 
at the low atmospheric temperatures characteristic of the giant planets, the large background H$_2$ abundance 
ensures that this scheme dominates the OH loss. 

The OH that does not get recycled back to water ends up predominantly in CO, with a lesser amount in CO$_2$.
On Jupiter, Saturn, and Neptune, the main chemical schemes for photochemically converting the H$_2$O to CO involve 
the addition of OH radicals with unsaturated hydrocarbons such as C$_2$H$_2$ and C$_2$H$_4$ \citep[e.g., see 
schemes (7), (8), (14), \& (15) of][]{moses00b}, with the resulting C$_2$H$_2$OH and C$_2$H$_4$OH molecules reacting 
with H and CH$_3$ to produce species that are eventually photolyzed or react with H or hydrocarbon radicals to 
form CO.  The most efficient of these schemes from a column-integrated standpoint in the stratosphere is
\begin{eqnarray}
\HtwoO \, + \, h\nu \, & \rightarrow & \, \Hatom \, + \, \OH \nonumber \\
\OH \, + \, \CtwoHtwo \, + \, \M \, & \rightarrow & \, \CtwoHtwoOH \, + \, \M \nonumber \\
\CtwoHtwoOH \, + \, \Hatom \, & \rightarrow & \, \HCO \, + \, \CHthree \nonumber \\
\HCO \, + \, \Hatom \, & \rightarrow & \, \CO \, + \, \Htwo \nonumber \\
\Hatom \, + \, \CHthree \, + \, \M \, & \rightarrow & \, \CHfour \, + \, \M \nonumber \\
\noalign{\vglue -10pt}
\multispan3\hrulefill \nonumber \cr
\Net \ \ \HtwoO \, + \, \CtwoHtwo \, + \, 2\, \Hatom \, & \rightarrow & \, \CO \, + \, \CHfour \, + \, \Htwo  , \\
\end{eqnarray}
with M representing any third atmospheric molecule or atom.  However, these schemes represent a small 
percentage of the overall loss of H$_2$O and OH, with reactions that recycle the water dominating by more 
than an order of magnitude.

Other schemes that convert H$_2$O to CO and that involve the CH$_3$ radical (e.g., via the reactions O + CH$_3$ 
$\rightarrow$ H$_2$CO + H and OH + CH$_3$ $\rightarrow$ H$_2$CO + H$_2$) also occur and can be especially 
important in the upper stratospheres of Jupiter, Saturn, and Neptune.  On Uranus, where the column abundance of 
C$_2$H$_x$ species is small, schemes such as (2) above are relatively unimportant, and the dominant process 
converting the H$_2$O to CO from a column-integrated standpoint is
\begin{eqnarray}
\HtwoO \, + \, h\nu \, & \rightarrow & \, 2\, \Hatom \, + \, \Oatom \nonumber \\
\CHfour \, + \, h\nu \, & \rightarrow & \, \CHthree \, + \, \Hatom \nonumber \\
\Oatom \, + \, \CHthree \, & \rightarrow & \, \HtwoCO \, + \, \Hatom \nonumber \\
\HtwoCO \, + \, h\nu \, & \rightarrow & \, \Htwo \, + \, \CO \nonumber \\
\noalign{\vglue -10pt}
\multispan3\hrulefill \nonumber \cr
\Net \ \ \CHfour \, + \, \HtwoO \, & \rightarrow & \, \CO \, + \, \Htwo \, + \, 4\, \Hatom  . \\
\end{eqnarray}

Another important loss process for the water on all the giant planets is photolysis to produce OH, followed 
by reaction of the resulting OH with CO to form CO$_2$:
\begin{eqnarray}
\HtwoO \, + \, h\nu \, & \rightarrow & \, \Hatom \, + \, \OH \nonumber \\
\OH \, + \, \CO \, & \rightarrow & \, \COtwo \, + \, \Hatom \nonumber \\
\noalign{\vglue -10pt}
\multispan3\hrulefill \nonumber \cr
\Net \ \ \HtwoO \, + \, \CO \, & \rightarrow & \, \COtwo \, + \, 2\, \Hatom  . \\
\end{eqnarray}
This scheme is an important loss process for both CO and H$_2$O, and provides the dominant mechanism for producing 
CO$_2$ in these atmospheres.  Note that this scheme provides a photochemical source of carbon dioxide for the 
giant-planet stratospheres even if CO$_2$ is not directly released from the ablating grains or thermochemically 
produced during cometary impacts.

Carbon dioxide is lost by photolysis, with CO and oxygen atoms (either excited O($^1 D$) or ground state) as the 
products.  The bulk of the O($^1 D$) reacts with background H$_2$ to produce OH and eventually water, while the bulk of
the ground-state O atoms can react with CH$_3$ radicals to produce H$_2$CO and eventually lead back to CO.  
On Uranus and Neptune, the CO$_2$ is also lost through condensation in the lower stratosphere.

Carbon monoxide can be destroyed by photolysis, but only at very high altitudes, as it is shielded to a large 
extent by the more abundant H$_2$ and CH$_4$.  The dominant loss reaction for carbon monoxide is 
H + CO + M $\rightarrow$ HCO + M, 
but the bulk of the HCO produced this way ends up back as CO through reaction of HCO with atomic H or through 
photolysis.  Given the efficiency of CO recycling and the lack of permanent effective loss processes, carbon monoxide 
is very long lived in the giant-planet stratospheres.  

Other oxygen species, such as methanol, formaldehyde, ketene, and acetaldehyde are produced from the coupled 
oxygen-hydrocarbon photochemistry, but in amounts that are currently unobservable.  See \citet{moses00b} for 
their dominant production and loss reactions.

Scheme (1) above and others involving C$_2$H$_4$ and OH \citep[see][]{moses00b} 
can cause a slight reduction in the abundance of C$_2$H$_2$, C$_2$H$_4$, and other unsaturated hydrocarbons 
whose abundance depends on C$_2$H$_2$ and/or C$_2$H$_4$.  However, for the water influx rates we derive for 
the giant planets, this reduction is negligible for all planets, even Saturn, with its relatively large 
inferred H$_2$O influx rate (see Fig.~\ref{figsatoxy}).
The large CO abundance on Neptune, on the other hand, appears to affect the hydrocarbon photochemical 
products (see Fig.~\ref{fignepoxy}), reducing the abundance of C$_2$H$_2$, C$_2$H$_4$, and higher-order 
hydrocarbons whose production depends on C$_2$H$_2$ and C$_2$H$_4$.  This reduction in hydrocarbon mixing ratios 
is caused by shielding of C$_2$H$_6$ from ultraviolet radiation in the $\sim$1450--1550 \AA\ range by the 
fourth positive band system of CO, leading to a reduced photolysis rate for C$_2$H$_6$ in the middle and lower 
stratosphere.  However, this result is largely an artifact of the low spectral resolution of our UV cross 
sections and solar flux (i.e., 5-nm resolution in the relevant wavelength region).  The fourth positive system of 
CO has a lot of fine structure not captured in our model \citep[see][]{myer70}, with a fairly low continuum 
cross section between strong peaks that will not be as effective in shielding the ethane.  Future models should 
test the effect of higher-resolution cross sections.

On Uranus, the methane homopause is so deep within the atmosphere that the bulk of the ablation of icy grains occurs 
above the region where methane resides.  The coupled oxygen-carbon chemistry then occurs through CO and H$_2$O, 
or through CO alone, not through H$_2$O and CH$_4$.  In fact, the photolysis of CO actually provides a source of CH$_4$ 
in the thermosphere, which shows up as the extra bulge in the 10$^{-4}$--10$^{-2}$ mbar region in Fig.~\ref{figuranoxy}.  
The dominant mechanism for producing the CH$_4$ in this region is 
\begin{eqnarray}
\CO \, + \, h\nu \, & \rightarrow & \, \C \, + \, \Oatom \nonumber \\
\C \, + \, \Htwo \, + \, \M \, & \rightarrow & \, \threeCHtwo \, + \, \M \nonumber \\
\Hatom \, + \, \threeCHtwo \, & \rightarrow & \, \CH \, + \, \Htwo \nonumber \\
\CH \, + \, \Htwo \, + \, \M \, & \rightarrow & \, \CHthree \, + \, \M \nonumber \\
\Hatom \, + \, \CHthree \, + \, \M \, & \rightarrow & \, \CHfour \, + \, \M \nonumber \\
\noalign{\vglue -10pt}
\multispan3\hrulefill \nonumber \cr
\Net \ \ \CO \, + \, \Htwo \, + \, 2\, \Hatom \, & \rightarrow & \, \CHfour \, + \, \Oatom  . \\
\end{eqnarray}

\section{Other potential chemical processing of the ablated vapor}\label{sec:thermo}

The observations described in sections 3.2--3.5 indicate that CO is more abundant than H$_2$O in the stratospheres 
of the giant planets.  Large cometary impacts may explain the high CO/H$_2$O ratios on Neptune \citep{lellouch05}, 
Saturn \citep{cavalie10}, and in the southern hemisphere (at least) of Jupiter \citep{lellouch97,lellouch02}, with 
the oxygen from the comet being thermochemically converted to CO in high-temperature shocks during a plume-splashback 
phase of the impact \citep{zahnle96}.  However, our predicted magnitude of the ``background'' oxygen influx 
to Jupiter and Uranus from the ablation of ice-rich dust grains is consistent with the amount needed to explain the 
CO in lower stratospheres of these planets.  In fact, on Jupiter, our predicted oxygen ablation rate from icy grains 
is a factor of $\sim$250 larger than is needed to explain the low background amount of water not related to the recent 
Comet Shoemaker-Levy 9 impacts \citep{lellouch02}.  Why, then, is the background water abundance on Jupiter so low?  
We demonstrate 
in section \ref{sec:oxychem} that photochemistry cannot efficiently convert ablated water vapor to CO in giant-planet 
stratospheres.  Does some other chemical processing occur during the meteoroid entry phase or immediately after that can 
explain the relatively large CO/H$_2$O ratios in the stratospheres of Jupiter and Uranus?

\begin{figure*}[!htb]
\begin{center}
\includegraphics[clip=t,width=5.4in]{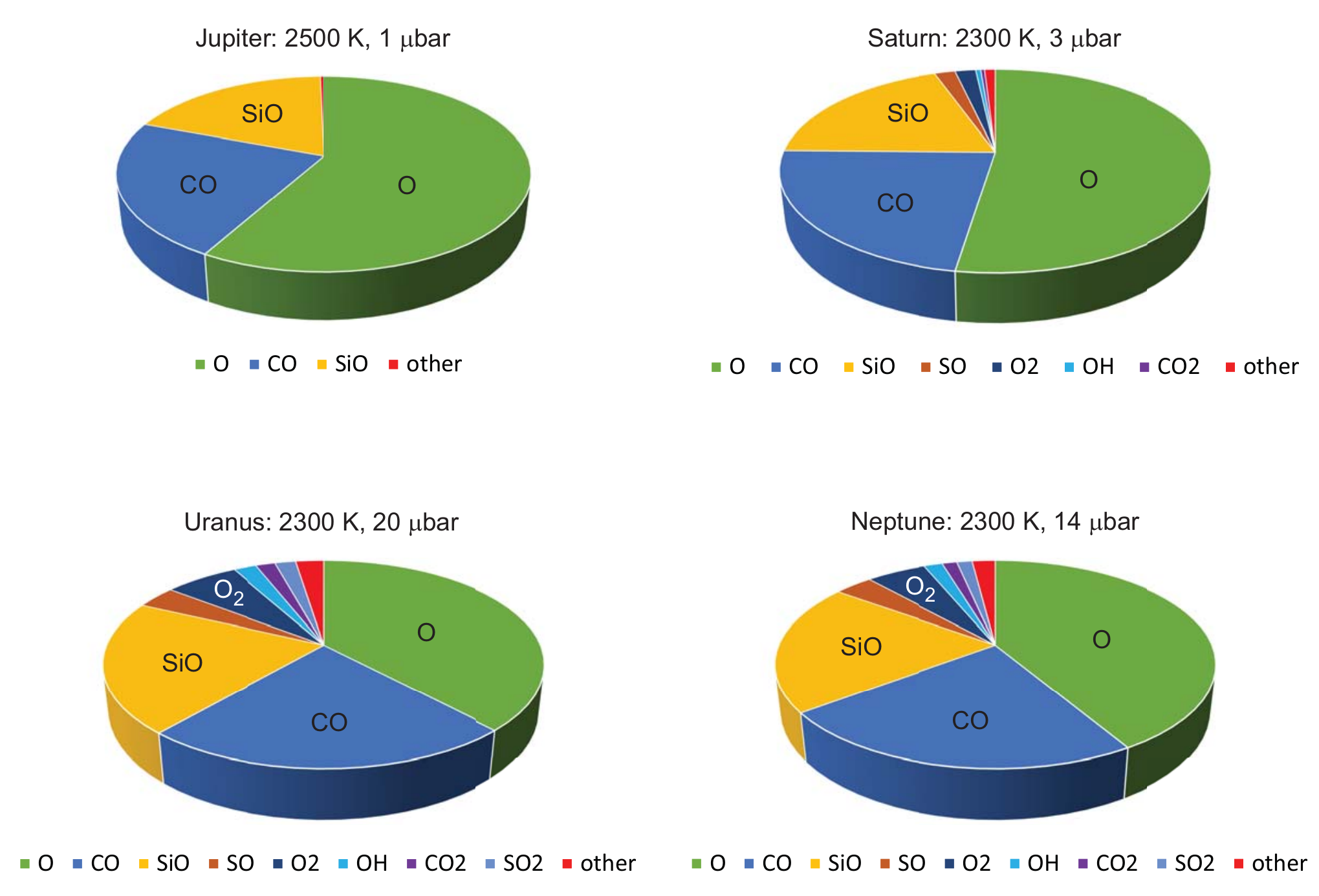}
\end{center}
\vspace{-0.5cm}
\caption{Partitioning of the oxygen-bearing vapor species in thermochemical equilibrium for a cometary 
composition grain \citep[][see text]{lisse07} at the temperatures and pressures corresponding to the 
peak of the silicate ablation profile for Jupiter (Top left), Saturn (Top right), Uranus (Bottom left), 
Neptune (Bottom right).\label{figthermo}}
\end{figure*}

One potential source of the CO is reaction of the oxygen in ices with the carbon from within the grains 
themselves.  From a thermochemical equilibrium standpoint, O and CO are the favored forms of the oxygen at the high 
temperatures and low pressures relevant to the silicate ablation process, with H$_2$O being a very minor component.  If 
reactions between vapor species within the grain itself occur as the grain heats up, or if reactions occur within 
the meteor trail, the tendency of these kinetic reactions could be to drive the oxygen toward O and CO if there is 
sufficient time for these reactions to occur (and note that the grains typically remain at the relevant high 
temperatures for tens of seconds or less).  Thus, if differential ablation is not very effective during meteoroid 
entry and the silicate phases dominate the grain heating behavior, then water would be a minor species being released 
from the grains.  For example, Fig.~\ref{figthermo} shows the oxygen vapor partitioning in thermochemical equilibrium  
at pressure and temperature conditions relative to the peak of the silicate ablation profile for grain material of 
the same composition as the coma of Comet Hale Bopp \citep{lisse07}.  Here, we assume a solar ratio for elements not 
considered by \citet{lisse07}, with P/C and N/C remaining in solar proportions, and other elements 
remaining in solar proportions relative to Si, and thermochemical equilibrium is calculated as described in 
\citet{moses13gj436}.  Note from Fig.~\ref{figthermo} that O, CO, and SiO dominate the 
oxygen vapor under these conditions, with H$_2$O coming in at $\ll$ 1\%.

Another possibility is dissociation of any molecular vapor phases following ablation, simply resulting from the high 
velocities of entry.  According to the dust dynamical modeling results of \citet{poppe16}, grain velocities at 
atmospheric entry are in the range 59.5--70.5 km s$^{-1}$ for Jupiter, 35.5--44.5 km s$^{-1}$ for Saturn, 20.5--29.5 
km s$^{-1}$ for Uranus, and 22.5--28.5 km s$^{-1}$ for Neptune.  From the point of view of the just-ablated 
oxygen-bearing molecules, they are experiencing collisions with H$_2$ molecules that have incoming energies of 37--52 eV 
for Jupiter, 13--21 eV for Saturn, 4.4--9.1 eV for Uranus, and 5.3--8.5 eV for Neptune.  The ionization energy for 
H$_2$O is 12.62 eV \citep{reutt86}, so the ablated water molecules can be ionized on Jupiter and Saturn, but not 
Uranus and Neptune.  The ionization energies for O, CO, and SiO are 13.6, 14.0, and 11.6 eV \citep{linstrom17,hildenbrand69}, 
so these species can also be potentially ionized on Jupiter and Saturn, but not Uranus and Neptune.  The H--OH bond 
energy is 5.1 eV, and that of O--H is 4.4 eV \citep{okabe78}, so dissociation of the H$_2$O and OH can occur on all 
the planets while the water is still being decelerated.  The bond energies of C--O and Si--O are 11.09 and 7.93 
eV, respectively \citep{okabe78,hildenbrand69}, so while collisions with H$_2$ may have sufficient energy to 
dissociate CO and SiO on Jupiter and Saturn, that is not necessarily the case on Uranus and Neptune.  The final ionization 
and energy states of the oxygen products will be important in determining the ultimate fate of the oxygen from 
further chemical reactions; unfortunately, we could not find experimental or theoretical data on collisions of H$_2$ 
with H$_2$O, CO, SiO at relevant energies.  

If these collisions lead to the formation of OH, O$^+$, or excited oxygen atoms such as O($^1 D$), the main final 
product will be water, due to the effectiveness of reactions of these species with H$_2$.  If the collisions primarily 
produce atomic O, which seems likely based on the above arguments, then the oxygen can end up in either CO or H$_2$O,
depending on whether the ablation is occurring above or below the methane homopause.  Below the CH$_4$ homopause, atomic 
oxygen will largely end up in CO through the reactions O + CH$_3$ $\rightarrow$ H$_2$CO + H (followed by 
photolysis and subsequent reactions to form CO) and O + CH$_3$ $\rightarrow$ CO + H$_2$ + H.  Above the homopause, 
the atomic O is longer lived and reacts relatively slowly with H$_2$ to form OH and eventually H$_2$O, but the vapor 
diffuses downward sufficiently slowly on all the giant planets that any atomic O ablated above the methane homopause 
will largely end up as H$_2$O.

\begin{figure*}[!htb]
\begin{center}
\includegraphics[clip=t,width=5.4in]{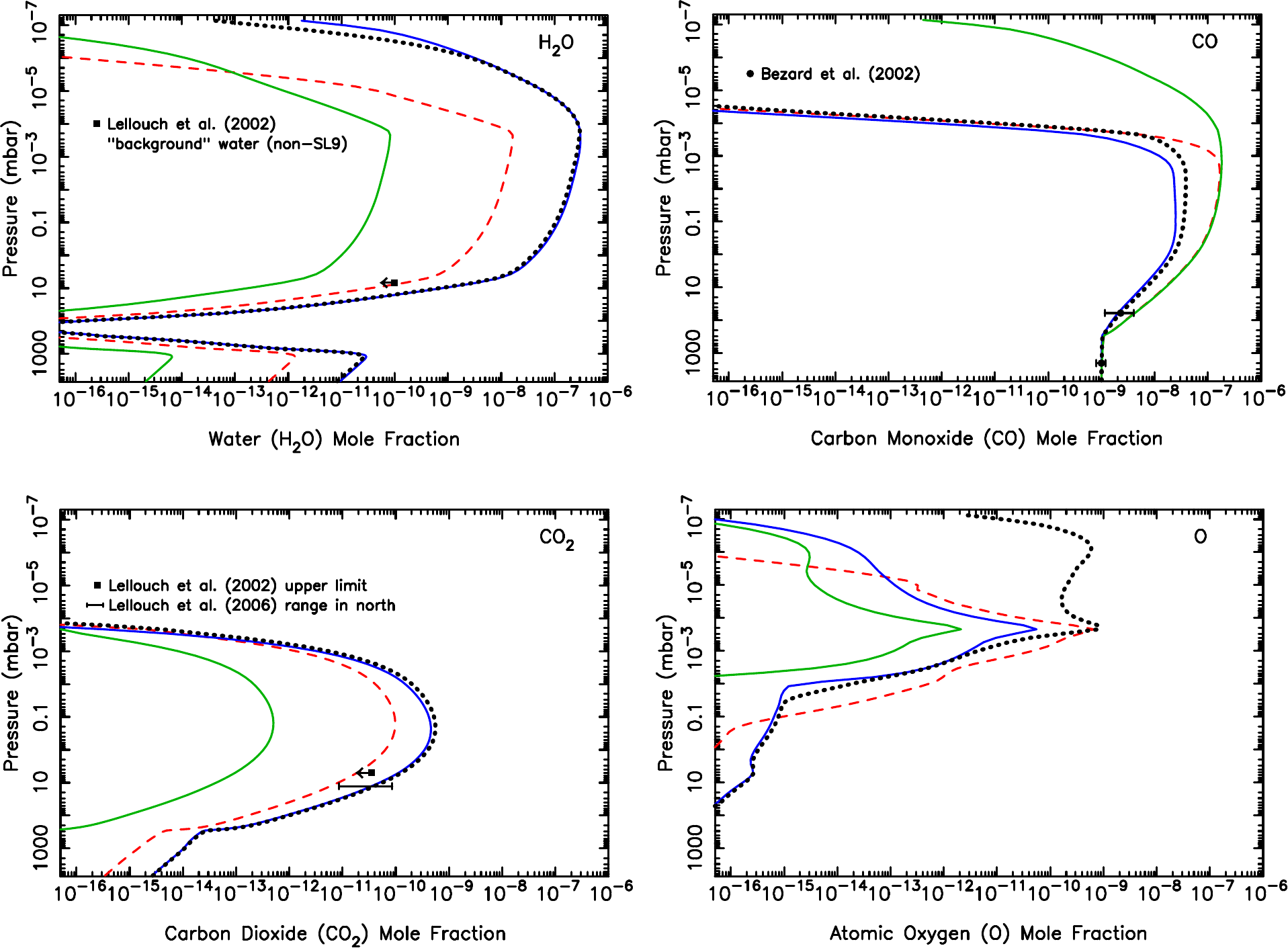}
\end{center}
\vspace{-0.5cm}
\caption{Mixing ratio profiles for H$_2$O (Top left), CO (Top right), CO$_2$ (Bottom left), and O (Bottom 
right) on Jupiter under the assumption that all the oxygen in the ices arrives in the atmosphere as (1) CO, 
with the icy grain ablation profile shown in Fig.~\ref{prodmjup} (green curves); (2) H$_2$O, with the 
icy grain ablation profile shown in Fig.~\ref{prodmjup} (blue curves); (3) O, with the icy grain ablation 
profile shown in Fig.~\ref{prodmjup} (dotted black curves); (4) O, with the silicate grain ablation
profile shown in Fig.~\ref{prodmjup} (dashed red curves).  Above the methane homopause, chemical reactions 
largely convert the atomic O to H$_2$O, while below the homopause, the O is converted to CO.
(For interpretation of the references to color in this figure legend, the reader is referred to the web 
version of this article.)
\label{figjupdusto}}
\end{figure*}

Figure \ref{figjupdusto} demonstrates the fate of the volatile oxygen in the incoming grains on Jupiter if it 
were all quickly converted to atomic O once released from the grains.  Two different cases are shown.  The first 
case (black dotted curves) assumes that differential ablation dominates, such that the oxygen from the icy component 
of the grains is released first, at high altitudes, and the second case (red dashed curves) assumes that simple 
ablation dominates, such that the oxygen from icy grains is released only when the silicate phases ablate 
at lower altitudes (see Fig.~\ref{prodmjup}).  Because a significant fraction of the silicate ablation occurs 
below the methane homopause, a significant fraction of the O released in the latter case ends up in CO, while 
the O in the first case overwhelmingly ends up as H$_2$O.  Note that the first case results in too much H$_2$O 
in Jupiter's atmosphere compared with the observationally inferred ``background'' water abundance (i.e., the 
H$_2$O not related to the SL9 impacts, \citealt{lellouch02}), while the second case provides a decent fit to 
the H$_2$O, CO, and CO$_2$ observations for the background abundances not related to SL9.  

This better fit from the low-altitude ablation case could be used as an argument in favor of the simple ablation 
process for Jupiter.  Alternatively, it is possible that the available oxygen in the incoming grains is tied up 
in silicates and/or hydrated silicates in the first place, rather than in ices.  Jupiter-family comet grains spend 
more of their lifetime at smaller heliocentric distances than their Oort-cloud or Edgeworth-Kuiper belt counterparts 
\citep{poppe16}, and it is possible that sublimation of the ice phases has occurred before the Jupiter-family comet 
grains enter the Jovian atmosphere.  However, studies subjecting carbonaceous chondrite samples to stepped pyrolysis 
\citep{court14} find that H$_2$O is released at a variety of temperatures, ranging from less than 600 K (perhaps due 
to desorption of terrestrial water contamination) to 700--900 K (dehydration of hydrated minerals) to $>$ 1100 K 
(mineral decomposition), so it might be expected that the ablation of hydrated-mineral phases would release water 
at temperatures intermediate between our assumed organic and silicate cases, rather than requiring much higher 
magnesium-silicate vaporization temperatures.  In any event, more sophisticated ablation models that consider 
realistic particle compositions and structures, as well as the immediate fate of the ablated vapor, will be needed
to shed more light on the interesting puzzle of the low background water abundance in the Jovian stratosphere. 

\begin{figure*}[!htb]
\begin{center}
\includegraphics[clip=t,width=5.4in]{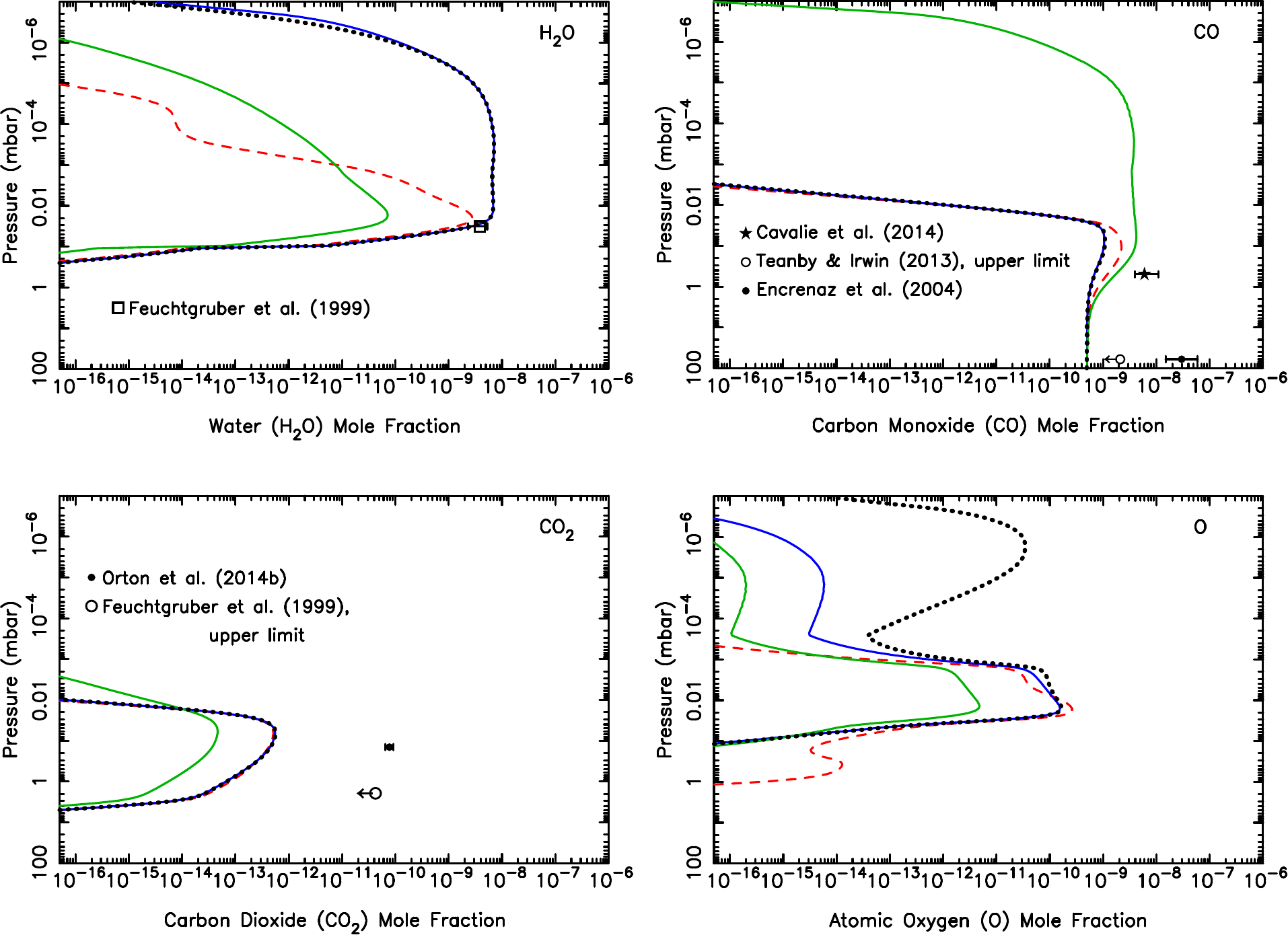}
\end{center}
\vspace{-0.5cm}
\caption{Mixing ratio profiles for H$_2$O (Top left), CO (Top right), CO$_2$ (Bottom left), and O (Bottom 
right) on Uranus under the assumption that all the oxygen in the ices arrives in the atmosphere as (1) CO, 
with the icy grain ablation profile shown in Fig.~\ref{prodmuran} (green curves); (2) H$_2$O, with the 
icy grain ablation profile shown in Fig.~\ref{prodmuran} (blue curves); (3) O, with the icy grain ablation 
profile shown in Fig.~\ref{prodmuran} (dotted black curves); (4) O, with the silicate grain ablation
profile shown in Fig.~\ref{prodmuran} (dashed red curves).  Above the methane homopause, chemical reactions 
largely convert the atomic O to H$_2$O, while below the homopause, the O is converted to CO.
(For interpretation of the references to color in this figure legend, the reader is referred to the web 
version of this article.)
\label{figurandusto}}
\end{figure*}

Fig.~\ref{figurandusto} illustrates the same cases for Uranus.  Because the methane homopause is so deep in the 
atmosphere on Uranus (due to weak atmospheric mixing), the dust grains ablate largely in the methane-free thermosphere,
particularly for the differential-ablation case where the ice ablates high in the atmosphere (black dotted curves). 
Therefore, the O is converted to H$_2$O in that case, and the results are not too different from the assumption of 
the oxygen arriving purely as H$_2$O from icy grain ablation.  On the other hand, a portion of the silicate ablation 
curve for Uranus (see Fig.~\ref{prodmuran}) falls within and below the methane homopause in the 0.01--0.1 mbar 
region, so more of the O is converted to CO under the assumption of simple ablation, for which the silicate phases 
dominate the gas release.  Even in that case, however, there is insufficient CO and CO$_2$ being produced to 
explain the observations of \citet{cavalie14} and \citet{orton14chem}.  As discussed in section \ref{sec:uran}, uncertainties 
in our meteoroid influx and ablation calculations could potentially be the cause of the model-data mismatch, 
combined with the possibility of any CO being released during the ablation process remaining as CO due to the 
lower entry energies at Uranus.  However, the fact that CO$_2$ is grossly underpredicted in the models suggests 
that either CO$_2$ is also being released from the grains and survives the ablation process and collisional aftermath 
(CO--O bond energy is 5.453 eV \citealt{okabe78}; see the solid line model shown in Fig.~\ref{figuranoxy}), or that 
Uranus has also experienced a large cometary impact within the 
last few hundred years that deposited both CO and CO$_2$.  Note that because H$_2$O condenses at such high altitudes 
on Uranus, it is possible that a non-trivial fraction of oxygen from any putative cometary impact could be tied up in 
water-ice hazes in the stratosphere.

\section{Conclusions}\label{sec:conclusions}

Small interplanetary dust grains are continually bombarding the upper atmospheres of Jupiter, Saturn, 
Uranus, and Neptune.  Ablation of these grains during high-velocity atmospheric entry can deliver oxygen, 
silicates, metals, and other material to the thermospheres and stratospheres of these planets --- material 
that is otherwise not intrinsically present at such altitudes.  Using the \citet{poppe16} dynamical model 
predictions for the dust populations in the outer solar system, we have modeled the meteoroid ablation 
process on the giant planets, along with the subsequent oxygen-hydrocarbon neutral stratospheric photochemistry 
that results from meteoroid ablation.  

We find that dust ablation occurs over a broad altitude range within the upper atmospheres of the giant planets.  
Under the assumption that differential ablation dominates \citep{mcneil98}, whereby different materials within 
the grain ablate at different points as the grain heats up, our models predict that the volatile icy 
components within the grains will ablate at the highest altitudes, beginning well above the methane homopause 
but extending down into the upper stratosphere where methane is present.  The more refractory components take 
longer to reach temperatures high enough to vaporize the material, so the ablation of these components (i.e., silicates, 
metals) extends well below the methane homopause and into the middle stratospheres of the planets.  Organic 
components will ablate at altitudes intermediate between those of the ice and silicate components.  The larger impact 
velocities at Jupiter and Saturn lead to higher grain temperatures during atmospheric entry for any particular 
particle size and material properties, leading to more efficient ablation than on Uranus and Neptune.  In 
fact, impact velocities at Uranus and Neptune are low enough that not all the silicates within the grains will 
completely ablate before the grains are decelerated and cool.  The lower the initial mass of a grain coming 
into Uranus and Neptune, the smaller the mass fraction that is lost during the ablation process.

We tracked the fate of the ablated oxygen-bearing species with a 1D photochemical model.  The molecular 
and/or atomic makeup of the ablated vapor is not obvious from first principles, so we tested the sensitivity 
of the photochemical results to the form of the ablated vapor.  Nominal best-fit models were created by adjusting 
the relative fraction of ablated molecules such as H$_2$O, CO, and CO$_2$ to produce the best fits to observations 
of these species on the giant planets.

The following main results are obtained from the theoretical calculations and model-data comparisons:

\begin{itemize}
\item The effective oxygen influx rate to Jupiter from the ablation of the icy components within the grains 
is 1.0$^{+2.2}_{-0.7} \, \scinot7.$ O atoms cm$^{-2}$ s$^{-1}$.  Dust from Jupiter-family comets dominates this 
influx rate, and the magnitude is consistent with what is needed to explain the large ``background'' CO
abundance in the Jovian lower stratosphere and upper troposphere that is unrelated to the recent Comet 
Shoemaker-Levy 9 impacts \citep{bezard02}.  However, because the background (non-SL9) H$_2$O abundance is 
inferred to be quite small in Jupiter's stratosphere \citep{lellouch02}, we conclude that the oxygen ablating 
from the incoming grains must be released directly in the form of CO, must be thermochemically converted 
to CO soon after ablation, or must be released as atomic O below the methane homopause, with subsequent 
photochemistry involving reactions of atomic O with CH$_3$ converting the oxygen to CO.  If the ablated oxygen 
were released as H$_2$O or OH, or as O or O$^+$ above the methane homopause, with only photochemical reactions 
shaping the speciation, then the predicted background water abundance in the Jovian stratosphere would be much 
larger than is observed.
\item Given the large expected oxygen influx rates to Jupiter from the ablation of interplanetary dust, and 
the requirement that the ablated oxygen predominantly end up as CO rather than H$_2$O, we conclude that 
dust ablation plays a major role in delivering external oxygen to Jupiter.  The delivery of oxygen by small 
comets, invoked by \citet{bezard02}, may still be important on Jupiter but is not required to 
explain the observations.
\item The effective oxygen influx rate to Saturn from the ablation of the ices within the incoming interplanetary
dust particles is 7.4$^{+16}_{-5.1} \, \scinot4.$ O atoms cm$^{-2}$ s$^{-1}$.  The interplanetary dust at Saturn 
derives from a mix of Jupiter-family comet, Oort-cloud comet, and Edgeworth-Kuiper belt dust populations.  Our 
predicted oxygen influx rate from interplanetary dust is more than an order of magnitude too small to explain the 
observed stratospheric H$_2$O, CO, and CO$_2$ abundances on Saturn
\citep{feuchtgruber97,moses00b,bergin00,cavalie09,cavalie10,abbas13}, so we conclude that interplanetary dust 
plays a minor role in delivering external oxygen to Saturn.  Instead, Saturn likely gets its external oxygen 
from water ejected from Enceladus \citep[e.g.,][]{jurac07,cassidy10,hartogh11,fleshman12}, from water and other 
oxygen from the rings \citep[e.g.,][]{connerney84,luhmann06,tseng10,moore15}, and/or from CO deposited from 
a large cometary impact that occurred a couple hundred years ago \citep[e.g.,][]{cavalie10}.
\item Because of the Enceladus and ring sources of H$_2$O, Saturn likely has a larger thermospheric 
H$_2$O/CO ratio than all the other giant planets (cf. Figs.~9, 11, 12, 14).  The incoming water will affect 
ionospheric chemistry and structure \citep[see][]{connerney84,mosesbass00,moore15}, but our models demonstrate
that at the H$_2$O influx rate needed from these sources to explain the observed stratospheric water abundance 
on Saturn, coupled oxygen-hydrocarbon photochemistry does not have a notable affect on the abundances of 
neutral hydrocarbon photochemical products \citep[see also][]{moses00b}.
\item The effective oxygen influx rate to Uranus from the ablation of the icy grain components 
is 8.9$^{+19}_{-6.1} \, \scinot4.$ O atoms cm$^{-2}$ s$^{-1}$.  Edgeworth-Kuiper belt grains supply the largest 
fraction at Uranus, but Oort-cloud comet dust and Jupiter-family comet dust populations also contribute.  Our 
predicted oxygen influx rate here is somewhat smaller than is needed to explain the observed amount of H$_2$O, 
CO, and CO$_2$ in the stratosphere of Uranus \citep{feuchtgruber99,cavalie14,orton14chem}, which our best-fit 
models indicate is of the order $\sim$4$\scinot5.$ O atoms cm$^{-2}$ s$^{-1}$.  This underprediction could simply 
be a consequence of uncertainties in our dust population or ablation calculations/assumptions, or it could 
suggest an additional source of external oxygen to Uranus, such as from cometary impacts.  Impact rate studies 
for the outer solar system \citep{levison97,zahnle03} suggest that kilometer and sub-kilometer size comets may 
deliver an oxygen amount similar to what we derive from interplanetary dust \citep{poppe16}.  This result, along 
with the fact that photochemical models that just consider ablated vapor in the form of CO or H$_2$O fail to 
reproduce the CO$_2$ abundance, and the fact that most of the ablation occurs above the methane homopause and 
so favors H$_2$O over CO as a final photochemical product from the ablated oxygen, combine to suggest that a 
cometary impact within the past few hundred years may contribute notably to the stratospheric CO and CO$_2$ 
currently seen on Uranus.
\item The comparatively high inferred H$_2$O/CO influx-rate ratio on Uranus in comparison to Jupiter and Neptune
may be a consequence of the weak Uranian atmospheric mixing and resulting low-altitude methane homopause.  Most of the 
dust ablation takes place at altitudes where atmospheric methane is not present, so the photochemistry that occurs 
after the oxygen is ablated favors H$_2$O production over CO production.  Alternatively, the high H$_2$O/CO ratio 
may indicate a local source of water from the Uranian rings or satellites.
\item The effective oxygen influx rate to Neptune from the ablation of the ices within the incoming interplanetary
dust particles is 7.5$^{+16}_{-5.1} \, \scinot5.$ O atoms cm$^{-2}$ s$^{-1}$, with the bulk of the dust deriving 
from the Edgeworth-Kuiper belt.  Although this dust influx rate can support the observed amount of H$_2$O in 
Neptune's atmosphere \citep{feuchtgruber97,feuchtgruber99}, it is more than an order of magnitude too small to explain
the huge amount of CO seen in Neptune's middle stratosphere \citep[e.g.,][]{lellouch05,hesman07,fletcher10,luszczcook13}. 
This model-data mismatch, along with the CO vertical profile that requires an external source and the fact that HCN 
was also discovered in Neptune's stratosphere \citep[e.g.,][]{marten93}, strongly suggests that Neptune experienced 
a large cometary impact within the past few hundred years \citep[see][]{lellouch05}.  Based on our dust ablation 
calculations, the bulk of the stratospheric oxygen on Neptune at the present time derives from that cometary impact, 
and not from interplanetary dust.
\item Coupled oxygen-hydrocarbon photochemistry is not very effective in the stratospheres of the giant 
planets.  Although H$_2$O is lost readily by UV photolysis, the main product, OH, reacts with H$_2$ to reform 
the water, so the water is quickly recycled and remains stable.  A small fraction of oxygen originally in water can 
be converted to CO through addition reactions of OH with C$_2$H$_2$ and C$_2$H$_4$, followed by further reactions that 
eventually produce CO (see section \ref{sec:oxychem}). At the inferred H$_2$O influx rates for the giant planets, these 
reactions have very little effect on hydrocarbon abundances.  By the same token, CO is shielded to some extent from 
photolysis below the methane homopause, and any atomic oxygen thus formed ends up largely reacting with CH$_3$ to 
eventually reform the CO.  At the inferred CO influx rates for the giant planets, we only see CO affecting hydrocarbon 
abundances on Neptune --- through the shielding of C$_2$H$_6$ from photolysis --- and that result is probably an 
artifact of the low-resolution UV cross sections in our model.  Coupled CO-H$_2$O photochemistry produces CO$_2$,
through the reaction CO + OH $\rightarrow$ CO$_2$ + H.
\item Cometary impacts naturally deliver a much greater amount of CO than H$_2$O to giant-planet stratospheres 
\citep[e.g.,][]{zahnle96,lellouch96}, but the high observed CO/H$_2$O ratio on the giant planets could also be 
a consequence of CO and O being the favored forms of volatile oxygen in thermochemical equilibrium at the high temperatures 
and low pressures encountered during the ablation of cometary-composition grains.  If simple ablation dominates over 
differential ablation, such that the volatile oxygen in the grains is released by ablation below the homopause, and if 
subsequent collisions with ambient H$_2$ molecules dissociate molecular species to produce atomic O, then further 
photochemical reactions will strongly favor CO over H$_2$O in giant-planet atmospheres.  Note that the CO abundance 
in giant-planet stratospheres rivals that of the major hydrocarbon photochemical products, a fact that is not widely 
appreciated (see Figs.~9, 11, 12, 14).
\item The apparently large external source of CO on the giant planets from both comets and the ablation of 
interplanetary dust complicates the determination of the deep oxygen abundance on the giant planets
\citep[e.g.][]{prinn77,lewis84,fegley85,fegley86,fegley94,lodders02,visscher05,visscher10co,visscher11,wang15,wang16,cavalie17}.  
The externally supplied CO is transported downward into the troposphere, adding to the quenched component upwelling 
from the interior.  Therefore, it may be difficult to uniquely separate the CO mixing ratio resulting 
from the interior source, which then makes it difficult to indirectly determine the deep H$_2$O abundance on 
the giant planets from this method.
\end{itemize}

The inferred influx rates from our photochemical models that provide the best overall fits to the available 
H$_2$O, CO, and CO$_2$ observations are shown in Table~\ref{tab:bestfit}.  Note that these solutions are not unique, 
so the influx values should be considered as representative only.  The ``effective'' oxygen influx rates ---
independent of the form in which the oxygen was introduced --- given by \citet{poppe16} provide a robust measure 
of the amount of total oxygen delivered to the giant planets from interplanetary dust impacts, independent of 
any photochemical modeling uncertainties.

\begin{table}[h]
 \caption{Dust-Supplied Oxygen Influx Rates (cm$^{-2}$ s$^{-1}$) From Our Best-Fit Photochemical Models}
  \vspace{6pt}
 \begin{tabular}{lccccc}
  \hline
                 &                 &                  &                   & extra source  & extra source  \\
                 &  CO             & H$_2$O           & CO$_2$            & of H$_2$O needed?  & of CO needed? \\
  \hline
Jupiter          & 7$\scinot6.$    &  4$\scinot4.$    &  1$\scinot5.$     & Yes (from SL9)      & Yes (from SL9) \\
Saturn$^{\ast}$  & 6.7$\scinot4.$  &  7.2$\scinot3.$  &  $<$ 1$\scinot3.$ & Yes (Enceladus)     & Yes (comet?) \\
Uranus           & 2.7$\scinot5.$  &  1.2$\scinot5.$  &  3$\scinot3.$     & No?                 & No?    \\
Neptune$^{\ast}$ & 5$\scinot5.$    &  2$\scinot5.$    &  2.3$\scinot4.$   & No                  & Yes (comet)    \\
  \hline
 \end{tabular}
 $^{\ast}$Using the cometary models shown in Figs.~\ref{figsatcomet} \& \ref{fignepcomet}.
 \label{tab:bestfit}
\end{table}

The ablation of interplanetary dust will have other interesting consequences for giant-planet atmospheres.  The 
addition of oxygen to the thermosphere --- potentially in the form of both CO, which has typically been ignored 
to date, as well as H$_2$O --- will affect ionospheric chemistry and structure through reactions with the main ions H$^+$ 
(for H$_2$O) and H$_3$$^+$ (for CO), with both H$_3$O$^+$ and HCO$^+$ becoming important components in the lower 
ionosphere.  Other vapor phases released from ablation could also affect ionospheric chemistry and structure, with 
long-lived atomic metal ions being particularly important in the lower ionosphere. Residual unablated grains or 
tiny recondensed meteoric debris particles could also affect ionospheric structure by becoming a sink for electrons and 
could provide a source of condensation nuclei that aids nucleation and condensation of other gas-phase species in 
the lower stratosphere.  Water introduced to the stratosphere from meteoroid ablation will condense in the lower 
stratosphere of all the giant planets, and CO$_2$ will condense on Uranus and Neptune.  The additional aerosols 
provided from the ablation of interplanetary dust can potentially affect hydrocarbon condensation at lower altitudes, 
as well as atmospheric temperatures, atmospheric transmission in the UV, and heterogeneous chemistry on grain 
surfaces.  

The ablation models presented here contain a lot of simplifying assumptions, such as uniform composition, uniform 
heating throughout the grain, and spherical particles.  More sophisticated future models could consider mixed 
compositions, diffusion of gases through the grain, sputtering, fragmentation, thermochemical reactions within the heated grain, 
and other more realistic treatments of the meteor stage.  Future models should also more realistically track the 
immediate fate of the energetic gases released during ablation, in terms of the consequences of further energetic 
collisions with atmospheric gases before the ablated vapor becomes thermalized.  We have focused here on the 
consequences to neutral atmospheric chemistry, but as mentioned above, the consequences of the ablation of 
interplanetary dust for ionospheric chemistry and the electron density profiles on the giant planets could be quite 
interesting.  It would also be worthwhile to investigate how the aerosols that result from the ablation process 
influence other physical and chemical processes in the atmosphere as the particles rain down through the atmosphere 
\citep[e.g.,][]{frankland16}.

Although interplanetary dust particles represent only a tiny incremental mass addition to the giant planets, the 
consequences of the continual dust bombardment can have important observable consequences for the upper atmospheres 
of these planets.

\section{Acknowledgments}\label{sec:acknow}

This material is based on research supported by the National Aeronautics and Space Administration (NASA) Science 
Mission Directorate under grant NNX13AG55G from the now-defunct Planetary Atmospheres Research Program.  We thank 
M. J. Wolff for assistance with the Mie scattering calculations, and Thibault Cavali{\'e} and an anonymous reviewer 
for useful comments that improved the manuscript.



\bibliographystyle{elsarticle-harv} 
\bibliography{references}

\begin{thebibliography}{183}
\expandafter\ifx\csname natexlab\endcsname\relax\def\natexlab#1{#1}\fi
\expandafter\ifx\csname url\endcsname\relax
  \def\url#1{\texttt{#1}}\fi
\expandafter\ifx\csname urlprefix\endcsname\relax\def\urlprefix{URL }\fi

\bibitem[{{Abbas} et~al.(2013){Abbas}, {LeClair}, {Woodard}, {Young},
  {Stanbro}, {Flasar}, {Kunde}, {Achterberg}, {Bjoraker}, {Brasunas},
  {Jennings}, and {the Cassini/CIRS Team}}]{abbas13}
{Abbas}, M.~M., {LeClair}, A., {Woodard}, E., {Young}, M., {Stanbro}, M.,
  {Flasar}, F.~M., {Kunde}, V.~G., {Achterberg}, R.~K., {Bjoraker}, G.,
  {Brasunas}, J., {Jennings}, D.~E., {the Cassini/CIRS Team}, 2013.
  {Distribution of CO$_{2}$ in Saturn's atmosphere from Cassini/CIRS infrared
  observations}. \apj 776, 73.

\bibitem[{{Allen} et~al.(1981){Allen}, {Yung}, and {Waters}}]{allen81}
{Allen}, M., {Yung}, Y.~L., {Waters}, J.~W., 1981. Vertical transport and
  photochemistry in the terrestrial mesosphere and lower thermosphere (50-120
  km). \jgr 86, 3617--3627.

\bibitem[{{Atkinson} et~al.(2012){Atkinson}, {Spilker}, {Lunine},
  {Simon-Miller}, {Atreya}, {Colaprete}, {Coustenis}, {Reh}, and
  {Spilker}}]{atkinson12}
{Atkinson}, D.~H., {Spilker}, T.~R., {Lunine}, J.~I., {Simon-Miller}, A.~A.,
  {Atreya}, S.~K., {Colaprete}, A., {Coustenis}, A., {Reh}, K.~R., {Spilker},
  L.~J., 2012. {Science from a Saturn Entry Probe Mission}. AGU Fall Meeting
  Abstracts, \#P13C--1960.

\bibitem[{{Baulch} et~al.(2005){Baulch}, {Bowman}, {Cobos}, {Cox}, {Just},
  {Kerr}, {Pilling}, {Stocker}, {Troe}, {Tsang}, {Walker}, and
  {Warnatz}}]{baulch05}
{Baulch}, D.~L., {Bowman}, C.~T., {Cobos}, C.~J., {Cox}, R.~J., {Just}, T.,
  {Kerr}, J.~A., {Pilling}, M.~J., {Stocker}, D., {Troe}, J., {Tsang}, W.,
  {Walker}, R.~W., {Warnatz}, J., 2005. {Evaluated kinetic data for combustion
  modeling: Supplement II}. J. Phys. Chem. Ref. Data 34, 757--1397.

\bibitem[{{Beer}(1975)}]{beer75}
{Beer}, R., 1975. Detection of carbon monoxide in {Jupiter}. \apjl 200,
  L167--L169.

\bibitem[{{Beer} and {Taylor}(1975)}]{beer78}
{Beer}, R., {Taylor}, F.~W., 1975. The abundance of carbon monoxide in
  {Jupiter}. \apjl 200, L167--L169.

\bibitem[{{Bergin} et~al.(2000){Bergin}, {Lellouch}, {Harwit}, {Gurwell},
  {Melnick}, {Ashby}, {Chin}, {Erickson}, {Goldsmith}, {Howe}, {Kleiner},
  {Koch}, {Neufeld}, {Patten}, {Plume}, {Schieder}, {Snell}, {Stauffer},
  {Tolls}, {Wang}, {Winnewisser}, and {Zhang}}]{bergin00}
{Bergin}, E.~A., {Lellouch}, E., {Harwit}, M., {Gurwell}, M.~A., {Melnick},
  G.~J., {Ashby}, M.~L.~N., {Chin}, G., {Erickson}, N.~R., {Goldsmith}, P.~F.,
  {Howe}, J.~E., {Kleiner}, S.~C., {Koch}, D.~G., {Neufeld}, D.~A., {Patten},
  B.~M., {Plume}, R., {Schieder}, R., {Snell}, R.~L., {Stauffer}, J.~R.,
  {Tolls}, V., {Wang}, Z., {Winnewisser}, G., {Zhang}, Y.~F., 2000.
  {Submillimeter Wave Astronomy Satellite} observations of {Jupiter} and
  {Saturn}: {Detection} of 557 {GHz} water emission from the upper atmosphere.
  \apjl 539, L147--L150.

\bibitem[{{B{\'e}zard} et~al.(2002){B{\'e}zard}, {Lellouch}, {Strobel},
  {Maillard}, and {Drossart}}]{bezard02}
{B{\'e}zard}, B., {Lellouch}, E., {Strobel}, D., {Maillard}, J.-P., {Drossart},
  P., 2002. {Carbon monoxide on Jupiter: Evidence for both internal and
  external sources}. Icarus 159, 95--111.

\bibitem[{{B{\'e}zard} et~al.(2001){B{\'e}zard}, {Moses}, {Lacy}, {Greathouse},
  {Richter}, and {Griffith}}]{bezard01dps}
{B{\'e}zard}, B., {Moses}, J.~I., {Lacy}, J., {Greathouse}, T., {Richter}, M.,
  {Griffith}, C., 2001. {Detection of ethylene (C$_{2}$H$_{4}$) on Jupiter and
  Saturn in non-auroral regions}. In: Bulletin of the American Astronomical
  Society. p. 1079.

\bibitem[{{B{\'e}zard} et~al.(1991){B{\'e}zard}, {Romani}, {Conrath}, and
  {Maguire}}]{bezard91}
{B{\'e}zard}, B., {Romani}, P.~N., {Conrath}, B.~J., {Maguire}, W.~C., 1991.
  {Hydrocarbons in Neptune's stratosphere from Voyager infrared observations}.
  J. Geophys. Res. 96, 18,961--18,975.

\bibitem[{{Bishop} et~al.(1992){Bishop}, {Atreya}, {Romani}, {Sandel}, and
  {Herbert}}]{bishop92}
{Bishop}, J., {Atreya}, S.~K., {Romani}, P.~N., {Sandel}, B.~R., {Herbert}, F.,
  1992. {Voyager 2 ultraviolet spectrometer solar occultations at Neptune:
  Constraints on the abundance of methane in the stratosphere}. \jgr 97,
  11681--11694.

\bibitem[{{Bjoraker} et~al.(1986){Bjoraker}, {Larson}, and
  {Kunde}}]{bjoraker86}
{Bjoraker}, G.~L., {Larson}, H.~P., {Kunde}, V.~G., 1986. {The gas composition
  of Jupiter derived from 5 micron airborne spectroscopic observations}. Icarus
  66, 579--609.

\bibitem[{{Bockel{\'e}e-Morvan}(2011)}]{bockeleemorvan11}
{Bockel{\'e}e-Morvan}, D., 2011. An overview of comet composition. In:
  {Cernicharo}, J., {Bachiller}, R. (Eds.), The Molecular Universe. Vol. 280 of
  IAU Symposium. pp. 261--274.

\bibitem[{{Burgdorf} et~al.(2006){Burgdorf}, {Orton}, {van Cleve}, {Meadows},
  and {Houck}}]{burgdorf06}
{Burgdorf}, M., {Orton}, G., {van Cleve}, J., {Meadows}, V., {Houck}, J., 2006.
  {Detection of new hydrocarbons in Uranus' atmosphere by infrared
  spectroscopy}. Icarus 184, 634--637.

\bibitem[{{Burns} et~al.(1979){Burns}, {Lamy}, and {Soter}}]{burns79}
{Burns}, J.~A., {Lamy}, P.~L., {Soter}, S., 1979. {Radiation forces on small
  particles in the solar system}. Icarus 40, 1--48.

\bibitem[{{Caldwell} et~al.(1988){Caldwell}, {Wagener}, and
  {Fricke}}]{caldwell88}
{Caldwell}, J., {Wagener}, R., {Fricke}, K.-H., 1988. {Observations of Neptune
  and Uranus below 2000 A with the IUE}. Icarus 74, 133--140.

\bibitem[{{Cassidy} and {Johnson}(2010)}]{cassidy10}
{Cassidy}, T.~A., {Johnson}, R.~E., 2010. Collisional spreading of {Enceladus'}
  neutral cloud. Icarus 209, 696--703.

\bibitem[{{Cavali{\'e}} et~al.(2008{\natexlab{a}}){Cavali{\'e}}, {Billebaud},
  {Biver}, {Dobrijevic}, {Lellouch}, {Brillet}, {Lecacheux}, {Hjalmarson},
  {Sandqvist}, {Frisk}, {Olberg}, {Bergin}, and {Odin Team}}]{cavalie08}
{Cavali{\'e}}, T., {Billebaud}, F., {Biver}, N., {Dobrijevic}, M., {Lellouch},
  E., {Brillet}, J., {Lecacheux}, A., {Hjalmarson}, {\AA}., {Sandqvist}, A.,
  {Frisk}, U., {Olberg}, M., {Bergin}, E.~A., {Odin Team}, 2008{\natexlab{a}}.
  {Observation of water vapor in the stratosphere of Jupiter with the Odin
  space telescope}. Planet. Space Sci. 56, 1573--1584.

\bibitem[{{Cavali{\'e}} et~al.(2009){Cavali{\'e}}, {Billebaud}, {Dobrijevic},
  {Fouchet}, {Lellouch}, {Encrenaz}, {Brillet}, {Moriarty-Schieven},
  {Wouterloot}, and {Hartogh}}]{cavalie09}
{Cavali{\'e}}, T., {Billebaud}, F., {Dobrijevic}, M., {Fouchet}, T.,
  {Lellouch}, E., {Encrenaz}, T., {Brillet}, J., {Moriarty-Schieven}, G.~H.,
  {Wouterloot}, J.~G.~A., {Hartogh}, P., 2009. {First observation of CO at 345
  GHz in the atmosphere of Saturn with the JCMT: New constraints on its
  origin}. Icarus 203, 531--540.

\bibitem[{{Cavali{\'e}} et~al.(2008{\natexlab{b}}){Cavali{\'e}}, {Billebaud},
  {Fouchet}, {Lellouch}, {Brillet}, and {Dobrijevic}}]{cavalie08co}
{Cavali{\'e}}, T., {Billebaud}, F., {Fouchet}, T., {Lellouch}, E., {Brillet},
  J., {Dobrijevic}, M., 2008{\natexlab{b}}. {Observations of CO on Saturn and
  Uranus at millimeter wavelengths: new upper limit determinations}. \aap 484,
  555--561.

\bibitem[{{Cavali{\'e}} et~al.(2012){Cavali{\'e}}, {Biver}, {Hartogh},
  {Dobrijevic}, {Billebaud}, {Lellouch}, {Sandqvist}, {Brillet}, {Lecacheux},
  {Hjalmarson}, {Frisk}, {Olberg}, and {Odin Team}}]{cavalie12}
{Cavali{\'e}}, T., {Biver}, N., {Hartogh}, P., {Dobrijevic}, M., {Billebaud},
  F., {Lellouch}, E., {Sandqvist}, A., {Brillet}, J., {Lecacheux}, A.,
  {Hjalmarson}, {\AA}., {Frisk}, U., {Olberg}, M., {Odin Team}, 2012. {Odin
  space telescope monitoring of water vapor in the stratosphere of Jupiter}.
  Planet. Space Sci. 61, 3--14.

\bibitem[{{Cavali{\'e}} et~al.(2013){Cavali{\'e}}, {Feuchtgruber}, {Lellouch},
  {de Val-Borro}, {Jarchow}, {Moreno}, {Hartogh}, {Orton}, {Greathouse},
  {Billebaud}, {Dobrijevic}, {Lara}, {Gonz{\'a}lez}, and {Sagawa}}]{cavalie13}
{Cavali{\'e}}, T., {Feuchtgruber}, H., {Lellouch}, E., {de Val-Borro}, M.,
  {Jarchow}, C., {Moreno}, R., {Hartogh}, P., {Orton}, G., {Greathouse}, T.~K.,
  {Billebaud}, F., {Dobrijevic}, M., {Lara}, L.~M., {Gonz{\'a}lez}, A.,
  {Sagawa}, H., 2013. {Spatial distribution of water in the stratosphere of
  Jupiter from Herschel HIFI and PACS observations}. \aap 553, 21.

\bibitem[{{Cavali{\'e}} et~al.(2010){Cavali{\'e}}, {Hartogh}, {Billebaud},
  {Dobrijevic}, {Fouchet}, {Lellouch}, {Encrenaz}, {Brillet}, and
  {Moriarty-Schieven}}]{cavalie10}
{Cavali{\'e}}, T., {Hartogh}, P., {Billebaud}, F., {Dobrijevic}, M., {Fouchet},
  T., {Lellouch}, E., {Encrenaz}, T., {Brillet}, J., {Moriarty-Schieven},
  G.~H., 2010. {A cometary origin for CO in the stratosphere of Saturn?} \aap
  510, A88.

\bibitem[{{Cavali{\'e}} et~al.(2014){Cavali{\'e}}, {Moreno}, {Lellouch},
  {Hartogh}, {Venot}, {Orton}, {Jarchow}, {Encrenaz}, {Selsis}, {Hersant}, and
  {Fletcher}}]{cavalie14}
{Cavali{\'e}}, T., {Moreno}, R., {Lellouch}, E., {Hartogh}, P., {Venot}, O.,
  {Orton}, G.~S., {Jarchow}, C., {Encrenaz}, T., {Selsis}, F., {Hersant}, F.,
  {Fletcher}, L.~N., 2014. {The first submillimeter observation of CO in the
  stratosphere of Uranus}. \aap 562, A33.

\bibitem[{{Cavali{\'e}} et~al.(2017){Cavali{\'e}}, {Venot}, {Selsis},
  {Hersant}, {Hartogh}, and {Leconte}}]{cavalie17}
{Cavali{\'e}}, T., {Venot}, O., {Selsis}, F., {Hersant}, F., {Hartogh}, P.,
  {Leconte}, J., 2017. {Thermochemistry and vertical mixing in the tropospheres
  of Uranus and Neptune: How convection inhibition can affect the derivation of
  deep oxygen abundances}. Icarus 291, 1--16.

\bibitem[{{Chen} et~al.(1991){Chen}, {Judge}, {Wu}, {Caldwell}, {White}, and
  {Wagener}}]{chen91}
{Chen}, F., {Judge}, D.~L., {Wu}, C.~Y.~R., {Caldwell}, J., {White}, H.~P.,
  {Wagener}, R., 1991. {High-resolution, low-temperature photoabsorption cross
  sections of C$_2$H$_2$, PH$_3$, AsH$_3$, and GeH$_4$, with application to
  Saturn's atmosphere}. \jgr 96, 17.

\bibitem[{{Connerney}(1986)}]{connerney86}
{Connerney}, J.~E.~P., 1986. {Magnetic connection for Saturn's rings and
  atmosphere}. \grl 13, 773--776.

\bibitem[{{Connerney} and {Waite}(1984)}]{connerney84}
{Connerney}, J.~E.~P., {Waite}, J.~H., 1984. {New model of Saturn's ionosphere
  with an influx of water from the rings}. Nature 312, 136--138.

\bibitem[{{Court} and {Sephton}(2014)}]{court14}
{Court}, R.~W., {Sephton}, M.~A., 2014. {New estimates of the production of
  volatile gases from ablating carbonaceous micrometeoroids at Earth and Mars
  during an E-belt-type Late Heavy Bombardment}. Geochim. Cosmochim. Acta 145,
  175--205.

\bibitem[{{Courtin} et~al.(1984){Courtin}, {Gautier}, {Marten}, {B{\'e}zard},
  and {Hanel}}]{courtin84}
{Courtin}, R., {Gautier}, D., {Marten}, A., {B{\'e}zard}, B., {Hanel}, R.,
  1984. {The composition of Saturn's atmosphere at northern temperate latitudes
  from Voyager IRIS spectra - NH$_3$, PH$_3$, C$_2$H$_2$, C$_2$H$_6$, CH$_3$D,
  CH$_4$, and the Saturnian D/H isotopic ratio}. \apj 287, 899--916.

\bibitem[{{Courtin} et~al.(1996){Courtin}, {Gautier}, and
  {Strobel}}]{courtin96}
{Courtin}, R., {Gautier}, D., {Strobel}, D., 1996. {The CO abundance on Neptune
  from HST observations}. Icarus 123, 37--55.

\bibitem[{{Cravens}(1994)}]{cravens94}
{Cravens}, T.~E., 1994. {Comet Shoemaker-Levy-9 impact with Jupiter:
  Aeronomical predictions}. \grl 21, 1075--1078.

\bibitem[{{de Graauw} et~al.(1997){de Graauw}, {Feuchtgruber}, {Bezard},
  {Drossart}, {Encrenaz}, {Beintema}, {Griffin}, {Heras}, {Kessler}, {Leech},
  {Lellouch}, {Morris}, {Roelfsema}, {Roos-Serote}, {Salama}, {Vandenbussche},
  {Valentijn}, {Davis}, and {Naylor}}]{degraauw97}
{de Graauw}, T., {Feuchtgruber}, H., {Bezard}, B., {Drossart}, P., {Encrenaz},
  T., {Beintema}, D.~A., {Griffin}, M., {Heras}, A., {Kessler}, M., {Leech},
  K., {Lellouch}, E., {Morris}, P., {Roelfsema}, P.~R., {Roos-Serote}, M.,
  {Salama}, A., {Vandenbussche}, B., {Valentijn}, E.~A., {Davis}, G.~R.,
  {Naylor}, D.~A., 1997. First results of {ISO-SWS} observations of {Saturn}:
  detection of {CO$_2$}, {CH$_3$C$_2$H}, {C$_4$H$_2$}, and tropospheric
  {H$_2$O}. \aap 321, L13--L16.

\bibitem[{{Dello Russo} et~al.(2016){Dello Russo}, {Kawakita}, {Vervack}, and
  {Weaver}}]{dellorusso16}
{Dello Russo}, N., {Kawakita}, H., {Vervack}, R.~J., {Weaver}, H.~A., 2016.
  Emerging trends and a comet taxonomy based on the volatile chemistry measured
  in thirty comets with high-resolution infrared spectroscopy between 1997 and
  2013. Icarus 278, 301--332.

\bibitem[{{Dougherty} et~al.(2006){Dougherty}, {Khurana}, {Neubauer},
  {Russell}, {Saur}, {Leisner}, and {Burton}}]{dougherty06}
{Dougherty}, M.~K., {Khurana}, K.~K., {Neubauer}, F.~M., {Russell}, C.~T.,
  {Saur}, J., {Leisner}, J.~S., {Burton}, M.~E., 2006. {Identification of a
  dynamic atmosphere at Enceladus with the Cassini Magnetometer}. Science 311,
  1406--1409.

\bibitem[{{Encrenaz} et~al.(2004){Encrenaz}, {Lellouch}, {Drossart},
  {Feuchtgruber}, {Orton}, and {Atreya}}]{encrenaz04}
{Encrenaz}, T., {Lellouch}, E., {Drossart}, P., {Feuchtgruber}, H., {Orton},
  G.~S., {Atreya}, S.~K., 2004. {First detection of CO in Uranus}. \aap 413,
  L5--L9.

\bibitem[{{Encrenaz} et~al.(1996){Encrenaz}, {Serabyn}, and
  {Weisstein}}]{encrenaz96}
{Encrenaz}, T., {Serabyn}, E., {Weisstein}, E.~W., 1996. {Millimeter
  spectroscopy of Uranus and Neptune: Constraints on CO and PH $_{3}$
  tropospheric abundances}. Icarus 124, 616--624.

\bibitem[{{Esposito} and {Colwell}(1989)}]{esposito89}
{Esposito}, L.~W., {Colwell}, J.~E., 1989. {Creation of the Uranus rings and
  dust bands}. Nature 339, 605--607.

\bibitem[{{Fegley} and {Lodders}(1994)}]{fegley94}
{Fegley}, Jr., B., {Lodders}, K., 1994. Chemical models of the deep atmospheres
  of {Jupiter} and {Saturn}. Icarus 110, 117--154.

\bibitem[{{Fegley} and {Prinn}(1985)}]{fegley85}
{Fegley}, Jr., B., {Prinn}, R.~G., 1985. {Equilibrium and nonequilibrium
  chemistry of Saturn's atmosphere: Implications for the observability of
  PH$_3$, N$_2$, CO, and GeH$_4$}. \apj 299, 1067--1078.

\bibitem[{{Fegley} and {Prinn}(1986)}]{fegley86}
{Fegley}, Jr., B., {Prinn}, R.~G., 1986. Chemical models of the deep atmosphere
  of {Uranus}. \apj 307, 852--865.

\bibitem[{{Festou} and {Atreya}(1982)}]{festou82}
{Festou}, M.~C., {Atreya}, S.~K., 1982. {Voyager ultraviolet stellar
  occultation measurements of the composition and thermal profiles of the
  Saturnian upper atmosphere}. \grl 9, 1147--1150.

\bibitem[{{Feuchtgruber} et~al.(1997){Feuchtgruber}, {Lellouch}, {de Graauw},
  {B{\'e}zard}, {Encrenaz}, and {Griffin}}]{feuchtgruber97}
{Feuchtgruber}, H., {Lellouch}, E., {de Graauw}, T., {B{\'e}zard}, B.,
  {Encrenaz}, T., {Griffin}, M., 1997. External supply of oxygen to the
  atmospheres of the giant planets. Nature 389, 159--162.

\bibitem[{{Feuchtgruber} et~al.(1999){Feuchtgruber}, {Lellouch}, {Encrenaz},
  {Bezard}, {Coustenis}, {Drossart}, {Salama}, {de Graauw}, and
  {Davis}}]{feuchtgruber99}
{Feuchtgruber}, H., {Lellouch}, E., {Encrenaz}, T., {Bezard}, B., {Coustenis},
  A., {Drossart}, P., {Salama}, A., {de Graauw}, T., {Davis}, G.~R., 1999.
  Oxygen in the stratospheres of the giant planets and {Titan}. In: {Cox}, P.,
  {Kessler}, M. (Eds.), The Universe as Seen by ISO. Vol. SP-427 of ESA Special
  Publication. pp. 133--136.

\bibitem[{{Fleshman} et~al.(2012){Fleshman}, {Delamere}, {Bagenal}, and
  {Cassidy}}]{fleshman12}
{Fleshman}, B.~L., {Delamere}, P.~A., {Bagenal}, F., {Cassidy}, T., 2012. The
  roles of charge exchange and dissociation in spreading {Saturn's} neutral
  clouds. \jgr 117, E05007.

\bibitem[{{Fletcher} et~al.(2010{\natexlab{a}}){Fletcher}, {Achterberg},
  {Greathouse}, {Orton}, {Conrath}, {Simon-Miller}, {Teanby}, {Guerlet},
  {Irwin}, and {Flasar}}]{fletcher10}
{Fletcher}, L.~N., {Achterberg}, R.~K., {Greathouse}, T.~K., {Orton}, G.~S.,
  {Conrath}, B.~J., {Simon-Miller}, A.~A., {Teanby}, N., {Guerlet}, S.,
  {Irwin}, P.~G.~J., {Flasar}, F.~M., 2010{\natexlab{a}}. {Seasonal change on
  Saturn from Cassini/CIRS observations, 2004-2009}. Icarus 208, 337--352.

\bibitem[{{Fletcher} et~al.(2010{\natexlab{b}}){Fletcher}, {Drossart},
  {Burgdorf}, {Orton}, and {Encrenaz}}]{fletcher10akari}
{Fletcher}, L.~N., {Drossart}, P., {Burgdorf}, M., {Orton}, G.~S., {Encrenaz},
  T., 2010{\natexlab{b}}. {Neptune's atmospheric composition from AKARI
  infrared spectroscopy}. Astron. Astrophys. 514, A17.

\bibitem[{{Fletcher} et~al.(2009){Fletcher}, {Orton}, {Teanby}, {Irwin}, and
  {Bjoraker}}]{fletcher09}
{Fletcher}, L.~N., {Orton}, G.~S., {Teanby}, N.~A., {Irwin}, P.~G.~J.,
  {Bjoraker}, G.~L., 2009. {Methane and its isotopologues on Saturn from
  Cassini/CIRS observations}. Icarus 199, 351--367.

\bibitem[{{Fletcher} et~al.(2012){Fletcher}, {Swinyard}, {Salji},
  {Polehampton}, {Fulton}, {Sidher}, {Lellouch}, {Moreno}, {Orton},
  {Cavali{\'e}}, {Courtin}, {Rengel}, {Sagawa}, {Davis}, {Hartogh}, {Naylor},
  {Walker}, and {Lim}}]{fletcher12spire}
{Fletcher}, L.~N., {Swinyard}, B., {Salji}, C., {Polehampton}, E., {Fulton},
  T., {Sidher}, S., {Lellouch}, E., {Moreno}, R., {Orton}, G., {Cavali{\'e}},
  T., {Courtin}, R., {Rengel}, M., {Sagawa}, H., {Davis}, G.~R., {Hartogh}, P.,
  {Naylor}, D., {Walker}, H., {Lim}, T., 2012. {Sub-millimetre spectroscopy of
  Saturn's trace gases from Herschel/SPIRE}. Astron. Astrophys. 539, A44.

\bibitem[{{Fouchet} et~al.(2000){Fouchet}, {Lellouch}, {B{\'e}zard},
  {Feuchtgruber}, {Drossart}, and {Encrenaz}}]{fouchet00hc}
{Fouchet}, T., {Lellouch}, E., {B{\'e}zard}, B., {Feuchtgruber}, H.,
  {Drossart}, P., {Encrenaz}, T., 2000. {Jupiter's hydrocarbons observed with
  ISO-SWS: vertical profiles of C$_2$H$_6$ and C$_2$H$_2$, detection of
  CH$_3$C$_2$H}. \aap 355, L13--L17.

\bibitem[{{Frankland} et~al.(2016){Frankland}, {James}, {S{\'a}nchez},
  {Mangan}, {Willacy}, {Poppe}, and {Plane}}]{frankland16}
{Frankland}, V.~L., {James}, A.~D., {S{\'a}nchez}, J.~D.~C., {Mangan}, T.~P.,
  {Willacy}, K., {Poppe}, A.~R., {Plane}, J.~M.~C., 2016. {Uptake of acetylene
  on cosmic dust and production of benzene in Titan's atmosphere}. Icarus 278,
  88--99.

\bibitem[{{Gladstone} and {Yung}(1983)}]{gladstone83}
{Gladstone}, G.~R., {Yung}, Y.~L., 1983. {An analysis of the reflection
  spectrum of Jupiter from 1500 A to 1740 A}. \apj 266, 415--424.

\bibitem[{{Greathouse} et~al.(2010){Greathouse}, {Gladstone}, {Moses}, {Stern},
  {Retherford}, {Vervack}, {Slater}, {Versteeg}, {Davis}, {Young}, {Steffl},
  {Throop}, and {Parker}}]{greathouse10}
{Greathouse}, T.~K., {Gladstone}, G.~R., {Moses}, J.~I., {Stern}, S.~A.,
  {Retherford}, K.~D., {Vervack}, R.~J., {Slater}, D.~C., {Versteeg}, M.~H.,
  {Davis}, M.~W., {Young}, L.~A., {Steffl}, A.~J., {Throop}, H., {Parker},
  J.~W., 2010. {New Horizons Alice ultraviolet observations of a stellar
  occultation by Jupiter's atmosphere}. Icarus 208, 293--305.

\bibitem[{{Greathouse} et~al.(2005){Greathouse}, {Lacy}, {B{\'e}zard}, {Moses},
  {Griffith}, and {Richter}}]{greathouse05}
{Greathouse}, T.~K., {Lacy}, J.~H., {B{\'e}zard}, B., {Moses}, J.~I.,
  {Griffith}, C.~A., {Richter}, M.~J., 2005. {Meridional variations of
  temperature, C$_{2}$H$_{2}$ and C$_{2}$H$_{6}$ abundances in Saturn's
  stratosphere at southern summer solstice}. Icarus 177, 18--31.

\bibitem[{{Greathouse} et~al.(2006){Greathouse}, {Lacy}, {B{\'e}zard}, {Moses},
  {Richter}, and {Knez}}]{greathouse06}
{Greathouse}, T.~K., {Lacy}, J.~H., {B{\'e}zard}, B., {Moses}, J.~I.,
  {Richter}, M.~J., {Knez}, C., Mar. 2006. {The first detection of propane on
  Saturn}. Icarus 181, 266--271.

\bibitem[{{Greathouse} et~al.(2011){Greathouse}, {Richter}, {Lacy}, {Moses},
  {Orton}, {Encrenaz}, {Hammel}, and {Jaffe}}]{greathouse11}
{Greathouse}, T.~K., {Richter}, M., {Lacy}, J., {Moses}, J., {Orton}, G.,
  {Encrenaz}, T., {Hammel}, H.~B., {Jaffe}, D., Aug. 2011. {A spatially
  resolved high spectral resolution study of Neptune's stratosphere}. Icarus
  214, 606--621.

\bibitem[{{Grebowsky} et~al.(2002){Grebowsky}, {Moses}, and
  {Pesnell}}]{grebow02}
{Grebowsky}, J.~M., {Moses}, J.~I., {Pesnell}, W.~D., 2002. Meteoric
  material---an important component of planetary atmospheres. In: {Mendillo},
  M., {Nagy}, A., {Waite}, J.~H. (Eds.), Atmospheres in the Solar System:
  Comparative Aeronomy. American Geophysical Union Monograph Series,
  Washington, DC, pp. 235--244.

\bibitem[{{Greenberg} and {Li}(1999)}]{greenberg99}
{Greenberg}, J.~M., {Li}, A., 1999. Morphological structure and chemical
  composition of cometary nuclei and dust. Space Sci. Rev. 90, 149--161.

\bibitem[{{Gr\"un} et~al.(1985){Gr\"un}, {Zook}, {Fechtig}, and
  {Giese}}]{grun85}
{Gr\"un}, E., {Zook}, H.~A., {Fechtig}, H., {Giese}, R.~H., 1985. Collisional
  balance of the meteoritic complex. Icarus 62, 244--272.

\bibitem[{{Guerlet} et~al.(2010){Guerlet}, {Fouchet}, {B{\'e}zard}, {Moses},
  {Fletcher}, {Simon-Miller}, and {Flasar}}]{guerlet10}
{Guerlet}, S., {Fouchet}, T., {B{\'e}zard}, B., {Moses}, J.~I., {Fletcher},
  L.~N., {Simon-Miller}, A.~A., {Flasar}, F.~M., 2010. {Meridional distribution
  of CH$_{3}$C$_{2}$H and C$_{4}$H$_{2}$ in {Saturn's} stratosphere from
  {CIRS/Cassini} limb and nadir observations}. Icarus 209, 682--695.

\bibitem[{{Guerlet} et~al.(2009){Guerlet}, {Fouchet}, {B{\'e}zard},
  {Simon-Miller}, and {Flasar}}]{guerlet09}
{Guerlet}, S., {Fouchet}, T., {B{\'e}zard}, B., {Simon-Miller}, A.~A.,
  {Flasar}, F.~M., 2009. Vertical and meridional distribution of ethane,
  acetylene and propane in {Saturn's} stratosphere from {CIRS/Cassini} limb
  observations. Icarus 203, 214--232.

\bibitem[{{Guilloteau} et~al.(1993){Guilloteau}, {Dutrey}, {Marten}, and
  {Gautier}}]{guilloteau93}
{Guilloteau}, S., {Dutrey}, A., {Marten}, A., {Gautier}, D., 1993. {CO in the
  troposphere of Neptune: Detection of the J = 1-0 line in absorption}. \aap
  279, 661--667.

\bibitem[{{Gustafson}(1994)}]{gustafson94}
{Gustafson}, B.~A.~S., 1994. Physics of zodiacal dust. Annual Review of Earth
  and Planetary Sciences 22, 553--595.

\bibitem[{{Hartogh} et~al.(2011){Hartogh}, {Lellouch}, {Moreno},
  {Bockel{\'e}e-Morvan}, {Biver}, {Cassidy}, {Rengel}, {Jarchow},
  {Cavali{\'e}}, {Crovisier}, {Helmich}, and {Kidger}}]{hartogh11}
{Hartogh}, P., {Lellouch}, E., {Moreno}, R., {Bockel{\'e}e-Morvan}, D.,
  {Biver}, N., {Cassidy}, T., {Rengel}, M., {Jarchow}, C., {Cavali{\'e}}, T.,
  {Crovisier}, J., {Helmich}, F.~P., {Kidger}, M., 2011. Direct detection of
  the {Enceladus} water torus with {Herschel}. \aap 532, L2.

\bibitem[{{H{\'e}brard} et~al.(2013){H{\'e}brard}, {Dobrijevic}, {Loison},
  {Bergeat}, {Hickson}, and {Caralp}}]{hebrard13}
{H{\'e}brard}, E., {Dobrijevic}, M., {Loison}, J.~C., {Bergeat}, A., {Hickson},
  K.~M., {Caralp}, F., 2013. {Photochemistry of C$_{3}$H$_{p}$ hydrocarbons in
  Titan's stratosphere revisited}. Astron. Astrophys. 552, A132.

\bibitem[{{Herbert} et~al.(1987){Herbert}, {Sandel}, {Yelle}, {Holberg},
  {Broadfoot}, {Shemansky}, {Atreya}, and {Romani}}]{herbert87}
{Herbert}, F., {Sandel}, B.~R., {Yelle}, R.~V., {Holberg}, J.~B., {Broadfoot},
  A.~L., {Shemansky}, D.~E., {Atreya}, S.~K., {Romani}, P.~N., 1987. {The upper
  atmosphere of Uranus - EUV occultations observed by Voyager 2}. \jgr 92,
  15093--15109.

\bibitem[{{Hesman} et~al.(2007){Hesman}, {Davis}, {Matthews}, and
  {Orton}}]{hesman07}
{Hesman}, B.~E., {Davis}, G.~R., {Matthews}, H.~E., {Orton}, G.~S., 2007. {The
  abundance profile of CO in Neptune's atmosphere}. Icarus 186, 342--353.

\bibitem[{{Hildenbrand} and {Murad}(1969)}]{hildenbrand69}
{Hildenbrand}, D.~L., {Murad}, E., 1969. Dissociation energy and ionization
  potential of silicon monoxide. J. Chem. Phys.

\bibitem[{{Horanyi}(1996)}]{horanyi96}
{Horanyi}, M., 1996. Charged dust dynamics in the solar system. Annual Review
  of Astronomy and Astrophysics 34, 383--418.

\bibitem[{{Irwin} et~al.(2014){Irwin}, {Lellouch}, {de Bergh}, {Courtin},
  {B{\'e}zard}, {Fletcher}, {Orton}, {Teanby}, {Calcutt}, {Tice}, {Hurley}, and
  {Davis}}]{irwin14}
{Irwin}, P.~G.~J., {Lellouch}, E., {de Bergh}, C., {Courtin}, R., {B{\'e}zard},
  B., {Fletcher}, L.~N., {Orton}, G.~S., {Teanby}, N.~A., {Calcutt}, S.~B.,
  {Tice}, D., {Hurley}, J., {Davis}, G.~R., 2014. {Line-by-line analysis of
  Neptune's near-IR spectrum observed with Gemini/NIFS and VLT/CRIRES}. Icarus
  227, 37--48.

\bibitem[{{Janches} et~al.(2009){Janches}, {Dyrud}, {Broadley}, and
  {Plane}}]{janches09}
{Janches}, D., {Dyrud}, L.~P., {Broadley}, S.~L., {Plane}, J.~M.~C., 2009.
  First observation of micrometeoroid differential ablation in the atmosphere.
  \grl 36, L06101.

\bibitem[{{Jurac} and {Richardson}(2007)}]{jurac07}
{Jurac}, S., {Richardson}, J.~D., 2007. Neutral cloud interaction with
  {Saturn's} main rings. \grl 34, L08102.

\bibitem[{{Kim} et~al.(2014){Kim}, {Sim}, {Sohn}, and {Moses}}]{kim14}
{Kim}, S.~J., {Sim}, C.~K., {Sohn}, M.~R., {Moses}, J.~I., 2014. {CH$_{4}$
  mixing ratios at microbar pressure levels of Jupiter as constrained by
  3-micron ISO data}. Icarus 237, 42--51.

\bibitem[{{Kim} et~al.(2001){Kim}, {Pesnell}, {Grebowsky}, and {Fox}}]{kim01}
{Kim}, Y.~H., {Pesnell}, W.~D., {Grebowsky}, J.~M., {Fox}, J.~L., 2001.
  {Meteoric ions in the ionosphere of Jupiter}. Icarus 150, 261--278.

\bibitem[{{Kostiuk} et~al.(1987){Kostiuk}, {Espenak}, {Mumma}, {Deming}, and
  {Zipoy}}]{kostiuk87}
{Kostiuk}, T.~., {Espenak}, F., {Mumma}, M.~J., {Deming}, D., {Zipoy}, D.,
  1987. {Variability of ethane on Jupiter}. Icarus 72, 394--410.

\bibitem[{{Kostiuk} et~al.(1992){Kostiuk}, {Romani}, {Espenak}, and
  {Bezard}}]{kostiuk92}
{Kostiuk}, T., {Romani}, P., {Espenak}, F., {Bezard}, B., 1992. {Stratospheric
  ethane on Neptune - Comparison of groundbased and Voyager IRIS retrievals}.
  Icarus 99, 353--362.

\bibitem[{{Landgraf} et~al.(2002){Landgraf}, {Liou}, {Zook}, and
  {Gr{\"u}n}}]{landgraf02}
{Landgraf}, M., {Liou}, J.-C., {Zook}, H.~A., {Gr{\"u}n}, E., 2002. {Origins of
  Solar System dust beyond Jupiter}. Astron. J. 123, 2857--2861.

\bibitem[{{Larson} et~al.(1978){Larson}, {Fink}, and {Treffers}}]{larson78}
{Larson}, H.~P., {Fink}, U., {Treffers}, R.~C., 1978. {Evidence for CO in
  Jupiter's atmosphere from airborne spectroscopic observations at 5 microns}.
  \apj 219, 1084--1092.

\bibitem[{{Lellouch}(1996)}]{lellouch96}
{Lellouch}, E., 1996. Chemistry induced by the impacts: {Observations}. In:
  {Noll}, K.~S., {Weaver}, H.~A., {Feldman}, P.~D. (Eds.), IAU Colloq. 156: The
  Collision of Comet Shoemaker-Levy 9 and Jupiter. Cambridge Univ. Press,
  Cambridge, pp. 213--242.

\bibitem[{{Lellouch} et~al.(1997){Lellouch}, {B{\'e}zard}, {Moreno},
  {Bockel{\'e}e-Morvan}, {Colom}, {Crovisier}, {Festou}, {Gautier}, {Marten},
  and {Paubert}}]{lellouch97}
{Lellouch}, E., {B{\'e}zard}, B., {Moreno}, R., {Bockel{\'e}e-Morvan}, D.,
  {Colom}, P., {Crovisier}, J., {Festou}, M., {Gautier}, D., {Marten}, A.,
  {Paubert}, G., 1997. {Carbon monoxide in Jupiter after the impact of comet
  Shoemaker-Levy 9}. Planet. Space Sci. 45, 1203--1212.

\bibitem[{{Lellouch} et~al.(2002){Lellouch}, {B{\'e}zard}, {Moses}, {Davis},
  {Drossart}, {Feuchtgruber}, {Bergin}, {Moreno}, and {Encrenaz}}]{lellouch02}
{Lellouch}, E., {B{\'e}zard}, B., {Moses}, J.~I., {Davis}, G.~R., {Drossart},
  P., {Feuchtgruber}, H., {Bergin}, E.~A., {Moreno}, R., {Encrenaz}, T., 2002.
  The origin of water vapor and carbon dioxide in {Jupiter's} stratosphere.
  Icarus 159, 112--131.

\bibitem[{{Lellouch} et~al.(2006){Lellouch}, {B{\'e}zard}, {Strobel},
  {Bjoraker}, {Flasar}, and {Romani}}]{lellouch06}
{Lellouch}, E., {B{\'e}zard}, B., {Strobel}, D.~F., {Bjoraker}, G.~L.,
  {Flasar}, F.~M., {Romani}, P.~N., 2006. {On the HCN and CO$_{2}$ abundance
  and distribution in Jupiter's stratosphere}. Icarus 184, 478--497.

\bibitem[{{Lellouch} et~al.(2010){Lellouch}, {Hartogh}, {Feuchtgruber},
  {Vandenbussche}, {de Graauw}, {Moreno}, {Jarchow}, {Cavali{\'e}}, {Orton},
  {Banaszkiewicz}, {Blecka}, {Bockel{\'e}e-Morvan}, {Crovisier}, {Encrenaz},
  {Fulton}, {K{\"u}ppers}, {Lara}, {Lis}, {Medvedev}, {Rengel}, {Sagawa},
  {Swinyard}, {Szutowicz}, {Bensch}, {Bergin}, {Billebaud}, {Biver}, {Blake},
  {Blommaert}, {Cernicharo}, {Courtin}, {Davis}, {Decin}, {Encrenaz},
  {Gonzalez}, {Jehin}, {Kidger}, {Naylor}, {Portyankina}, {Schieder}, {Sidher},
  {Thomas}, {de Val-Borro}, {Verdugo}, {Waelkens}, {Walker}, {Aarts}, {Comito},
  {Kawamura}, {Maestrini}, {Peacocke}, {Teipen}, {Tils}, and
  {Wildeman}}]{lellouch10}
{Lellouch}, E., {Hartogh}, P., {Feuchtgruber}, H., {Vandenbussche}, B., {de
  Graauw}, T., {Moreno}, R., {Jarchow}, C., {Cavali{\'e}}, T., {Orton}, G.,
  {Banaszkiewicz}, M., {Blecka}, M.~I., {Bockel{\'e}e-Morvan}, D., {Crovisier},
  J., {Encrenaz}, T., {Fulton}, T., {K{\"u}ppers}, M., {Lara}, L.~M., {Lis},
  D.~C., {Medvedev}, A.~S., {Rengel}, M., {Sagawa}, H., {Swinyard}, B.,
  {Szutowicz}, S., {Bensch}, F., {Bergin}, E., {Billebaud}, F., {Biver}, N.,
  {Blake}, G.~A., {Blommaert}, J.~A.~D.~L., {Cernicharo}, J., {Courtin}, R.,
  {Davis}, G.~R., {Decin}, L., {Encrenaz}, P., {Gonzalez}, A., {Jehin}, E.,
  {Kidger}, M., {Naylor}, D., {Portyankina}, G., {Schieder}, R., {Sidher}, S.,
  {Thomas}, N., {de Val-Borro}, M., {Verdugo}, E., {Waelkens}, C., {Walker},
  H., {Aarts}, H., {Comito}, C., {Kawamura}, J.~H., {Maestrini}, A.,
  {Peacocke}, T., {Teipen}, R., {Tils}, T., {Wildeman}, K., 2010. {First
  results of Herschel-PACS observations of Neptune}. \aap 518, L152.

\bibitem[{{Lellouch} et~al.(2015){Lellouch}, {Moreno}, {Orton}, {Feuchtgruber},
  {Cavali{\'e}}, {Moses}, {Hartogh}, {Jarchow}, and {Sagawa}}]{lellouch15}
{Lellouch}, E., {Moreno}, R., {Orton}, G.~S., {Feuchtgruber}, H.,
  {Cavali{\'e}}, T., {Moses}, J.~I., {Hartogh}, P., {Jarchow}, C., {Sagawa},
  H., 2015. {New constraints on the CH$_{4}$ vertical profile in Uranus and
  Neptune from Herschel observations}. \aap 579, A121.

\bibitem[{{Lellouch} et~al.(2005){Lellouch}, {Moreno}, and
  {Paubert}}]{lellouch05}
{Lellouch}, E., {Moreno}, R., {Paubert}, G., 2005. {A dual origin for Neptune's
  carbon monoxide?} \aap 430, L37--L40.

\bibitem[{{Lellouch} et~al.(1994){Lellouch}, {Romani}, and
  {Rosenqvist}}]{lellouch94}
{Lellouch}, E., {Romani}, P.~N., {Rosenqvist}, J., 1994. {The vertical
  Distribution and Origin of HCN in Neptune's Atmosphere}. Icarus 108,
  112--136.

\bibitem[{{Levison} and {Duncan}(1997)}]{levison97}
{Levison}, H.~F., {Duncan}, M.~J., 1997. {From the Kuiper Belt to
  Jupiter-family comets: The spatial distribution of ecliptic comets}. Icarus
  127, 13--32.

\bibitem[{{Lewis} and {Fegley}(1984)}]{lewis84}
{Lewis}, J.~S., {Fegley}, Jr., M.~B., 1984. Vertical distribution of
  disequilibrium species in {Jupiter's} troposphere. Space Sci. Rev. 39,
  163--192.

\bibitem[{{Li} and {Greenberg}(1997)}]{ligreen97}
{Li}, A., {Greenberg}, J.~M., 1997. A unified model of interstellar dust. \aap
  323, 566--584.

\bibitem[{{Lindal} et~al.(1987){Lindal}, {Lyons}, {Sweetnam}, {Eshleman}, and
  {Hinson}}]{lindal87}
{Lindal}, G.~F., {Lyons}, J.~R., {Sweetnam}, D.~N., {Eshleman}, V.~R.,
  {Hinson}, D.~P., Dec. 1987. {The atmosphere of Uranus - Results of radio
  occultation measurements with Voyager 2}. Journal of Geophysical Research 92,
  14987--15001.

\bibitem[{{Linstrom} and {Mallard}(2017)}]{linstrom17}
{Linstrom}, P.~J., {Mallard}, W.~G., E., 2017. NIST Chemistry WebBook, NIST
  Standard Reference Database Number 69. National Institute of Standards and
  Technology, Gaithersburg, MD.

\bibitem[{{Liou} and {Zook}(1997)}]{liou97}
{Liou}, J.-C., {Zook}, H.~A., 1997. Evolution of interplanetary dust particles
  in mean motion resonances with planets. Icarus 128, 354--367.

\bibitem[{{Lisse} et~al.(2007){Lisse}, {Kraemer}, {Nuth}, {Li}, and
  {Joswiak}}]{lisse07}
{Lisse}, C.~M., {Kraemer}, K.~E., {Nuth}, J.~A., {Li}, A., {Joswiak}, D., 2007.
  {Comparison of the composition of the Tempel 1 ejecta to the dust in Comet
  C/Hale Bopp 1995 O1 and YSO HD 100546}. Icarus 191, 223--240.

\bibitem[{{Lisse} et~al.(2006){Lisse}, {VanCleve}, {Adams}, {A'Hearn},
  {Fern{\'a}ndez}, {Farnham}, {Armus}, {Grillmair}, {Ingalls}, {Belton},
  {Groussin}, {McFadden}, {Meech}, {Schultz}, {Clark}, {Feaga}, and
  {Sunshine}}]{lisse06}
{Lisse}, C.~M., {VanCleve}, J., {Adams}, A.~C., {A'Hearn}, M.~F.,
  {Fern{\'a}ndez}, Y.~R., {Farnham}, T.~L., {Armus}, L., {Grillmair}, C.~J.,
  {Ingalls}, J., {Belton}, M.~J.~S., {Groussin}, O., {McFadden}, L.~A.,
  {Meech}, K.~J., {Schultz}, P.~H., {Clark}, B.~C., {Feaga}, L.~M., {Sunshine},
  J.~M., 2006. Spitzer spectral observations of the {Deep} {Impact} ejecta.
  Science 313, 635--640.

\bibitem[{{Lodders} and {Fegley}(2002)}]{lodders02}
{Lodders}, K., {Fegley}, B., 2002. Atmospheric chemistry in giant planets,
  brown dwarfs, and low-mass dwarf stars. {I.} carbon, nitrogen, and oxygen.
  Icarus 155, 393--424.

\bibitem[{{Lodders} and {Fegley}(1994)}]{lodders94}
{Lodders}, K., {Fegley}, Jr., B., 1994. {The origin of carbon monoxide in
  Neptunes's atmosphere}. Icarus 112, 368--375.

\bibitem[{{Luhmann} et~al.(2006){Luhmann}, {Johnson}, {Tokar}, {Ledvina}, and
  {Cravens}}]{luhmann06}
{Luhmann}, J.~G., {Johnson}, R.~E., {Tokar}, R.~L., {Ledvina}, S.~A.,
  {Cravens}, T.~E., 2006. {A model of the ionosphere of Saturn's rings and its
  implications}. Icarus 181, 465--474.

\bibitem[{{Luszcz-Cook} and {de Pater}(2013)}]{luszczcook13}
{Luszcz-Cook}, S.~H., {de Pater}, I., 2013. {Constraining the origins of
  Neptune's carbon monoxide abundance with CARMA millimeter-wave observations}.
  Icarus 222, 379--400.

\bibitem[{{Lyons}(1995)}]{lyons95}
{Lyons}, J.~R., 1995. {Metal ions in the atmosphere of Neptune}. Science 267,
  648--651.

\bibitem[{{Majeed} and {McConnell}(1991)}]{majeed91}
{Majeed}, T., {McConnell}, J.~C., 1991. {The upper ionospheres of Jupiter and
  Saturn}. Planetary and Space Sciences 39, 1715--1732.

\bibitem[{{Malhotra} and {Mathews}(2011)}]{malhotra11}
{Malhotra}, A., {Mathews}, J.~D., 2011. {A statistical study of meteoroid
  fragmentation and differential ablation using the Resolute Bay Incoherent
  Scatter Radar}. \jgr 116, A04316.

\bibitem[{{Marten} et~al.(1993){Marten}, {Gautier}, {Owen}, {Sanders},
  {Matthews}, {Atreya}, {Tilanus}, and {Deane}}]{marten93}
{Marten}, A., {Gautier}, D., {Owen}, T., {Sanders}, D.~B., {Matthews}, H.~E.,
  {Atreya}, S.~K., {Tilanus}, R.~P.~J., {Deane}, J.~R., 1993. {First
  observations of CO and HCN on Neptune and Uranus at millimeter wavelengths
  and the implications for atmospheric chemistry}. \apj 406, 285--297.

\bibitem[{{Marten} et~al.(2005){Marten}, {Matthews}, {Owen}, {Moreno},
  {Hidayat}, and {Biraud}}]{marten05}
{Marten}, A., {Matthews}, H.~E., {Owen}, T., {Moreno}, R., {Hidayat}, T.,
  {Biraud}, Y., 2005. {Improved constraints on Neptune's atmosphere from
  submillimetre-wavelength observations}. \aap 429, 1097--1105.

\bibitem[{{McNeil} et~al.(1998){McNeil}, {Lai}, and {Murad}}]{mcneil98}
{McNeil}, W.~J., {Lai}, S.~T., {Murad}, E., 1998. Differential ablation of
  cosmic dust and implications for the relative abundances of atmospheric
  metals. \jgr 103, 10899--10912.

\bibitem[{{Meadows} et~al.(2008){Meadows}, {Orton}, {Line}, {Liang}, {Yung},
  {van Cleve}, and {Burgdorf}}]{meadows08}
{Meadows}, V.~S., {Orton}, G., {Line}, M., {Liang}, M.-C., {Yung}, Y.~L., {van
  Cleve}, J., {Burgdorf}, M.~J., 2008. {First Spitzer observations of Neptune:
  Detection of new hydrocarbons}. Icarus 197, 585--589.

\bibitem[{{Molina-Cuberos} et~al.(2008){Molina-Cuberos}, {L{\'o}pez-Moreno},
  and {Arnold}}]{molinacuberos08}
{Molina-Cuberos}, J.~G., {L{\'o}pez-Moreno}, J.~J., {Arnold}, F., 2008.
  Meteoric layers in planetary atmospheres. Space Sci. Rev. 137, 175--191.

\bibitem[{{Moore} et~al.(2015){Moore}, {O'Donoghue}, {M{\"u}ller-Wodarg},
  {Galand}, and {Mendillo}}]{moore15}
{Moore}, L., {O'Donoghue}, J., {M{\"u}ller-Wodarg}, I., {Galand}, M.,
  {Mendillo}, M., 2015. {Saturn ring rain: Model estimates of water influx into
  Saturn's atmosphere}. Icarus 245, 355--366.

\bibitem[{{Moore} et~al.(2004){Moore}, {Mendillo}, {M{\"u}ller-Wodarg}, and
  {Murr}}]{moore04}
{Moore}, L.~E., {Mendillo}, M., {M{\"u}ller-Wodarg}, I.~C.~F., {Murr}, D.~L.,
  2004. {Modeling of global variations and ring shadowing in Saturn's
  ionosphere}. Icarus 172, 503--520.

\bibitem[{{Moreno} et~al.(2003){Moreno}, {Marten}, {Matthews}, and
  {Biraud}}]{moreno03}
{Moreno}, R., {Marten}, A., {Matthews}, H.~E., {Biraud}, Y., 2003. {Long-term
  evolution of CO, CS and HCN in Jupiter after the impacts of comet
  Shoemaker-Levy 9}. Planet. Space Sci. 51, 591--611.

\bibitem[{{Morrissey} et~al.(1995){Morrissey}, {Feldman}, {McGrath}, {Wolven},
  and {Moos}}]{morrissey95}
{Morrissey}, P.~F., {Feldman}, P.~D., {McGrath}, M.~A., {Wolven}, B.~C.,
  {Moos}, H.~W., 1995. {The ultraviolet reflectivity of Jupiter at 3.5 Angstrom
  resolution from Astro-1 and Astro-2}. \apjl 454, L65.

\bibitem[{{Moses}(1992)}]{moses92abl}
{Moses}, J.~I., 1992. Meteoroid ablation in {Neptune}'s atmosphere. Icarus 99,
  368--383.

\bibitem[{{Moses}(1997)}]{moses97}
{Moses}, J.~I., 1997. {Dust ablation during the Shoemaker-Levy 9 impacts}. \jgr
  102, 21619--21644.

\bibitem[{{Moses}(2001)}]{moses01}
{Moses}, J.~I., 2001. Meteoroid ablation on the outer planets. In: Lunar and
  Planetary Science Conference. Vol.~32. p. 1161.

\bibitem[{{Moses} et~al.(1992){Moses}, {Allen}, and {Yung}}]{moses92nucl}
{Moses}, J.~I., {Allen}, M., {Yung}, Y.~L., 1992. Hydrocarbon nucleation and
  aerosol formation in {Neptune}'s atmosphere. Icarus 99, 318--346.

\bibitem[{{Moses} et~al.(2015){Moses}, {Armstrong}, {Fletcher}, {Friedson},
  {Irwin}, {Sinclair}, and {Hesman}}]{moses15}
{Moses}, J.~I., {Armstrong}, E.~S., {Fletcher}, L.~N., {Friedson}, A.~J.,
  {Irwin}, P.~G.~J., {Sinclair}, J.~A., {Hesman}, B.~E., 2015. {Evolution of
  stratospheric chemistry in the Saturn storm beacon region}. Icarus 261,
  149--168.

\bibitem[{{Moses} and {Bass}(2000)}]{mosesbass00}
{Moses}, J.~I., {Bass}, S.~F., 2000. {The effects of external material on the
  chemistry and structure of Saturn's ionosphere}. \jgr 105, 7013--7052.

\bibitem[{{Moses} et~al.(2000{\natexlab{a}}){Moses}, {B{\'e}zard}, {Lellouch},
  {Gladstone}, {Feuchtgruber}, and {Allen}}]{moses00a}
{Moses}, J.~I., {B{\'e}zard}, B., {Lellouch}, E., {Gladstone}, G.~R.,
  {Feuchtgruber}, H., {Allen}, M., 2000{\natexlab{a}}. Photochemistry of
  {Saturn}'s atmosphere. {I.} {Hydrocarbon} chemistry and comparisons with
  {ISO} observations. Icarus 143, 244--298.

\bibitem[{{Moses} et~al.(2005){Moses}, {Fouchet}, {B{\'e}zard}, {Gladstone},
  {Lellouch}, and {Feuchtgruber}}]{moses05}
{Moses}, J.~I., {Fouchet}, T., {B{\'e}zard}, B., {Gladstone}, G.~R.,
  {Lellouch}, E., {Feuchtgruber}, H., 2005. Photochemistry and diffusion in
  {Jupiter}'s stratosphere: {Constraints} from {ISO} observations and
  comparisons with other giant planets. \jgr 110, E08001.

\bibitem[{{Moses} et~al.(2000{\natexlab{b}}){Moses}, {Lellouch}, {B{\'e}zard},
  {Gladstone}, {Feuchtgruber}, and {Allen}}]{moses00b}
{Moses}, J.~I., {Lellouch}, E., {B{\'e}zard}, B., {Gladstone}, G.~R.,
  {Feuchtgruber}, H., {Allen}, M., 2000{\natexlab{b}}. Photochemistry of
  {Saturn}'s atmosphere. {II.} {Effects} of an influx of external oxygen.
  Icarus 145, 166--202.

\bibitem[{{Moses} et~al.(2013){Moses}, {Line}, {Visscher}, {Richardson},
  {Nettelmann}, {Fortney}, {Barman}, {Stevenson}, and
  {Madhusudhan}}]{moses13gj436}
{Moses}, J.~I., {Line}, M.~R., {Visscher}, C., {Richardson}, M.~R.,
  {Nettelmann}, N., {Fortney}, J.~J., {Barman}, T.~S., {Stevenson}, K.~B.,
  {Madhusudhan}, N., 2013. Compositional diversity in the atmospheres of hot
  {Neptunes}, with application to {GJ} 436b. \apj 777, 34.

\bibitem[{{Moses} et~al.(1995){Moses}, {Rages}, and {Pollack}}]{moses95c}
{Moses}, J.~I., {Rages}, K., {Pollack}, J.~B., 1995. An analysis of {Neptune}'s
  stratospheric haze using high-phase-angle {Voyager} images. Icarus 113,
  232--266.

\bibitem[{{Mousis} et~al.(2016){Mousis}, {Atkinson}, {Spilker}, {Venkatapathy},
  {Poncy}, {Frampton}, {Coustenis}, {Reh}, {Lebreton}, {Fletcher}, {Hueso},
  {Amato}, {Colaprete}, {Ferri}, {Stam}, {Wurz}, {Atreya}, {Aslam}, {Banfield},
  {Calcutt}, {Fischer}, {Holland}, {Keller}, {Kessler}, {Leese}, {Levacher},
  {Morse}, {Mu{\~n}oz}, {Renard}, {Sheridan}, {Schmider}, {Snik}, {Waite},
  {Bird}, {Cavali{\'e}}, {Deleuil}, {Fortney}, {Gautier}, {Guillot}, {Lunine},
  {Marty}, {Nixon}, {Orton}, and {S{\'a}nchez-Lavega}}]{mousis16}
{Mousis}, O., {Atkinson}, D.~H., {Spilker}, T., {Venkatapathy}, E., {Poncy},
  J., {Frampton}, R., {Coustenis}, A., {Reh}, K., {Lebreton}, J.-P.,
  {Fletcher}, L.~N., {Hueso}, R., {Amato}, M.~J., {Colaprete}, A., {Ferri}, F.,
  {Stam}, D., {Wurz}, P., {Atreya}, S., {Aslam}, S., {Banfield}, D.~J.,
  {Calcutt}, S., {Fischer}, G., {Holland}, A., {Keller}, C., {Kessler}, E.,
  {Leese}, M., {Levacher}, P., {Morse}, A., {Mu{\~n}oz}, O., {Renard}, J.-B.,
  {Sheridan}, S., {Schmider}, F.-X., {Snik}, F., {Waite}, J.~H., {Bird}, M.,
  {Cavali{\'e}}, T., {Deleuil}, M., {Fortney}, J., {Gautier}, D., {Guillot},
  T., {Lunine}, J.~I., {Marty}, B., {Nixon}, C., {Orton}, G.~S.,
  {S{\'a}nchez-Lavega}, A., 2016. {The Hera Saturn entry probe mission}.
  Planet. Space Sci. 130, 80--103.

\bibitem[{{Mousis} et~al.(2014){Mousis}, {Fletcher}, {Lebreton}, {Wurz},
  {Cavali{\'e}}, {Coustenis}, {Courtin}, {Gautier}, {Helled}, {Irwin}, {Morse},
  {Nettelmann}, {Marty}, {Rousselot}, {Venot}, {Atkinson}, {Waite}, {Reh},
  {Simon}, {Atreya}, {Andr{\'e}}, {Blanc}, {Daglis}, {Fischer}, {Geppert},
  {Guillot}, {Hedman}, {Hueso}, {Lellouch}, {Lunine}, {Murray}, {O`Donoghue},
  {Rengel}, {S{\'a}nchez-Lavega}, {Schmider}, {Spiga}, {Spilker}, {Petit},
  {Tiscareno}, {Ali-Dib}, {Altwegg}, {Bolton}, {Bouquet}, {Briois}, {Fouchet},
  {Guerlet}, {Kostiuk}, {Lebleu}, {Moreno}, {Orton}, and {Poncy}}]{mousis14}
{Mousis}, O., {Fletcher}, L.~N., {Lebreton}, J.-P., {Wurz}, P., {Cavali{\'e}},
  T., {Coustenis}, A., {Courtin}, R., {Gautier}, D., {Helled}, R., {Irwin},
  P.~G.~J., {Morse}, A.~D., {Nettelmann}, N., {Marty}, B., {Rousselot}, P.,
  {Venot}, O., {Atkinson}, D.~H., {Waite}, J.~H., {Reh}, K.~R., {Simon}, A.~A.,
  {Atreya}, S., {Andr{\'e}}, N., {Blanc}, M., {Daglis}, I.~A., {Fischer}, G.,
  {Geppert}, W.~D., {Guillot}, T., {Hedman}, M.~M., {Hueso}, R., {Lellouch},
  E., {Lunine}, J.~I., {Murray}, C.~D., {O`Donoghue}, J., {Rengel}, M.,
  {S{\'a}nchez-Lavega}, A., {Schmider}, F.-X., {Spiga}, A., {Spilker}, T.,
  {Petit}, J.-M., {Tiscareno}, M.~S., {Ali-Dib}, M., {Altwegg}, K., {Bolton},
  S.~J., {Bouquet}, A., {Briois}, C., {Fouchet}, T., {Guerlet}, S., {Kostiuk},
  T., {Lebleu}, D., {Moreno}, R., {Orton}, G.~S., {Poncy}, J., 2014.
  {Scientific rationale for Saturn's in situ exploration}. Planet. Space Sci.
  104, 29--47.

\bibitem[{{Murray} et~al.(1974){Murray}, {Pottie}, and {Pupp}}]{murray74}
{Murray}, J.~J., {Pottie}, R.~F., {Pupp}, C., 1974. {The vapor pressures and
  enthalpies of sublimation of five polycyclic aromatic hydrocarbons}. Can. J.
  Chem. 52, 557--563.

\bibitem[{{Myer} and {Samson}(1970)}]{myer70}
{Myer}, J.~A., {Samson}, J.~A.~R., 1970. Vacuum-ultraviolet absorption cross
  sections of {CO}, {HCl}, and {ICN} between 1050 and 2100 \aa. J. Chem. Phys.
  52, 266--271.

\bibitem[{{Nagy} et~al.(2009){Nagy}, {Kliore}, {Mendillo}, {Miller}, {Moore},
  {Moses}, {M{\"u}ller-Wodarg}, and {Shemansky}}]{nagy09}
{Nagy}, A.~F., {Kliore}, A.~J., {Mendillo}, M., {Miller}, S., {Moore}, L.,
  {Moses}, J.~I., {M{\"u}ller-Wodarg}, I., {Shemansky}, D., 2009. {Upper
  Atmosphere and Ionosphere of Saturn}. In: {Dougherty}, M.~K., {Esposito},
  L.~W., {Krimigis}, S.~M. (Eds.), Saturn from Cassini-Huygens. Springer, pp.
  181--201.

\bibitem[{{Naylor} et~al.(1994){Naylor}, {Davis}, {Griffin}, {Clark},
  {Gautier}, and {Marten}}]{naylor94}
{Naylor}, D.~A., {Davis}, G.~R., {Griffin}, M.~J., {Clark}, T.~A., {Gautier},
  D., {Marten}, A., 1994. {Broad-band spectrscopic detection of the CO J=3-2
  tropospheric absorption in the atmosphere of Neptune}. \aap 291, L51--L53.

\bibitem[{{Nixon} et~al.(2010){Nixon}, {Achterberg}, {Romani}, {Allen},
  {Zhang}, {Teanby}, {Irwin}, and {Flasar}}]{nixon10}
{Nixon}, C.~A., {Achterberg}, R.~K., {Romani}, P.~N., {Allen}, M., {Zhang}, X.,
  {Teanby}, N.~A., {Irwin}, P.~G.~J., {Flasar}, F.~M., 2010. {Abundances of
  Jupiter's trace hydrocarbons from Voyager and Cassini}. Planet. Space Sci.
  58, 1667--1680.

\bibitem[{{Noll} et~al.(1997){Noll}, {Gilmore}, {Knacke}, {Womack}, {Griffith},
  and {Orton}}]{noll97}
{Noll}, K.~S., {Gilmore}, D., {Knacke}, R.~F., {Womack}, M., {Griffith}, C.~A.,
  {Orton}, G., 1997. {Carbon monoxide in Jupiter after Comet Shoemaker-Levy 9}.
  Icarus 126, 324--335.

\bibitem[{{Noll} et~al.(1986{\natexlab{a}}){Noll}, {Knacke}, {Geballe}, and
  {Tokunaga}}]{noll86co}
{Noll}, K.~S., {Knacke}, R.~F., {Geballe}, T.~R., {Tokunaga}, A.~T.,
  1986{\natexlab{a}}. {Detection of carbon monoxide in Saturn}. \apjl 309,
  L91--L94.

\bibitem[{{Noll} et~al.(1988){Noll}, {Knacke}, {Geballe}, and
  {Tokunaga}}]{noll88co}
{Noll}, K.~S., {Knacke}, R.~F., {Geballe}, T.~R., {Tokunaga}, A.~T., 1988. {The
  origin and vertical distribution of carbon monoxide in Jupiter}. \apj 324,
  1210--1218.

\bibitem[{{Noll} et~al.(1986{\natexlab{b}}){Noll}, {Knacke}, {Tokunaga},
  {Lacy}, {Beck}, and {Serabyn}}]{noll86}
{Noll}, K.~S., {Knacke}, R.~F., {Tokunaga}, A.~T., {Lacy}, J.~H., {Beck}, S.,
  {Serabyn}, E., 1986{\natexlab{b}}. {The abundances of ethane and acetylene in
  the atmospheres of Jupiter and Saturn}. Icarus 65, 257--263.

\bibitem[{{Noll} and {Larson}(1990)}]{noll90}
{Noll}, K.~S., {Larson}, H.~P., 1990. The spectrum of {Saturn} from 1990--2230
  cm$^{-1}$: Abundances of {A}s{H}$_3$, {CH}$_3${D}, {CO}, {G}e{H}$_4$, and
  {PH}$_3$. Icarus 89, 168--189.

\bibitem[{{Okabe}(1978)}]{okabe78}
{Okabe}, H., 1978. Photochemistry of Small Molecules. Wiley, New York.

\bibitem[{{Ollivier} et~al.(2000){Ollivier}, {Dobrij{\'e}vic}, and
  {Parisot}}]{ollivier00}
{Ollivier}, J.~L., {Dobrij{\'e}vic}, M., {Parisot}, J.~P., 2000. {New
  photochemical model of Saturn's atmosphere}. Planet. Space Sci. 48, 699--716.

\bibitem[{{\"Opik}(1958)}]{opik58}
{\"Opik}, E.~J., 1958. Physics of meteor flight in the atmosphere. Interscience
  Publishers, New York.

\bibitem[{{Orton} et~al.(2014{\natexlab{a}}){Orton}, {Fletcher}, {Moses},
  {Mainzer}, {Hines}, {Hammel}, {Martin-Torres}, {Burgdorf}, {Merlet}, and
  {Line}}]{orton14temp}
{Orton}, G.~S., {Fletcher}, L.~N., {Moses}, J.~I., {Mainzer}, A.~K., {Hines},
  D., {Hammel}, H.~B., {Martin-Torres}, F.~J., {Burgdorf}, M., {Merlet}, C.,
  {Line}, M.~R., 2014{\natexlab{a}}. {Mid-infrared spectroscopy of Uranus from
  the Spitzer Infrared Spectrometer: 1. Determination of the mean temperature
  structure of the upper troposphere and stratosphere}. Icarus 243, 494--513.

\bibitem[{{Orton} et~al.(1992){Orton}, {Lacy}, {Achtermann}, {Parmar}, and
  {Blass}}]{orton92}
{Orton}, G.~S., {Lacy}, J.~H., {Achtermann}, J.~M., {Parmar}, P., {Blass},
  W.~E., Dec. 1992. {Thermal spectroscopy of Neptune - The stratospheric
  temperature, hydrocarbon abundances, and isotopic ratios}. Icarus 100,
  541--555.

\bibitem[{{Orton} et~al.(2014{\natexlab{b}}){Orton}, {Moses}, {Fletcher},
  {Mainzer}, {Hines}, {Hammel}, {Martin-Torres}, {Burgdorf}, {Merlet}, and
  {Line}}]{orton14chem}
{Orton}, G.~S., {Moses}, J.~I., {Fletcher}, L.~N., {Mainzer}, A.~K., {Hines},
  D., {Hammel}, H.~B., {Martin-Torres}, J., {Burgdorf}, M., {Merlet}, C.,
  {Line}, M.~R., 2014{\natexlab{b}}. {Mid-infrared spectroscopy of Uranus from
  the Spitzer infrared spectrometer: 2. Determination of the mean composition
  of the upper troposphere and stratosphere}. Icarus 243, 471--493.

\bibitem[{{Poppe}(2015)}]{poppe15}
{Poppe}, A.~R., 2015. {Interplanetary dust influx to the Pluto-Charon system}.
  Icarus 246, 352--359.

\bibitem[{{Poppe}(2016)}]{poppe16}
{Poppe}, A.~R., 2016. {An improved model for interplanetary dust fluxes in the
  outer Solar System}. Icarus 264, 369--386.

\bibitem[{{Porco} et~al.(2006){Porco}, {Helfenstein}, {Thomas}, {Ingersoll},
  {Wisdom}, {West}, {Neukum}, {Denk}, {Wagner}, {Roatsch}, {Kieffer}, {Turtle},
  {McEwen}, {Johnson}, {Rathbun}, {Veverka}, {Wilson}, {Perry}, {Spitale},
  {Brahic}, {Burns}, {Del Genio}, {Dones}, {Murray}, and {Squyres}}]{porco06}
{Porco}, C.~C., {Helfenstein}, P., {Thomas}, P.~C., {Ingersoll}, A.~P.,
  {Wisdom}, J., {West}, R., {Neukum}, G., {Denk}, T., {Wagner}, R., {Roatsch},
  T., {Kieffer}, S., {Turtle}, E., {McEwen}, A., {Johnson}, T.~V., {Rathbun},
  J., {Veverka}, J., {Wilson}, D., {Perry}, J., {Spitale}, J., {Brahic}, A.,
  {Burns}, J.~A., {Del Genio}, A.~D., {Dones}, L., {Murray}, C.~D., {Squyres},
  S., 2006. {Cassini observes the active south pole of Enceladus}. Science 311,
  1393--1401.

\bibitem[{{Prather} et~al.(1978){Prather}, {Logan}, and {McElroy}}]{prather78}
{Prather}, M.~J., {Logan}, J.~A., {McElroy}, M.~B., 1978. {Carbon monoxide in
  Jupiter's upper atmosphere - an extraplanetary source}. \apj 223, 1072--1081.

\bibitem[{{Press} et~al.(1992){Press}, {Teukolsky}, {Vetterling}, and
  {Flannery}}]{press92}
{Press}, W.~H., {Teukolsky}, S.~A., {Vetterling}, W.~T., {Flannery}, B.~P.,
  1992. Numerical Recipes in {FORTRAN}, 2nd Edition. Cambridge University
  Press, New York.

\bibitem[{{Prinn} and {Barshay}(1977)}]{prinn77}
{Prinn}, R.~G., {Barshay}, S.~S., 1977. Carbon monoxide on {Jupiter} and
  implications for atmospheric convection. Science 198, 1031--1034.

\bibitem[{{Pryor} et~al.(1994){Pryor}, {Na}, and {Gladstone}}]{pryor94}
{Pryor}, W.~R., {Na}, C.~Y., {Gladstone}, G.~R., 1994. {How will dust from
  Shoemaker-Levy 9 alter Jupiter's stratospheric aerosol populations?} \grl 21,
  1079--1082.

\bibitem[{{Reutt} et~al.(1986){Reutt}, {Wang}, {Lee}, and {Shirley}}]{reutt86}
{Reutt}, J.~E., {Wang}, L.~S., {Lee}, Y.~T., {Shirley}, D.~A., 1986. {Molecular
  beam photoelectron spectroscopy and femtosecond intramolecular dynamics of
  H$_2$O$^{+}$ and D$_2$O$^{+}$}. J. Chem. Phys. 85, 6928--6939.

\bibitem[{{Rezac} et~al.(2014){Rezac}, {de Val-Borro}, {Hartogh},
  {Cavali{\'e}}, {Jarchow}, {Rengel}, and {Dobrijevic}}]{rezac14}
{Rezac}, L., {de Val-Borro}, M., {Hartogh}, P., {Cavali{\'e}}, T., {Jarchow},
  C., {Rengel}, M., {Dobrijevic}, M., 2014. {New determination of the HCN
  profile in the stratosphere of Neptune from millimeter-wave spectroscopy}.
  \aap 563, A4.

\bibitem[{{Rizk} and {Hunten}(1990)}]{rizk90}
{Rizk}, B., {Hunten}, D.~M., 1990. {Solar heating of the Uranian mesopause by
  dust of ring origin}. Icarus 88, 429--447.

\bibitem[{{Rizk} et~al.(1991){Rizk}, {Hunten}, and {Engel}}]{rizk91}
{Rizk}, B., {Hunten}, D.~M., {Engel}, S., 1991. {Effects of size-dependent
  emissivity on maximum temperatures during micrometeorite entry}. \jgr 96,
  1303--1314.

\bibitem[{{Romani} et~al.(2008){Romani}, {Jennings}, {Bjoraker}, {Sada},
  {McCabe}, and {Boyle}}]{romani08}
{Romani}, P.~N., {Jennings}, D.~E., {Bjoraker}, G.~L., {Sada}, P.~V., {McCabe},
  G.~H., {Boyle}, R.~J., 2008. {Temporally varying ethylene emission on
  Jupiter}. Icarus 198, 420--434.

\bibitem[{{Rosenqvist} et~al.(1992){Rosenqvist}, {Lellouch}, {Romani},
  {Paubert}, and {Encrenaz}}]{rosenqvist92}
{Rosenqvist}, J., {Lellouch}, E., {Romani}, P.~N., {Paubert}, G., {Encrenaz},
  T., 1992. {Millimeter-wave observations of Saturn, Uranus, and Neptune: CO
  and HCN on Neptune}. \apjl 392, L99--L102.

\bibitem[{{Roux} et~al.(2008){Roux}, {Temprado}, {Chickos}, and
  {Nagano}}]{roux08}
{Roux}, M.~V., {Temprado}, M., {Chickos}, J.~S., {Nagano}, Y., 2008. Critically
  evaluated thermochemical properties of polycyclic aromatic hydrocarbons.
  Journal of Physical and Chemical Reference Data 37, 1855--1996.

\bibitem[{{Sada} et~al.(1998){Sada}, {Bjoraker}, {Jennings}, {McCabe}, and
  {Romani}}]{sada98}
{Sada}, P.~V., {Bjoraker}, G.~L., {Jennings}, D.~E., {McCabe}, G.~H., {Romani},
  P.~N., 1998. {Observations of CH$_{4}$, C$_{2}$H$_{6}$, and C$_{2}$H$_{2}$ in
  the stratosphere of Jupiter}. Icarus 136, 192--201.

\bibitem[{{Sada} et~al.(2005){Sada}, {Bjoraker}, {Jennings}, {Romani}, and
  {McCabe}}]{sada05}
{Sada}, P.~V., {Bjoraker}, G.~L., {Jennings}, D.~E., {Romani}, P.~N., {McCabe},
  G.~H., 2005. {Observations of C$_{2}$H$_{6}$ and C$_{2}$H$_{2}$ in the
  stratosphere of Saturn}. Icarus 173, 499--507.

\bibitem[{{Sada} et~al.(1996){Sada}, {McCabe}, {Bjoraker}, {Jennings}, and
  {Reuter}}]{sada96c2h6}
{Sada}, P.~V., {McCabe}, G.~H., {Bjoraker}, G.~L., {Jennings}, D.~E., {Reuter},
  D.~C., 1996. {$^{13}$C-Ethane in the atmospheres of Jupiter and Saturn}. \apj
  472, 903--907.

\bibitem[{{Sander} et~al.(2011){Sander}, {Friedl}, {Abbatt}, {Barker},
  {Burkholder}, {Golden}, {Kolb}, J., {Moortgat}, {Wine}, {Huie}, and
  {Orkin}}]{sander11}
{Sander}, S.~P., {Friedl}, R.~R., {Abbatt}, J.~P.~D., {Barker}, J.~R.,
  {Burkholder}, J.~B., {Golden}, D.~M., {Kolb}, C.~E., J., K.~M., {Moortgat},
  G.~K., {Wine}, P.~H., {Huie}, R.~E., {Orkin}, V.~L., 2011. Chemical kinetics
  and photochemical data for use in atmospheric studies. {JPL} Publication
  10-6.

\bibitem[{{Schulz} et~al.(1999){Schulz}, {Encrenaz}, {B{\'e}zard}, {Romani},
  {Lellouch}, and {Atreya}}]{schulz99}
{Schulz}, B., {Encrenaz}, T., {B{\'e}zard}, B., {Romani}, P.~N., {Lellouch},
  E., {Atreya}, S.~K., Oct. 1999. {Detection of C\_2H\_4 in Neptune from
  ISO/PHT-S observations}. Astron. Astrophys 350, L13--L17.

\bibitem[{{Shinagawa} and {Waite}(1989)}]{shinagawa89}
{Shinagawa}, H., {Waite}, J.~H., 1989. {The ionosphere of Neptune}. \grl 16,
  945--947.

\bibitem[{{Sinclair} et~al.(2013){Sinclair}, {Irwin}, {Fletcher}, {Moses},
  {Greathouse}, {Friedson}, {Hesman}, {Hurley}, and {Merlet}}]{sinclair13}
{Sinclair}, J.~A., {Irwin}, P.~G.~J., {Fletcher}, L.~N., {Moses}, J.~I.,
  {Greathouse}, T.~K., {Friedson}, A.~J., {Hesman}, B., {Hurley}, J., {Merlet},
  C., 2013. {Seasonal variations of temperature, acetylene and ethane in
  Saturn's atmosphere from 2005 to 2010, as observed by Cassini-CIRS}. Icarus
  225, 257--271.

\bibitem[{{Smith} et~al.(1983){Smith}, {Shemansky}, {Holberg}, {Broadfoot},
  {Sandel}, and {McConnell}}]{smith83}
{Smith}, G.~R., {Shemansky}, D.~E., {Holberg}, J.~B., {Broadfoot}, A.~L.,
  {Sandel}, B.~R., {McConnell}, J.~C., 1983. {Saturn's upper atmosphere from
  the Voyager 2 EUV solar and stellar occultations}. \jgr 88, 8667--8678.

\bibitem[{{Stern}(1996)}]{stern96}
{Stern}, S.~A., 1996. {Signatures of collisions in the Kuiper Disk}. \aap 310,
  999--1010.

\bibitem[{{Strobel} et~al.(1990){Strobel}, {Summers}, {Herbert}, and
  {Sandel}}]{strobel90}
{Strobel}, D.~F., {Summers}, M.~E., {Herbert}, F., {Sandel}, B.~R., 1990. {The
  photochemistry of methane in the atmosphere of Triton}. \grl 17,
  1729--1731729--17322.

\bibitem[{{Strobel} and {Yung}(1979)}]{strobel79}
{Strobel}, D.~F., {Yung}, Y.~L., 1979. {The Galilean satellites as a source of
  CO in the Jovian upper atmosphere}. Icarus 37, 256--263.

\bibitem[{{Teanby} and {Irwin}(2013)}]{teanby13}
{Teanby}, N.~A., {Irwin}, P.~G.~J., 2013. {An external origin for carbon
  monoxide on Uranus from Herschel/SPIRE?} \apjl 775, L49.

\bibitem[{{Tseng} et~al.(2010){Tseng}, {Ip}, {Johnson}, {Cassidy}, and
  {Elrod}}]{tseng10}
{Tseng}, W.-L., {Ip}, W.-H., {Johnson}, R.~E., {Cassidy}, T.~A., {Elrod},
  M.~K., 2010. {The structure and time variability of the ring atmosphere and
  ionosphere}. Icarus 206, 382--389.

\bibitem[{{Visscher} and {Fegley}(2005)}]{visscher05}
{Visscher}, C., {Fegley}, Jr., B., 2005. Chemical constraints on the water and
  total oxygen abundances in the deep atmosphere of {Saturn}. \apj 623,
  1221--1227.

\bibitem[{{Visscher} and {Moses}(2011)}]{visscher11}
{Visscher}, C., {Moses}, J.~I., 2011. Quenching of carbon monoxide and methane
  in the atmospheres of cool brown dwarfs and hot {Jupiters}. \apj 738, 72.

\bibitem[{{Visscher} et~al.(2010b){Visscher}, {Moses}, and
  {Saslow}}]{visscher10co}
{Visscher}, C., {Moses}, J.~I., {Saslow}, S.~A., 2010b. The deep water
  abundance on {Jupiter}: {New} constraints from thermochemical kinetics and
  diffusion modeling. Icarus 209, 602--615.

\bibitem[{{von Zahn} et~al.(1999){von Zahn}, {Gerding}, {H{\"o}ffner},
  {McNeil}, and {Murad}}]{vonzahn99}
{von Zahn}, U., {Gerding}, M., {H{\"o}ffner}, J., {McNeil}, W.~J., {Murad}, E.,
  1999. {Iron, calcium, and potassium atom densities in the trails of Leonids
  and other meteors: Strong evidence for differential ablation}. Meteoritics
  and Planetary Science 34, 1017--1027.

\bibitem[{{Vondrak} et~al.(2008){Vondrak}, {Plane}, {Broadley}, and
  {Janches}}]{vondrak08}
{Vondrak}, T., {Plane}, J.~M.~C., {Broadley}, S., {Janches}, D., 2008. A
  chemical model of meteoric ablation. Atmospheric Chemistry \& Physics 8,
  7015--7031.

\bibitem[{{Wagener} et~al.(1985){Wagener}, {Caldwell}, {Owen}, {Kim},
  {Encrenaz}, and {Combes}}]{wagener85}
{Wagener}, R., {Caldwell}, J., {Owen}, T., {Kim}, S.-J., {Encrenaz}, T.,
  {Combes}, M., 1985. {The Jovian stratosphere in the ultraviolet}. Icarus 63,
  222--236.

\bibitem[{{Waite} and {Cravens}(1987)}]{waite87}
{Waite}, Jr., J.~H., {Cravens}, T.~E., 1987. {Current review of the Jupiter,
  Saturn, and Uranus ionospheres}. Advances in Space Research 7, 119--134.

\bibitem[{{Wang} et~al.(2015){Wang}, {Gierasch}, {Lunine}, and
  {Mousis}}]{wang15}
{Wang}, D., {Gierasch}, P.~J., {Lunine}, J.~I., {Mousis}, O., 2015. {New
  insights on Jupiter's deep water abundance from disequilibrium species}.
  Icarus 250, 154--164.

\bibitem[{{Wang} et~al.(2016){Wang}, {Lunine}, and {Mousis}}]{wang16}
{Wang}, D., {Lunine}, J.~I., {Mousis}, O., 2016. {Modeling the disequilibrium
  species for Jupiter and Saturn: Implications for Juno and Saturn entry
  probe}. Icarus 276, 21--38.

\bibitem[{{Warren}(1984)}]{warren84}
{Warren}, S.~G., 1984. Optical constants of ice from the ultraviolet to the
  microwave. Applied Optics 23, 1206--1225.

\bibitem[{{Yamamoto} and {Mukai}(1998)}]{yamamoto98}
{Yamamoto}, S., {Mukai}, T., 1998. {Dust production by impacts of interstellar
  dust on Edgeworth-Kuiper Belt objects}. \aap 329, 785--791.

\bibitem[{{Yelle} et~al.(2001){Yelle}, {Griffith}, and {Young}}]{yelle01}
{Yelle}, R.~V., {Griffith}, C.~A., {Young}, L.~A., 2001. {Structure of the
  Jovian stratosphere at the Galileo probe entry site}. Icarus 152, 331--346.

\bibitem[{{Yelle} et~al.(1993){Yelle}, {Herbert}, {Sandel}, {Vervack}, and
  {Wentzel}}]{yelle93}
{Yelle}, R.~V., {Herbert}, F., {Sandel}, B.~R., {Vervack}, Jr., R.~J.,
  {Wentzel}, T.~M., Jul. 1993. {The distribution of hydrocarbons in Neptune's
  upper atmosphere}. Icarus 104, 38--59.

\bibitem[{{Yelle} et~al.(1996){Yelle}, {Young}, {Vervack}, {Young}, {Pfister},
  and {Sandel}}]{yelle96}
{Yelle}, R.~V., {Young}, L.~A., {Vervack}, R.~J., {Young}, R., {Pfister}, L.,
  {Sandel}, B.~R., 1996. {Structure of Jupiter's upper atmosphere: Predictions
  for Galileo}. \jgr 101, 2149--2162.

\bibitem[{{Yung} et~al.(1984){Yung}, {Allen}, and {Pinto}}]{yung84}
{Yung}, Y.~L., {Allen}, M., {Pinto}, J.~P., 1984. Photochemistry of the
  atmosphere of {Titan}: Comparison between model and observations. \apjss 55,
  465--506.

\bibitem[{{Zahnle}(1996)}]{zahnle96}
{Zahnle}, K., 1996. {Dynamics and chemistry of SL9 plumes}. In: {K.~S.~Noll,
  H.~A.~Weaver, \& P.~D.~Feldman} (Ed.), IAU Colloq. 156: The Collision of
  Comet Shoemaker-Levy 9 and Jupiter. pp. 183--212.

\bibitem[{{Zahnle} et~al.(2003){Zahnle}, {Schenk}, {Levison}, and
  {Dones}}]{zahnle03}
{Zahnle}, K., {Schenk}, P., {Levison}, H., {Dones}, L., 2003. {Cratering rates
  in the outer Solar System}. Icarus 163, 263--289.

\end{thebibliography}






\end{document}